\documentclass[a4paper,11pt]{article}
\usepackage{jheppub}
\usepackage{lineno}

\graphicspath{{./fig/}}

\usepackage{comment}
\usepackage{xcolor}
\newcommand{\RNum}[1]{\uppercase\expandafter{\romannumeral #1\relax}}

\usepackage{soul} % strike through

\newcommand{\BoldVec}[1]{\mathchoice%
  {\mbox{\boldmath $\displaystyle     #1$}}%
  {\mbox{\boldmath $\textstyle        #1$}}%
  {\mbox{\boldmath $\scriptstyle      #1$}}%
  {\mbox{\boldmath $\scriptscriptstyle#1$}}%
}

\def\half{{\textstyle{1\over2}}}

\def\onethird{{\textstyle{1\over3}}}

\def\threefourth{{\textstyle{3\over4}}}

\newcommand{\HH}{{\cal H}}
\newcommand{\RR}{{\cal R}}

\newcommand{\Mpl}{M_{\rm Pl}}

\newcommand{\OmGW}{\Omega_{\rm GW}}

\newcommand{\xx}{\BoldVec{x}{}}
\newcommand{\kk}{\BoldVec{k}{}}

\newcommand{\vv}{\BoldVec{v}{}}

\newcommand{\ee}{\BoldVec{e}{}}
\newcommand{\dd}{{\rm \, d}}

\newcommand{\tfin}{t_{\rm fin}}

\newcommand{\bra}[1]{\langle #1\rangle}

\newcommand{\Eq}[1]{Eq.~(\ref{#1})}
\newcommand{\EEq}[1]{Equation~(\ref{#1})}

\newcommand{\Eqs}[2]{Eqs.~(\ref{#1}) and (\ref{#2})}
\newcommand{\Eqss}[2]{Eqs.~(\ref{#1})--(\ref{#2})}
\newcommand{\Fig}[1]{Fig.~\ref{#1}}
\newcommand{\Tab}[1]{Tab.~\ref{#1}}
\newcommand{\Figs}[2]{Figs.~\ref{#1} and \ref{#2}}
\newcommand{\FFig}[1]{Figure~\ref{#1}}
\newcommand{\Sec}[1]{Sec.~\ref{#1}}
\newcommand{\Secs}[2]{Secs.~\ref{#1} and \ref{#2}}
\newcommand{\Secss}[2]{Secs.~\ref{#1}--\ref{#2}}
\newcommand{\App}[1]{App.~\ref{#1}}

\newcommand{\vw}{v_{\rm{w}}}
\newcommand{\cs}{c_{\rm{s}}}

\newcommand{\damp}{{e}}

\newcommand{\GW}{{\rm{GW}}}

\def\be{\begin{equation}}
\def\ee{\end{equation}}
\def\bea{\begin{eqnarray}}
\def\eea{\end{eqnarray}}
\usepackage[normalem]{ulem}

\arxivnumber{2409.03651} % Only if you have one
\subheader{KOBE-COSMO-24-03, TUM-HEP-1522/24, DESY-24-131, CERN-TH-2025-072}

\title{Gravitational waves from first-order phase transitions: from weak to strong}

\author[a,b]{Chiara Caprini,}
\author[c]{Ryusuke Jinno,}
\author[d]{Thomas Konstandin,}
\author[a,1]{Alberto Roper Pol,\note{Corresponding author: \href{mailto:alberto.roperpol@unige.ch}{alberto.roperpol@unige.ch}}}
\author[e]{Henrique Rubira,}
\author[d,f,2]{Isak Stomberg\note{Corresponding author: \href{mailto:isak.stomberg@desy.de}{isak.stomberg@ific.uv.es}}}

\affiliation[a]{D\'epartement de Physique Th\'eorique, Universit\'e de Gen\`eve,
CH-1211 Gen\`eve, Switzerland}
\affiliation[b]{Theoretical Physics Department, CERN, CH-1211 Gen\`eve, Switzerland}
\affiliation[c]{Department of Physics, Kobe University, Kobe 657-8501, Japan}
\affiliation[d]{Deutsches Elektronen-Synchrotron DESY, Notkestr. 85, 22607 Hamburg, Germany}
\affiliation[e]{Physik Department T31, Technische Universit\"at M\"unchen
James-Franck-Stra{\ss}e 1, D-85748 Garching, Germany}
\affiliation[f]{IFIC, Universitat de València-CSIC, C/ Catedràtico José Beltràn 2, E-46980, Paterna, Spain}

\emailAdd{chiara.caprini@cern.ch}
\emailAdd{jinno@phys.sci.kobe-u.ac.jp}
\emailAdd{thomas.konstandin@desy.de}
\emailAdd{alberto.roperpol@unige.ch}
\emailAdd{henrique.rubira@tum.de}
\emailAdd{isak.stomberg@ific.uv.es}

\begin{document}

\abstract{
We study the generation of gravitational waves (GWs) during a cosmological first-order phase transition (PT)
using the recently introduced Higgsless approach to numerically simulate the fluid motion induced by the PT.
We present for the first time GW spectra sourced by bulk fluid motion in the aftermath of
strong first-order PTs ($\alpha = 0.5$), alongside
weak ($\alpha = 0.0046$) and intermediate ($\alpha = 0.05$) PTs, previously considered in the literature.
We find that, for intermediate and strong
PTs, the kinetic energy in our simulations 
decays,
following a power law in time. 
The decay is potentially determined
by non-linear dynamics and hence related to the production of vorticity.
We show that the assumption that the source is stationary in time,
characteristic of compressional motion
in the linear regime (sound waves),
agrees with our numerical results 
for weak PTs, since in this case
the
kinetic energy does not decay with time.
We then provide a theoretical framework that
extends the stationary assumption to one
that accounts for the time evolution of the source: 
as a result, the GW energy density is no longer linearly increasing with
the source duration,
but proportional to the integral over time
of the squared kinetic energy fraction. 
This effectively reduces the linear growth rate of the GW energy density and allows to account for the period
of transition from the linear to the non-linear regimes of the fluid perturbations.
We validate the novel theoretical model with the results of simulations and
provide templates for the GW spectrum for a broad range of PT parameters. 
}
\maketitle
\flushbottom

\section{Introduction}

The door to a new era with the promise of groundbreaking discovery was opened with the inaugural direct detections, by the LIGO-Virgo collaboration,
of gravitational waves (GWs) emanating from mergers of black holes and neutron stars \cite{LIGOScientific:2016aoc,LIGOScientific:2016sjg,LIGOScientific:2017vwq}.
The forthcoming observing run by the LIGO-Virgo-KAGRA (LVK) collaboration are expected to accumulate more events~\cite{KAGRA:2021vkt}.
Efforts among Pulsar Timing Array (PTA) collaborations have furthermore unveiled convincing evidence of a stochastic gravitational wave background
(SGWB) at nano-Hertz frequencies~\cite{EPTA:2023fyk,NANOGrav:2023gor,Reardon:2023gzh,Xu:2023wog}.
While a compelling
candidate for the source of this radiation is the superposition of supermassive black hole mergers, implying an astrophysical origin,
it is important to point out that
primordial sources of cosmological origin
can also explain the observed signal
(see, e.g., Refs.~\cite{EPTA:2023xxk,NANOGrav:2023hvm}).
These breakthroughs in GW detection give us ears to astrophysical events and cosmological history inaccessible through other means of observation.

Looking into the 2030s, the launch of the Laser Interferometer Space Antenna (LISA) mission~\cite{LISA:2017pwj,Colpi:2024xhw}, designed to probe
GWs in the unexplored milli-Hertz frequency band,
is setting the stage for a potential
overhaul of modern cosmology~\cite{Seoane:2013qna,amaroseoane2017laser,LISACosmologyWorkingGroup:2022jok}.
Until the time of launch, joint strides in data analysis techniques and theoretical progress are necessary to leverage the full potential of the LISA mission once it flies.
It is in this spirit that
studies on cosmological SGWB production gain their motivation.

Cosmological sources of GWs that can be explored by LISA include inflation, particle production, topological defects, and primordial black holes throughout different stages of the expansion history of the Universe (see Refs.~\cite{Maggiore:2018sht,Caprini:2018mtu} and references therein).
Of particular interest is the phenomenon of first-order phase transitions (PTs)~\cite{Witten:1984rs} that could have occurred in the early Universe.
While the Standard Model (SM) predicts a crossover at the electroweak scale~\cite{Kajantie:1996mn,Gurtler:1997hr,Csikor:1998eu}, many theories beyond the Standard Model accommodate a first-order electroweak PT (see Ref.~\cite{Caprini:2019egz} and references therein).
During such a transition, the order parameter initially becomes trapped in a false vacuum expectation
value in the symmetric phase.
Subsequently, vacuum or thermal fluctuations locally induce a transition to the true vacuum in the broken phase, forming tiny seeds of bubbles~\cite{Coleman:1977py, Linde:1980tt, Steinhardt:1981ct}.
The released vacuum energy drives the expansion of these bubbles, which eventually collide with each other, generating anisotropic stresses in the energy distribution and thus sourcing GWs. This process is highly non-thermal, suggesting that the baryon asymmetry of the Universe might have its origin in it~\cite{Kuzmin:1985mm,Cohen:1993nk,Rubakov:1996vz,Riotto:1999yt,Morrissey:2012db}.
While bubble collisions themselves are an important source of GWs~\cite{Kosowsky:1991ua,Kosowsky:1992rz,Kosowsky:1992vn,Kamionkowski:1993fg,Caprini:2007xq,Huber:2008hg,Konstandin:2017sat,Jinno:2017fby,Cutting:2018tjt}, it has been shown in Ref.~\cite{Hindmarsh:2013xza} that  compressional
fluid motion in the linearized regime (i.e., sound waves)
induced in the primordial plasma by the scalar walls often dominates the GW production for PTs when the broken-phase bubbles do not run away,\footnote{These results have been found in the absence of magnetic fields
and for small fluid perturbations. A
primordial magnetic field generated or present during the
PT, and/or the production of non-linear fluid perturbations,
can efficiently induce vortical motion
\cite{Quashnock:1988vs,Brandenburg:1996fc} due to the
high conductivity and Reynolds number of the primordial plasma \cite{Ahonen:1996nq,Arnold:2000dr}.}
which is expected to be
the case unless the PT is dominated by vacuum, e.g., for supercooled PTs \cite{vonHarling:2017yew,Bodeker:2017cim,Kobakhidze:2017mru,Caprini:2019egz}.
This occurs when the friction exerted on the bubble
walls by the fluid particles is strong enough to balance
the vacuum energy released and, hence, the bubbles
reach a terminal velocity \cite{Ignatius:1993qn,Huber:2008hg, Espinosa:2010hh, Hindmarsh:2013xza,Bodeker:2017cim}.

The production of GWs from fluid perturbations can be
decomposed in two contributions: sound waves (or, for generality, acoustic/compressional turbulence) \cite{Hindmarsh:2013xza,Hindmarsh:2015qta,Hindmarsh:2016lnk,Hindmarsh:2017gnf,Niksa:2018ofa,Hindmarsh:2019phv,RoperPol:2019wvy,Jinno:2020eqg,Dahl:2021wyk,Jinno:2022mie,RoperPol:2023dzg,Sharma:2023mao,Dahl:2024eup} and vortical turbulence \cite{Kosowsky:1992rz,Kosowsky:2001xp,Gogoberidze:2007an,Caprini:2009fx,Caprini:2009yp,Niksa:2018ofa,RoperPol:2018sap,RoperPol:2019wvy,Kahniashvili:2020jgm,Brandenburg:2021tmp,Brandenburg:2021bvg,RoperPol:2021xnd,RoperPol:2022iel,Auclair:2022jod,Sharma:2022ysf}.
While analytical modeling of GW production
from fluid perturbations is important~\cite{Caprini:2009fx,Hindmarsh:2016lnk,Jinno:2016vai,Jinno:2017fby,Hindmarsh:2019phv,Niksa:2018ofa,RoperPol:2022iel,RoperPol:2023dzg}, numerical simulations are essential for a comprehensive understanding of the entire process.
It is believed that after a first-order PT, the fluid motion initially manifests as compressional motion
and then develops non-linearly,
allowing for the formation of shocks and vorticity \cite{Pen:2015qta,Dahl:2021wyk,Dahl:2024eup},
and the subsequent development of turbulence \cite{Brandenburg:2017rnt,Brandenburg:2017neh}. 
The non-linear evolution is inevitable due to  the large Reynolds number of the fluid in the early Universe \cite{Arnold:2000dr}.
The transition from the fully compressional
to the vortical turbulence regime can be especially important in strong transitions,  when non-linearities can play an important role.

Currently, large-scale simulations have been performed by the Helsinki-Sussex group, which numerically solve a coupled scalar field-fluid system~\cite{Hindmarsh:2013xza,Hindmarsh:2015qta,Hindmarsh:2017gnf,Cutting:2019zws}.
First-order PTs in the early Universe exhibit a significant hierarchy: this is the hierarchy between the typical scales inherent in the order parameter field (including the thickness of the walls) and those in the cosmological fluid (including the bubble size and the sound-shell thickness). For the electroweak PT, the hierarchical separation can be as large as $\Mpl / T \sim 10^{16}$,
being $\Mpl$ the Planck mass.
This fact naturally leads to the idea of the Higgsless scheme proposed by part of the authors of the present paper~\cite{Jinno:2020eqg,Jinno:2022mie}.
In this scheme, the microphysics of the wall is introduced as a non-dynamical (although space and time dependent)
energy-injecting boundary condition within the bag equation of state~\cite{Espinosa:2010hh,Giese:2020rtr,Giese:2020znk},
and the bubble walls are assumed to have reached a terminal velocity,
such that the fluid perturbations reach a self-similar
solution in a very short time scale (much 
shorter than the time scale for collisions) \cite{Espinosa:2010hh}.
See also Refs.~\cite{Bodeker:2017cim,Gouttenoire:2021kjv,Azatov:2023xem} for discussions on the bubble wall terminal velocity.
As a result, the scheme is able to capture the macroscopic dynamics necessary for GW production without being 
required to also solve for the hierarchically smaller scales.

In this paper, we explore the previously uncharted realm of GWs
sourced by fluid perturbations induced in
strong first-order PTs.\footnote{We clarify that hereby, by strong transitions, we refer to $\alpha = 0.5$ (see \Eq{alpha} for definition), still far from the supercooled regime where the scalar field potential energy dominates the energy content of the Universe, allowing for $\alpha \gg 1$ \cite{Cutting:2018tjt,Ellis:2019oqb}.}
We also update results for weak and intermediate transitions and compare with other results in the literature \cite{Hindmarsh:2017gnf,Hindmarsh:2019phv,Jinno:2022mie}.
We perform approximately 1000 simulations involving a parameter scan over wall 
velocities $\vw \in [0.32, 0.8]$ in increments of 0.04
for weak ($\alpha = 0.0046$), intermediate ($\alpha = 0.05$), and strong ($\alpha = 0.5$) PTs, 
using the Higgsless approach, to clarify the dependence on the underlying physical quantities of various characteristics of the GW spectrum, especially
focusing on its overall integrated amplitude.
We then assess the long-term evolution of the system, and
discuss indications of non-linearities in the fluid bulk motion
and their impact on the GW production.
A thorough assessment of the vorticity production possibly associated to the non-linearities deserves a dedicated study; here we simply
include an appendix presenting preliminary results on
the longitudinal and transverse components of the velocity field spectra.

The outline of the paper is as follows.
In \Sec{sec:setup}, we review the GW production in a PT.
We first describe the bubble nucleation history and the conservation
laws of the fluid perturbations in
\Secs{nucleation_hist}{hydro_eqs}.
Then, \Sec{GW_production} presents the
prescription used to compute the GW spectrum
at present time from the numerical results
(more detail is given in \App{sec:GW_prod}).
In \Sec{GW_sw}, we review the GW sourcing under the stationary
assumption,
characteristic of the sound-shell model (which assumes 
linear fluid perturbations) \cite{Hindmarsh:2016lnk,Hindmarsh:2019phv,RoperPol:2023dzg,Sharma:2023mao}, and used to interpret
numerical results of GW generation in weak and intermediate PTs
\cite{Hindmarsh:2013xza,Hindmarsh:2015qta,Hindmarsh:2017gnf,Jinno:2020eqg,Jinno:2022mie}.
We then
present
a novel model in \Sec{sw_extended} that extends the unequal time correlator (UETC)
to a locally stationary UETC, allowing us to account for
the effect of the source decay.
In \Sec{sec:expansion}, we extend this model 
to include the effect of the expansion
of the Universe.
In \Sec{sec:numerical_setup},
we discuss updates to the previous Higgsless simulations \cite{Jinno:2022mie}, and summarize
the physical and numerical setup of the simulations.
In \Sec{sec:results}, we discuss the numerical results.
We first present a convergence analysis for the kinetic
energy and the integrated GW amplitude in \Sec{sec:convergence}, which is combined with
a convergence analysis of
the potential effects of under-resolving the fluid perturbations at the level of the bubble profiles,
presented in \App{sec:kinetic_ed}.
The result of these convergence analyses is an estimate
of the kinetic energy fraction
when the entire volume is converted to the broken phase, which is then compared
to the single-bubble kinetic energy fraction commonly used in GW studies.
We then analyse in \Sec{decay_K2} the time evolution of the kinetic energy fraction showing that it can be fitted by a decaying power law in time,
accurately reproducing the simulations.
To investigate the origin of the observed decay, we
briefly discuss the development of vorticity
in \App{sec:vort}, but we leave a detailed treatment of the onset of vorticity for future work.
The growth of the GW integrated amplitude with the source
duration is presented in \Sec{GW_spec_time}, together with estimates
of the GW efficiency $\tilde \Omega_{\rm GW}$.
We investigate the GW spectral shape in \Sec{sec:shape} and
fit the numerical results to
a double broken power law, providing
estimates for the positions of the relevant
spectral scales.
We summarize our results in \Sec{sec:summary}, and present a template for the GW spectrum,
which we will make publicly available via \href{https://github.com/CosmoGW/cosmoGW}{\sc CosmoGW} \cite{cosmogw}.

%%%%%%%%%%%%%%%%%%%%%%%%%%%%%%%%%%%%%%%%%%%%%%%%%%%%%%%%%%%%%%%%%%%%%%%%%%
\section{Gravitational waves from a phase transition\label{sec:setup}}
%%%%%%%%%%%%%%%%%%%%%%%%%%%%%%%%%%%%%%%%%%%%%%%%%%%%%%%%%%%%%%%%%%%%%%%%%%

In this section, we describe the production of gravitational waves from the fluid
perturbations induced in the primordial plasma by the nucleation
of bubbles in a first-order PT.
In \Sec{nucleation_hist}, we review the nucleation history of
broken-phase bubbles that is used
in the Higgsless approach, and in \Sec{hydro_eqs}, we review the
the relativistic hydrodynamic equations.
In \Sec{GW_production}, we 
describe how the GW generation is tackled within the Higgsless simulations, and discuss its applicability (see also App~\ref{sec:GW_prod} for details).
In \Sec{GW_sw}, we 
review the production of GWs under the 
assumption that the anisotropic stress UETC is stationary. This is usually assumed in the literature 
when the source of GWs corresponds to sound waves
\cite{Hindmarsh:2013xza,Hindmarsh:2015qta,Hindmarsh:2016lnk,Caprini:2015zlo,Hindmarsh:2017gnf,Hindmarsh:2019phv,Caprini:2019egz,RoperPol:2023dzg}. In \Sec{sw_extended}, we propose an extension of the stationary model to a locally stationary UETC,
allowing to account for the dynamics of a source that is decaying in time.
We also provide a proxy to extend our results to the expanding Universe in \Sec{sec:expansion}.
We validate the applicability of the proposed model
to the integrated GW amplitude in \Sec{sec:results}
against the results from
our numerical simulations, described in \Sec{sec:numerical_setup}.

%%%%%%%%%%%%%%%%%%%%%%%%%%%%%%%%%%%%%%%%%%%%%%%%%%%%%%%%%%%%%%%%%%%%%%%%%%
\subsection{Bubble nucleation histories}
\label{nucleation_hist}
%%%%%%%%%%%%%%%%%%%%%%%%%%%%%%%%%%%%%%%%%%%%%%%%%%%%%%%%%%%%%%%%%%%%%%%%%%

In the Higgsless approach \cite{Jinno:2022mie}, a fundamental assumption is that the
broken-phase bubbles reach a terminal wall velocity $\vw$ due to the friction exerted by the fluid particles
\cite{Espinosa:2010hh,Bodeker:2017cim,Gouttenoire:2021kjv,Azatov:2023xem},
such that $\vw$ can be prescribed as an input of the
simulations. This enables the construction of \emph{bubble nucleation histories}, encompassing nucleation times and locations, as well as the \emph{predetermined} expansion of the bubbles.
We assume an exponentially increasing-in-time probability of bubble nucleation,
$P (t) \simeq P_n \exp \left[\beta\left(t-t_n \right)\right]$,
where $\beta$ determines the usual rate of bubble nucleation evaluated at the nucleation time $t_n$, such that the action $S$ has been Taylor-expanded around this time
\cite{Caprini:2019egz}.
The time dependence is hereby inherited from the 
temperature dependence of the tunnelling action $S/T$ and the fact that the temperature scales inversely with the scale factor
\be
\beta = \frac{d}{dt} \left(\frac{S}{T}\right)\biggr|_{t = t_n} = - H\, T \frac{d}{dT} \left(\frac{S}{T}\right)\biggr|_{T = T_n} \, ,
\ee
where $H$ is the Hubble scale.
A detailed description of how such bubble nucleation histories are constructed is found in Ref.~\cite{Jinno:2022mie},
and examples of how modified bubble nucleation histories can be constructed are found in Refs.~\cite{Jinno:2021ury,Blasi:2023rqi}.
We note that in previous numerical work
\cite{Hindmarsh:2015qta,Hindmarsh:2017gnf,Cutting:2019zws}, bubbles
are nucleated simultaneously.
This is expected to have an impact on the spectral peak and the amplitude, but not on the spectral
shape \cite{Weir:2016tov,Hindmarsh:2017gnf}.

\subsection{Relativistic hydrodynamic equations}
\label{hydro_eqs}

The relativistic hydrodynamic equations of motion are derived from
the conservation
of the energy-momentum tensor $T^{\mu \nu}$, which, in Minkowski space-time reads
\begin{equation}
\partial_\mu T^{\mu \nu}=0\,.
\end{equation}

These equations of motion hold during the PT under the assumption that the
duration of the transition is much shorter than a Hubble time, i.e., $\beta/ H_\ast \gg 1$, which allows to neglect the expansion of the Universe.
They can also be applied
to the fluid motion after the PT ends if the fluid is dominated
by radiation particles: indeed, the conservation laws then become conformally invariant and
hence reduce 
to those in Minkowski space-time by a conformal transformation \cite{Brandenburg:1996fc,Subramanian:1997gi,RoperPol:2025lgc}.
We take $T^{\mu \nu}$ to be that of a perfect fluid 
\begin{equation}
T^{\mu \nu}=u^\mu u^\nu w-  \eta^{\mu\nu}
p\rm \, ,
\end{equation}
where $u^\mu=\gamma(1, v^i)$ is the fluid four-velocity, $\gamma=1 / \sqrt{1-v^2}$ is the 
Lorentz factor, $w$ and $p$ are respectively the enthalpy and pressure in the system,
and $\eta^{\mu \nu} = {\rm diag} \{1, -1, -1, -1\}$ is the Minkowski metric tensor.
The equations of motion couple to the state of the vacuum through the bag equation of state, for which we take the
sound speed to be $\cs^2 = 1/3$,
\begin{equation}
p=\frac{1}{3} a T^4-\epsilon \, , \quad w=T \frac{d p}{d T}=\frac{4}{3} a T^4, \label{wp}
\end{equation}
with $T$ being the temperature.
The \emph{bag constant} $\epsilon$ \cite{Espinosa:2010hh}, defined
as the difference in vacuum energy density between the symmetric and broken phases, is thus promoted to a time- and space-dependent quantity 
\begin{equation}
\epsilon(t, \xx)=\left\{\begin{array}{ll}
0 & \,\text {inside bubbles\,, } \\
\epsilon & \,\text {outside bubbles\,, }
\end{array}\right.
\end{equation} 
whose time evolution is uniquely determined for each bubble nucleation history by the terminal wall velocity $\vw$.
We therefore neglect the model-dependent
possibility that heating in the broken phase
can slow down the expansion of the Higgs front
when the latter propagates as a deflagration~\cite{Cutting:2019zws},
and also that strong PTs can potentially
lead to runaway behavior~\cite{Bodeker:2009qy, Bodeker:2017cim, Krajewski:2024gma}. 

The relevant quantity for boundary conditions of the fluid at the Higgs interface is 
the difference in the trace of the energy-momentum tensor, $\theta = w-4p$,
normalized by the mean enthalpy
at the nucleation temperature~\cite{Giese:2020znk},
\be
\alpha = \frac{\Delta \theta}{3 w_n} = \frac{4\epsilon}{3 w_n} \,,  \label{alpha}
\ee
where the second equality holds within the bag equation of state.
We use this quantity in the following to parameterize the strength of the PT in the system \cite{Caprini:2019egz}.
The conservation laws for a relativistic perfect fluid are
\begin{equation}
\begin{aligned}
\partial_t T^{00}+\nabla_i T^{i0} & =0 \,,\\
\partial_t T^{j0} +\nabla_i T^{i j}(T^{\mu0}, \epsilon) & =0 \, .
\end{aligned}
\end{equation}
Note that $T^{i j}(T^{\mu 0}, \epsilon)$ depends on the state of the vacuum such that, effectively, the expanding bubbles perturb the fluid as the latent heat of the vacuum is (locally) deposited. For more details on these equations, we refer the reader to the original Higgsless reference \cite{Jinno:2022mie} and the review \cite{RoperPol:2025lgc}.

%%%%%%%%%%%%%%%%%%%%%%%%%%%%%%%%%%%%%%%%%%%%%%%%%%%%%%%%%%%%%%%%%%%%%%
\subsection{Gravitational wave production}
\label{GW_production}
%%%%%%%%%%%%%%%%%%%%%%%%%%%%%%%%%%%%%%%%%%%%%%%%%%%%%%%%%%%%%%%%%%%%%%

The GW spectrum as a present-time observable
is computed using the following relation
\begin{equation}
    \OmGW (k) = \frac{1}{\rho_{\mathrm{tot}}} \frac{d \rho_{\mathrm{GW}}}{d \ln k} = 
    3 \,  {\cal T}_{\rm GW} \,\, \left(\frac{H_\ast}{\beta}\right)^2 \,\, {\cal I} (k)\,,
    \label{OmGW_Ik}
\end{equation}
where  $\rho_{\text {tot }}$ is the total energy density of the Universe at present time,
${\cal T}_{\rm GW}$
is the transfer function given in \Eq{transf_func},
and $k$ is
the physical wave number,
which can be converted to the frequency as $f = k/(2 \pi)$.

The function ${\cal I}(k)$ represents the spectrum of the stochastic GW signal, and it is given by a double time integral of the anisotropic stress UETC, $E_\Pi(t_1,t_2,k)$ (see \Eq{twop_Pi} in \App{sec:GW_prod}), multiplied by the Green's function of the GW equation.
We have introduced the prefactor $(H_\ast/\beta)^2$ in \Eq{OmGW_Ik} to 
express ${\cal I}$ in normalized time and wave number units
$\tilde t \equiv t \beta$ and $\tilde k \equiv k/\beta$
(see \App{sec:GW_prod}, in particular \Eq{OmGW_aver}, for its derivation from the solution of the GW equation),
\begin{equation}
    {\cal I} (\tilde t_\ast, \tilde t_{\rm fin}, \tilde k) = \frac{k}{2} \, 
    \int_{\tilde t_\ast}^{\tilde t_{\rm fin}} \int_{\tilde t_\ast}^{\tilde t_{\rm fin}}
    E_\Pi (\tilde t_1, \tilde t_2, \tilde k) \cos (\tilde k \tilde t_-)
    \dd \tilde t_1 \dd \tilde t_2\,,\label{Ik}
\end{equation}
where $\tilde t_- = \tilde t_1 - \tilde t_2$.
Here $\tilde t_\ast$ and $\tilde t_{\rm fin}$ denote the initial and final times of action of the GW source. 
In \Eq{Ik}, the expansion of the Universe is not taken into account,
and it therefore holds only if the GW production process lasts for less than one Hubble time, $\tau_{\rm fin} \equiv t_{\rm fin} - t_\ast \ll H_\ast^{-1}$, where $H_\ast$ is the Hubble rate at the PT time $t_\ast$.
The fact that the duration of the PT, determined by the inverse of the nucleation rate $\beta^{-1}$, satisfies $\beta/H_\ast \geq 1$, does not imply that the GW sourcing time is also short. 
In particular, for GWs generated by fluid motion, the time scale of dissipation of the fluid kinetic energy is 
typically much longer than one Hubble time,
as it is set by the kinematic viscosity in the early Universe, which is very small \cite{Arnold:2000dr,Caprini:2009yp}.
However, the Higgless simulations are
performed
in Minkowski space-time, and therefore we neglect the Universe expansion in \Eq{Ik} and the following, to connect with the simulations formalism.
We discuss in \Sec{sec:expansion} how to extend our results to an expanding Universe.

The anisotropic stress UETC $E_\Pi (\tilde t_1, \tilde t_2, \tilde k)$
can be estimated numerically.
In particular, within the Higgsless simulations,
this numerical estimate leads to the following expression of the GW spectrum
(see \Eq{OmGW} and Refs.~\cite{Jinno:2020eqg,Jinno:2022mie}),
given in the normalized units of the simulation
\begin{equation}
    {\cal I}_{\rm sim} (\tilde t_\ast, \tilde t_{\rm fin}, \tilde k) = \frac{\tilde k^3}
    {4 \pi^2 \tilde V} \int_{\Omega_{\tilde k}}
    \frac{\dd \Omega_{\tilde k}}{4 \pi} \Lambda_{ijlm}
    (\hat \kk)
    \bigl[\tilde T_{ij} (\tilde t_\ast,
    \tilde t_{\rm fin}, \tilde q, \tilde \kk) \, \tilde T_{lm}^\ast (\tilde t_\ast,
    \tilde t_{\rm fin}, \tilde q, \tilde \kk) \bigr]_{\tilde q = \tilde k}\,,
    \label{Weinberg}
\end{equation}
being $\tilde V \equiv V \beta^3$ and $\Lambda_{ijlm}$
the transverse and traceless operator
\begin{equation}
    \Lambda_{ijlm} (\hat \kk) = P_{il} P_{jm} - \half P_{ij} P_{lm}\,, \quad {\rm with \ \ } P_{ij} = \delta_{ij} - \hat k_i \hat k_j\,.
\end{equation}
The function $\tilde T_{ij} (\tilde t_\ast,
    \tilde t_{\rm fin},\tilde q, \tilde \kk)$
is computed from the normalized 
stress-energy tensor $\tilde T_{ij} (t, \xx)$, sourcing the GWs, as [see \Eq{Tij_qk}]
\begin{equation}
\label{eq:Fourier transform}
\tilde T_{i j}(\tilde t_\ast, \tilde t_{\rm fin}, \tilde q, \tilde \kk)
= \int_{\tilde t_\ast}^{\tilde t_{\rm fin}} \dd \tilde t \,  e^{i \tilde q \tilde t} \int \dd^3 \tilde \xx \, e^{-i \tilde \kk \cdot \tilde \xx} 
\, \tilde T_{i j}(\tilde t, \tilde \xx) \,, 
\end{equation}
where $\tilde T_{ij} (t, \xx) = w \gamma^2 v_i v_j/\bar \rho$, and $\bar \rho = \threefourth \bar w + \epsilon = \threefourth \, (1 + \alpha) \, \bar w$ is the average total energy density [see \Eqs{wp}{alpha}].

As shown in \App{sec:GW_prod},
using \Eq{Weinberg}
one can directly obtain from the simulation a quantity equivalent to the GW spectrum at present time [see \Eq{OmGW_Ik}], under two assumptions: first,
that the statistical homogeneity and isotropy of the early Universe can be accounted for through an angular integration over shells with fixed wave number $k$ over the simulated
single realization of
the normalized source $\tilde T_{i j}(\tilde t_{\rm init}, \tilde t_{\rm end}, \tilde q, \tilde \kk)$; second, that the source has stopped operating by the
end of the simulation $\tilde t_{\rm end}$.
Indeed, \Eq{OmGW_Ik} implies a time average over oscillations at present time, which can be performed only when $\tilde t\gg 
\tilde t_{\rm fin}$.
However, it is often not computationally
affordable to run a simulation until $\tilde t_{\rm fin}$, especially for
slowly decaying sources. 
In practice, the quantity $\tilde T_{i j}(\tilde t_{\rm init}, \tilde t_{\rm end}, \tilde q, \tilde \kk)$ is computed in the
simulation over an interval of time $\tilde t \in (\tilde t_{\rm init}, \tilde t_{\rm end})$, where the sole criterion in choosing $\tilde t_{\rm end}$ is computational feasibility.
However,
the GW source has effectively stopped operating at a given wave
number $\tilde k$ also whenever the amplitude of the GW spectrum on that mode has reached its saturation amplitude, entering its free-propagation
regime (cf.~discussion in \App{sec:GW_prod}).
Therefore, the GW spectrum evaluated from the simulation is accurate
for any wave number $\tilde k$ that has already
reached its saturated amplitude by
the end of the simulation at $\tilde t_{\rm end}$,
 even if $\tilde t_{\rm end} < \tilde t_{\rm fin}$.
On the other hand, not all the modes in our simulations have reached this regime. 
As in previous numerical work \cite{Hindmarsh:2013xza,Hindmarsh:2015qta,Hindmarsh:2017gnf,Cutting:2019zws,Jinno:2020eqg,Jinno:2022mie},
one therefore needs to take into account this limitation
when interpreting the numerical results,
especially for the low-$k$ tail of the GW spectrum and for its peak amplitude \cite{RoperPol:2023dzg}, as
we discuss in \Sec{sec:results}.

In the following, we will propose extending the stationary UETC usually assumed for
sound waves (see \Sec{GW_sw})
to a locally stationary UETC that can incorporate
the decay of the source with time observed in the simulations of strong PTs
and of some intermediate ones (see \Sec{sw_extended}).
As we will see, the proposed model will allow us to extrapolate our results from the final time
of the simulations till the final time of GW production $t_{\rm fin}$, to estimate the amplitude of the present-day GW spectrum.
Note that also
numerical viscosity, which in the simulations can
dissipate
kinetic energy over time, could limit the time and $\tilde k$ range for which the simulations can actually be accurate.
This can potentially affect the transfer of energy from the fluid perturbations to GWs, especially at large $\tilde k$.

\subsection{Gravitational waves from stationary sound waves}
\label{GW_sw}

In this section, we review the theoretical modeling
of the GW production from sound waves from previous work. This is useful 
since we will present our numerical results 
in terms of quantities previously used in the literature,\footnote{The simulations concern the full
hydrodynamical system, from initial compressional motion to the possible development of
large fluid perturbations, which can occur 
in intermediate to strongly first order PTs.
On the other hand, when the PT is weak, the
linearized regime
of fluid perturbations holds, and indeed we will see that the simulations recover the sound-wave system described in previous work.}
and also as an introduction to the formalism used in the next section to model decaying sources.

In the context of GW production from sound waves, it has
extensively been assumed that the UETC is stationary, i.e., it only depends on the difference $t_- = t_1 - t_2$ \cite{Kosowsky:2001xp,Caprini:2009fx,Hindmarsh:2013xza,Hindmarsh:2015qta,Hindmarsh:2016lnk,Hindmarsh:2017gnf,Niksa:2018ofa,Hindmarsh:2019phv,Guo:2020grp,RoperPol:2023dzg,Caprini:2024ofd},
\begin{equation}
    E_\Pi (t_1, t_2, k) = 2 \, k^2 \, K^2 \, f(t_-, k)\,, \label{eq:Epistat}
\end{equation}
where $K = \rho_{\rm kin}/\bar \rho$ is the time-independent kinetic 
energy fraction, with $\rho_{\rm kin} = \bra{w \gamma^2 v^2}$ the mean kinetic energy density of the system. 
Note that the source is considered \emph{stationary} but only within a finite duration $\tilde \tau_{\rm fin} = \tilde t_{\rm fin} - \tilde t_\ast$.
Under the assumption of \Eq{eq:Epistat}, \Eq{Ik} becomes
\begin{equation}
    {\cal I} (\tilde t_\ast, \tilde t_{\rm fin}, \tilde k) = 
    k^3 \, K^2
    \int_{\tilde t_\ast}^{\tilde t_{\rm fin}} \dd \tilde t
    \int_{\tilde t_\ast - \tilde t}^{\tilde t_{\rm fin} - \tilde t}
    \cos (k t_-) \, f(\tilde t_-, \tilde k) \dd \tilde t_-\,.
    \label{OmGW_stat}
\end{equation}
In the sound-shell model of Refs.~\cite{Hindmarsh:2016lnk,Hindmarsh:2019phv},
the limits of the integral over $\tilde t_-$ get extended
to $\pm \infty$, allowing to separate the two integrals, such that the integral over $\tilde t$ 
becomes the source duration, $\tilde \tau_{\rm fin}$.
Therefore, within the sound-shell model, the GW amplitude
depends linearly on the source duration.
The proportionality to $\tilde \tau_{\rm fin}$ has
usually been assumed in the literature,
especially, but not exclusively, for GW production from sound waves
\cite{Kosowsky:2001xp,Hindmarsh:2013xza,Hindmarsh:2015qta,Hindmarsh:2016lnk,Hindmarsh:2017gnf,Hindmarsh:2019phv,Guo:2020grp,Caprini:2019egz}.
The duration of the sound-wave phase
can be assumed to correspond to the 
shock time, which determines
the time it takes for non-linearities and vorticity to develop in the fluid, $\tilde \tau_{\rm fin} \equiv \tau_{\rm fin} \beta \sim \tilde \tau_{\rm sh} = \beta R_\ast/\sqrt{K}$, with $\beta R_\ast \equiv (8\pi)^{1/3} \max (\vw, \cs)$ being the
mean separation of the bubbles at the end of the PT \cite{Caprini:2019egz}.
The aforementioned assumption that the integration limits get extended to $\pm \infty$ holds when the support of the UETC $f(\tilde t_-, \tilde k)$ is small enough.
In the sound-wave case, the UETC is highly oscillatory, and Ref.~\cite{RoperPol:2023dzg} has shown that, at the level of the GW spectrum, the assumption holds in the region $k \tau_{\rm fin} \sim kR_\ast/\sqrt{K} \gg 1$. 
When $\tau_{\rm fin}/R_\ast \sim 1/\sqrt{K} \gg 1$, i.e., when the kinetic energy is small,
this covers
all relevant wave numbers
$k R_\ast \gg \sqrt{K}$, including the spectral breaks occurring at $k\gtrsim 1/R_*$ (see \Sec{sec:shape}).
However, if the kinetic energy is large, the GW spectrum gets modified with respect to the sound-shell prediction \cite{RoperPol:2023dzg} of previous work \cite{Hindmarsh:2016lnk,Hindmarsh:2019phv}.
Finally, it can be shown that under the same
assumptions, the remaining integral over $\tilde t_-$ in \Eq{OmGW_stat} is proportional
to $(\beta R_\ast)/c_{\rm s}$ (see for instance App.~B of Ref.~\cite{RoperPol:2023dzg}), such that 
\begin{equation}
    {\cal I} (\tilde t_\ast, \tilde t_{\rm fin}, \tilde k) = \tilde \Omega_{\rm GW} \,
    K^2 \, \beta R_\ast \, \tilde \tau_{\rm fin} \,  S(kR_\ast)\,,
    \label{OmGW_sshell}
\end{equation}
where we have absorbed the $1/\cs$ factor\footnote{In this paper, we do not study the dependence of the GW signal on $\cs$.
For a study of the $\cs$ dependence in the sound-shell model see, e.g., Ref.~\cite{Giombi:2024kju}.} in $\tilde \Omega_{\rm GW}$,  which denotes the GW production
efficiency, and $S$ is a normalized spectral shape satisfying $\int \dd \ln k \, S(k) = 1$.
Using \Eq{OmGW_Ik}, the final GW spectrum
can be written as \cite{Hindmarsh:2013xza,Hindmarsh:2015qta,Hindmarsh:2016lnk,Hindmarsh:2017gnf,Hindmarsh:2019phv,Caprini:2019egz,RoperPol:2023dzg,RoperPol:2023bqa,Caprini:2024hue}
\begin{equation} 
    \Omega_{\rm GW} (k) = 3 \, {\cal T}_{\rm GW} \, \tilde \Omega_{\rm GW} \, K^2
    \ H_\ast R_\ast \ H_\ast \tau_{\rm fin} \ S(k R_\ast)\,.
    \label{OmGW_stat2}
\end{equation}

As done  in previous numerical studies \cite{Hindmarsh:2013xza,Hindmarsh:2015qta,Hindmarsh:2017gnf,Hindmarsh:2019phv}, the GW production
efficiency $\tilde \Omega_{\rm GW}$, in the situation in which the GW generation scales linearly with the source duration, can be estimated from the simulations via
\begin{equation}
    \tilde \Omega_{\rm GW}= \frac{{\cal I}_{\rm sim} ^{\rm int}}{ K^2 \, \beta R_\ast \, \tilde T_{\rm GW} }\,,\label{eq:OmGWtilde}
\end{equation}
where ${\cal I}_{\rm sim} ^{\rm int}=\int \dd \ln k \,{\cal I}_{\rm sim}$ [see \Eqs{Weinberg}{OmGW_sshell}], and $\tilde T_{\rm GW} = \tilde t_{\rm end} - \tilde t_{\rm init}$ is the time interval of the simulation in which
\Eq{Weinberg} is evaluated.
We will use similar expressions to \Eq{eq:OmGWtilde} in \Sec{sec:results} to analyze our results.
In particular, we will show that \Eq{OmGW_sshell} and in turn \eqref{OmGW_stat2}
well describe the simulation results as long as
the kinetic energy
remains approximately constant in time after
the PT ends, i.e., for weak PTs.
For intermediate and strong PTs, on the other hand, the simulations show a decay in time of the kinetic energy.
This requires a generalization of the stationary UETC assumption of \Eq{eq:Epistat},
which is described in \Sec{sw_extended}. 
We find that the GW spectrum \eqref{OmGW_stat2} in this case 
is proportional to the time integral of $K^2$, instead of to $K^2 \tau_{\mathrm{fin}}$.
This model is accurately validated by the numerical
results in \Sec{GW_spec_time}, and allows to still compute numerically the GW efficiency $\tilde \Omega_{\rm GW}$, and estimate the expected final GW amplitude at the final
time of GW production, even when the GW production does not scale linearly
with the source duration.

Alternatively,
previous work on Higgsless simulations presented
the results for the GW amplitude
in terms of the following parameterization
\cite{Jinno:2020eqg,Jinno:2022mie}:
\begin{equation}
    Q'(k) = \biggl(\frac{\bar \rho}{\bar w}\biggr)^2
    \, \frac{4 \pi^2}
    {\tilde T_{\rm GW}} 
    \, {\cal I}_{\rm sim} (\tilde t_{\rm init}, \tilde t_{\rm end}, \tilde k) \approx \frac{9 \pi^2}{4 \, \tilde T_{\rm GW}} \, (1 + \alpha)^2
    \,
    {\cal I}_{\rm sim} (\tilde t_\ast, \tilde t_{\rm fin}, \tilde k)\,,
    \label{Qprime_lin}
\end{equation}
where the prefactor $(\bar \rho/\bar w)^2$ in \Eq{Qprime_lin}
takes into account that the authors in Refs.~\cite{Jinno:2020eqg,Jinno:2022mie} used the mean
enthalpy to normalize $T_{ij}$ in the definition of $Q'$
instead of the total
energy density $\bar \rho$, as done in our case [see text below \Eq{eq:Fourier transform}].
References~\cite{Jinno:2020eqg,Jinno:2022mie}
found a strong dependence of the parameter $Q'/K^2 \sim {\cal I}_{\rm sim}/(K^2 \tilde T_{\rm GW})$ with $\vw$, 
and concluded that
${\cal I}_{\rm sim} \sim K_\xi^2\, \xi_{\rm shell} \, \tilde T_{\rm GW}$ [where $K_\xi$ is given in \Eq{Kxi} and $\xi_{\rm shell}$ in \Eq{eq:xi_shell}].
However,
in this paper we find
that inserting $R_\ast$ in \Eq{OmGW_sshell} to describe the GW amplitude, instead of using $Q'/K^2$,
provides the right GW efficiency
$\tilde \Omega_{\rm GW}$, which is virtually independent of both $\vw$
and the duration of the source $\tau_{\rm fin}$ (see \Sec{GW_spec_time}).
This is connected to the fact that the GW source is the bulk fluid motion and not the bubble collisions. 
Therefore, normalizing with quantities characteristic of the PT, such as $H_*/\beta$, instead of with quantities characteristic of the fluid, such as $H_*R_*$, can lead to misinterpretation \cite{Caprini:2009fx}. 
Indeed, the fluid is set into motion by the bubbles, whose characteristic scale is $R_*$ after the end of the PT, and evolves with its own dynamical timescale afterwards; the duration of the PT, $H_*/\beta$, on the other hand, does not directly influence the fluid dynamics. 
The same consideration applies to the breaks of the GW spectrum, which should also be normalized with $R_*$. This intuitive argument is  confirmed 
with simulations in the present analysis (cf.~\Sec{sec:results}).

The kinetic energy fraction $K$ is
usually taken to be the
one corresponding to a single bubble, which can be expressed
as
\begin{equation}
    K_\xi \equiv \frac{\kappa_\xi \, \alpha}{1 + \alpha}\,, \label{Kxi}
\end{equation}
where $\alpha$ characterizes the strength of the PT
[see \Eq{alpha}],
 and $\kappa_\xi \equiv \rho_{\rm kin}^\mathrm{sb}/\epsilon<1$ is the single-bubble
 efficiency factor, where $\rho_{\rm kin}^{\rm sb}$ denotes the kinetic energy density
 of a single bubble \cite{Espinosa:2010hh}.
We compare in \Sec{sec:results} the kinetic energy fraction found in the
simulations $K$ with the single-bubble one,
showing that the effect of collisions and non-linear dynamics
alters the kinetic energy fraction with respect to $K_\xi$.
Note that, in the sound-shell model, $K \neq K_\xi$ is time-independent but the exact value of
$K/K_\xi \sim {\cal O} (1)$ depends on the PT parameters \cite{Hindmarsh:2016lnk,Hindmarsh:2019phv,RoperPol:2023dzg}.
Similarly, Refs.~\cite{Hindmarsh:2017gnf,Cutting:2019zws} have also reported maximum
values of the kinetic energy fraction $K_{\rm max}$ in their simulations that differ from $K_\xi$.
Therefore, using $K_\xi$ for the GW production [cf.~\Eq{OmGW_stat2}] might lead to
a wrong estimate of the amplitude.

\subsection{Gravitational waves from decaying sources}
\label{sw_extended}

As we show in \Sec{sec:results}, for strong and intermediate PTs,
the kinetic energy starts to decay within the duration of the
simulations,
potentially due to the fact that the
system enters the non-linear regime.
Consequently,
the linear scaling of the GW amplitude  
with $\tau_{\rm fin}$ changes, impeding the use of \Eq{OmGW_sshell}
to estimate the GW efficiency.

In order to analyze such cases, we propose a  generalization of \Eq{eq:Epistat}, 
assuming a locally stationary UETC that allows to
account for the time dependence
of $K^2$:
\begin{equation}
    E_\Pi (t_1, t_2, k) = 2 \, k^2 \, K^2 (t_+) \, f(t_-, k)\,, \label{eq:Epilocstat}
\end{equation}
where $t_+ = \half (t_1 + t_2)$.
Note that locally-stationary kernels are acceptable as UETCs, since they give rise to
positive-definite GW spectra (see discussion in 
Refs.~\cite{Auclair:2022jod,Caprini:2009yp}).
Then, under the same approximation discussed above,
i.e., that the two integrals in $\tilde t$ and $\tilde t_-$ of \Eq{OmGW_stat} can be separated by sending the limits of the one in $\tilde t_-$ to $\pm\infty$, we find \cite{RoperPol:2023dzg}
\begin{equation}
    {\cal I} (\tilde t_\ast, \tilde t_{\rm fin}, \tilde k) = 
    k^3 \,
    \int_{\tilde t_\ast}^{\tilde t_{\rm fin}}  K^2 (\tilde t_+)
    \dd \tilde t_+
    \int_{-\infty}^{\infty}
    \cos (k t_-) \, f(\tilde t_-, \tilde k) \dd \tilde t_-\,,
    \label{OmGW_stat3}
\end{equation}
such that
$K^2 \, \tilde \tau_{\rm fin}$ in \Eq{OmGW_sshell} can be substituted by $K_{\rm int}^2$, 
\begin{equation}
    K_{\rm int}^2 (\tilde t_\ast, \tilde t_{\rm fin}) \equiv
   \int_{\tilde t_\ast}^{\tilde t_{\rm fin}} K^2 (\tilde t) \dd \tilde t
    \,. \label{K2rms}
\end{equation}
This yields
\begin{equation}
    {\cal I} (\tilde t_\ast, \tilde t_{\rm fin}, \tilde k) = \tilde \Omega_{\rm GW} \, K^2_{\rm int} (\tilde t_\ast, \tilde t_{\rm fin}) \,\, (\beta R_\ast) \,\, S(k R_\ast)\,, \label{OmGW_general}
\end{equation}
and the resulting GW spectrum is
\begin{equation} 
    \Omega_{\rm GW} (k) = 3 \, {\cal T}_{\rm GW} \, \tilde \Omega_{\rm GW} \, \frac{H_\ast}{\beta} \, K_{\rm int}^2
    \ H_\ast R_\ast \ S(k R_\ast)\,.
    \label{OmGW_stat4}
\end{equation}
Note that $K_{\rm int}^2$ reduces to $K^2 \, (\tilde t_{\rm fin} - \tilde t_\ast) = K^2 \, \tilde \tau_{\rm fin}$ when $K^2$ is constant in time and one recovers \Eq{OmGW_sshell} found
in the stationary assumption.
As already discussed in \Sec{GW_sw}, when
the UETC of the sound waves is provided solely by $f(t_-, k)$,
the two integrals of \Eq{OmGW_stat} can 
be separated only for $ k\tau_{\rm fin} \gg 1$.
{The same applies in this case as the introduction of $K(t_+)$
in the UETC does not affect the integral over $t_-$ in \Eq{OmGW_stat3}.}
These wave numbers include the peak of the GW spectrum when $\tau_{\rm fin} \gg R_\ast$, meaning~$\sqrt{K}\ll 1$.
This last condition might seem at odds with the fact that we will apply this formalism to intermediate and strong PTs.
Note, however, that the fluid kinetic 
energy fraction always remains small, of the order of the single-bubble one, $K_\xi$, even 
for intermediate and strong PTs [see \Eq{Kxi} and \Tab{tab:kappas}].
Only for PTs where the vacuum energy dominates and the radiation component can be neglected, i.e., $\alpha\gtrsim 1$, one can reach values $K\sim 1$.

A similar UETC to the one in \Eq{eq:Epilocstat}
has been considered, e.g., in Refs.~\cite{Niksa:2018ofa,Dahl:2024eup},
\begin{equation}
    E_v(t_1, t_2, k) =
    \sqrt{E_v(t_1, k) \, E_v(t_2, k)} \, \cos (k \cs t_-)\,,
\end{equation}
where $E_v (k)$ is the velocity spectrum at equal times $t_1 = t_2$, defined such that
\begin{equation}
     \bra{v^2} =  \int_0^\infty \dd k \, E_v (k)\,.
\end{equation}
The integral over $t_1$ and $t_2$ of this UETC can be
reduced to the integral in \Eq{K2rms} under the aforementioned assumptions
(see discussion in Sec.~\RNum{5} of Ref.~\cite{RoperPol:2023dzg})
and for small fluid perturbations.
Therefore, we expect both UETCs to have the same impact on the
integrated GW amplitude.
Further analyses are necessary to understand how different UETCs influence the GW signal spectral shape.

We show in \Sec{decay_K2} that the kinetic energy evolution
in the simulations can be fit with a decaying power law,
\begin{equation}
    K (\tilde t>\tilde t_0) = K_0 
\left(  \frac{\Delta \tilde t}{\Delta \tilde t_0}\right)^{-b}\,, \label{eq:Ktildegeneral}
\end{equation}
where $b \geq 0$ and $K_0$ are
parameters to be extracted from the numerical results, and $\tilde t_0$ corresponds to the time at which all the simulation box is in
the broken phase, and the decay of the kinetic energy is well in place (see \Sec{decay_K2}).
$\Delta \tilde t$ and $\Delta \tilde t_0$ in \Eq{eq:Ktildegeneral} are time
intervals with respect to the time-coordinate origin in the simulations, $\tilde t_{\rm ref}$.
Since the choice of $\tilde t_{\rm ref}$ is arbitrary, the value
of $\Delta \tilde t_0$ should be considered as a
parameter of the numerical fit.
Similarly, one has $\Delta \tilde t = \tilde t - \tilde t_0 + \Delta \tilde t_0$, where only the interval of time since the end of the PT, $\tilde t - \tilde t_0$, is physically relevant.

Note that the decay rate exponent $b$ could in principle be predicted within a physical model for the dynamical evolution of the fluid, and then compared to the simulation results, as done for instance in the case of  magnetohydrodynamic turbulence, for which theoretical predictions on the decay rate exponent exist (see, e.g., Ref.~\cite{Brandenburg:2017neh}).
However, this is still premature at the level of the present analysis, since we are missing such a model (even though a possible interpretation of the decay observed in simulations as due to the development of non-linearities is emerging in our results). 
For the kinetic energy fraction $K_0$, on the other hand, there is an available proxy, i.e., the single-bubble one, $K_\xi$. We indeed study the relation between $K_0$ and $K_\xi$ and analyze its dependence on the numerical
discretization in \Sec{sec:convergence} and \App{sec:kinetic_ed}.
We find that $K_0$ is not exactly $K_\xi$, as a consequence
of the effects due to collisions during the PT.
The simulations show that $K(\tilde t)$ grows initially, proportional to the volume of the simulation in the
broken phase (see \App{sec:kinetic_ed}).
Then, it reaches a peak value
and enters the decay stage at times that are practically the same across the tested values of $\alpha$ and $\vw$ (cf.~\Fig{fig:E_kin_evolution}).

From \Eq{K2rms}, using \Eq{eq:Ktildegeneral}, we find that the dependence of $K_{\rm int}^2$
with the source duration, $\tilde \tau_{\rm fin} = \tilde t_{\rm fin} - \tilde t_\ast$, is
\begin{equation}
     K^2_{\rm int} (\Delta \tilde t_\ast, \tilde \tau_{\rm fin}) =
     K_0^2 \, \Delta \tilde t_* \left(\frac{\Delta \tilde t_0}{\Delta \tilde t_*}\right)^{2b} \, \frac{(1 + \tilde \tau_{\rm fin}/\Delta \tilde t_*)^{1 - 2b} - 1}{1 - 2b}\,, \label{Kint_decay}
\end{equation}
where we have introduced $\Delta \tilde t_\ast = \tilde t_\ast - \tilde t_0 + \Delta \tilde t_0$,
and assumed that the GW production starts at $\tilde t_\ast \geq \tilde t_0$.
Again, only the time interval $\tilde t_\ast - \tilde t_0$ is physically meaningful.
When the source duration is very short $\tilde \tau_{\rm fin}/\Delta \tilde t_* \ll 1$, this expression reduces to
$K_{\rm int}^2 \to K_0^2 \, \tilde \tau_{\rm fin} (\Delta
\tilde t_0/\Delta \tilde t_*)^{2b}$
for any value of $b$:
indeed, in this case, the integral can be approximated as the product of the integrand evaluated at the lower boundary $\Delta \tilde t_*$
multiplied by the time interval itself, i.e.,
the source duration, and one goes back to the linear dependence with $\tilde \tau_{\rm fin}$
as in the stationary source case, in which the kinetic energy is assumed constant in time (see \Sec{GW_sw}).
For long duration
$\tilde \tau_{\rm fin}/\Delta \tilde t_* \gg 1$, on the other hand,
\Eq{Kint_decay} takes the following asymptotic limits:
\begin{align}
    \lim_{\tilde \tau_{\rm fin} \gg \Delta \tilde t_*}
    K_{\rm int}^2 (\tilde \tau_{\rm fin}) = 
    & \frac{K_0^2 \, \Delta \tilde t_0}{1 - 2 b} \, \biggl(\frac{\tilde \tau_{\rm fin}}{\Delta \tilde t_0}
    \biggr)^{1 - 2 b}\!\!\!\!\!, \qquad  & {\rm when \ \ } 0 \leq b < \half \,, \nonumber \\ 
    \lim_{\tilde \tau_{\rm fin} \gg \Delta \tilde t_*} K_{\rm int}^2 (\Delta \tilde t_\ast)
    =  &  \frac{K_0^2 \, \Delta \tilde t_*}{2 b - 1}\left(\frac{\Delta \tilde t_0}{\Delta \tilde t_*}\right)^{2b} \,,
   \qquad & {\rm when \ \ } b > \half \,. \label{growthrate_K}
\end{align}
Hence, when the decay is sufficiently slow, $0 \leq b < 1/2$, $K_{\rm int}^2$ is proportional to
$\tilde \tau_{\rm fin}^{1 - 2b}$,
thus generalizing
the linear dependence with $\tilde \tau_{\rm fin}$ obtained in the stationary assumption
to a shallower growth with the source duration in the case of a slowly decaying source.
On the other hand, when the decay rate is fast,
$b>1/2$, $K_{\rm int}^2$ saturates
to a fixed value, independent of the source duration.
If $b = 1/2$, the dependence with $\tilde \tau_{\rm fin}$ is logarithmic.

\EEq{OmGW_general} generalizes the description of the GW production to a decaying source, for which the assumption of local  stationarity in time holds.
It allows us to go beyond the usual assumption of the sound-shell model, valid for weak PTs, and it also
allows us to estimate the GW efficiency $\tilde \Omega_{\rm GW}$ when the kinetic energy
is decaying with time, cf.~\Eq{OmGW_numerical}.
In \Secs{decay_K2}{GW_spec_time}, we validate the proposed model
for the integrated GW amplitude and
\Eqs{OmGW_general}{Kint_decay} with the results of numerical simulations.

\subsection{Effect of the Universe expansion}
\label{sec:expansion}

Neglecting the expansion of the Universe, the sound-shell model finds that
a superposition of sound waves emits GWs with an amplitude 
proportional to the source duration $\tau_{\rm fin}$ (see \Eq{OmGW_stat2}
and Refs.~\cite{Kosowsky:2001xp,Gogoberidze:2007an,Caprini:2009fx,Hindmarsh:2013xza,Hindmarsh:2015qta,Hindmarsh:2016lnk,Hindmarsh:2017gnf,Hindmarsh:2019phv,RoperPol:2023bqa,RoperPol:2023dzg,Caprini:2024hue,Caprini:2024ofd}). 
This linear dependence on the source duration is typical of stationary sources when back-reaction (in this case, the decay of the source due to the GW production) is not taken into account.
In the case of sound waves, the unbounded linear increase is expected to be cut off by
the development 
of non-linearities.
The locally stationary UETC proposed in \Sec{sw_extended}
allows to account for a non-linear source that is decaying in time.
We have found that, for a power-law decay
with exponent $-b$,
when the decay rate of the kinetic energy is slow enough, $b < 1/2$,  
the GW amplitude still increases unbounded proportional to
$\tau_{\rm fin}^{1 - 2b}$ 
[see \Eq{growthrate_K}].
On the other hand, it saturates when $b > 1/2$.

Naturally, the expansion of the Universe inserts an extra decay component,
changing the dependence of the GW spectrum with the source duration.
In the stationary UETC case,
to take into account the expansion of the Universe,
the linear dependence with $H_*\tau_{\rm fin}$ of \Eq{OmGW_stat2}
can be substituted by the factor $\Upsilon (\mathcal{H}_*\delta\eta_{\rm fin}) = \mathcal{H}_*\delta\eta_{\rm fin}/(1 + \mathcal{H}_* \delta\eta_{\rm fin})$ \cite{Guo:2020grp,RoperPol:2023dzg}, 
where $\delta\eta_{\rm fin} \equiv \eta_{\rm fin} - \eta_\ast$ is
the source duration in conformal time,
and $\HH_\ast = H_\ast a_\ast$ is the conformal Hubble rate:
\begin{equation} 
    \Omega_{\rm GW,exp} (k) = 3 \, {\cal T}_{\rm GW} \, \tilde \Omega_{\rm GW} \, K^2 \,
    \frac{\HH_\ast \delta \eta_{\rm fin}}{1 + \HH_\ast \delta \eta_{\rm fin}}
    \ \HH_\ast \RR_\ast
    \ S_{\rm exp} ({k_c} \RR_\ast)\,.
    \label{OmGW_exp}
\end{equation}
$\RR_\ast = R_\ast/a_\ast$ and $k_c=a_*k$
are the comoving mean-bubble separation
and wave number, such that
$\HH_\ast \RR_\ast = H_\ast R_\ast$ {and $k_c \RR_\ast = kR_\ast$}.

It is important to remark that, when including expansion, the results are no longer
invariant under time translations, so we need to choose
absolute values for conformal times.
Assuming that the PT completes faster than one Hubble time,
and that the GW sourcing occurs during radiation domination, we can
set the initial and final conformal times of GW production to be
$\mathcal{H}_\ast \eta_\ast = 1$ and $\mathcal{H}_\ast \eta_{\rm fin} = 1 + \mathcal{H}_\ast \delta\eta_{\rm fin}$
(one could also then
normalize to $a_\ast = 1$, such that the conformal
Hubble rate is $\HH_\ast =  H_\ast$).
Note that Ref.~\cite{Guo:2020grp} writes the factor $\Upsilon$ in physical time as $\Upsilon=1-1/y$ with $y=\sqrt{1+2(t-t_s)H_s}$, and then associates the source lifetime in cosmic time $t-t_s$ to the time of onset of non-linearities,
estimated to correspond to the shock formation time,
$\tau_{\rm fin}\sim \tau_{\rm sh} \equiv
R_\ast/\sqrt{K}$ (see also Refs.~\cite{Hindmarsh:2020hop,Gowling:2021gcy,Gowling:2022pzb}).
However, the conformal invariance of the
fluid equations when the fluid is radiation-dominated only holds if they are expressed in comoving lengths and conformal times \cite{Brandenburg:1996fc,Subramanian:1997gi,RoperPol:2025lgc}. 
Therefore, any characteristic scale or time inferred from the fluid dynamics
in flat space-time, like $\tau_{\rm sh}$,
can be used in the expanding Universe \emph{only in terms of conformal time}.
Consequently,
it is the interval in conformal time
instead of the interval in cosmic time that should be
associated to the shock time
when evaluating the expected time for
non-linearities to develop.
In summary, when 
extending the stationary UETC case, and therefore the sound-shell model, to an expanding Universe, one should set the duration of the GW sourcing to $\delta\eta_{\rm fin} 
\sim \delta \eta_{\rm sh} \equiv \RR_\ast/\sqrt{K}$ in \Eq{OmGW_exp}, so that
$\Upsilon  \sim \HH_* \RR_*/(\sqrt{K}+\HH_* \RR_*)$.
Concerning the spectral shape $S_{\rm exp} (k_c \RR_*)$, the Universe expansion can influence the spectrum when the GW production is comparable or larger than the Hubble time, and the simulation results need to be complemented by analytical studies (cf.~Ref.~\cite{RoperPol:2023dzg}), or simulations accounting for the Universe expansion need to be run (cf.~Ref.~\cite{RoperPol:2019wvy}).
In this work, we only discuss the spectral shape arising in the context of our simulations (see \Sec{sec:shape}), and we defer a complete evaluation of the Universe expansion effect on the spectral shape to a future study.

In the locally stationary UETC case, extending
\Eq{OmGW_stat3} to apply in an expanding Universe \cite{RoperPol:2023dzg},
an effective integrated $K^2$ that can be used in \Eq{OmGW_stat4} to
estimate the effect of expansion is the following
\begin{equation}
    K^2_{\rm int, exp} \equiv \int_{1}^{\tilde \eta_{\rm fin}} \frac{\dd \tilde \eta}{\tilde \eta^2} K^2 (\tilde \eta)
    \,, \label{Kexp_general}
\end{equation}
where normalized conformal times refer to $\tilde \eta = \eta/\eta_\ast =\eta \HH_\ast$.
Note that this is different than the normalization used when ignoring the expansion
of the Universe for cosmic times, $\tilde t = t \beta$.
When $K^2$ is constant in time, this reduces to
\begin{equation}
    K^2_{\rm int, exp} = K^2 \int_1^{\tilde \eta_{\rm fin}} \frac{\dd \tilde \eta}
    {\tilde \eta^2} = K^2 \, \Upsilon (\delta \tilde
    \eta_{\rm fin})\,, \quad {\rm for}~K~{\rm constant}\,.
\end{equation}
Then, the resulting GW spectrum can be expressed as
\begin{equation} 
    \Omega_{\rm GW,exp} (k) = 3 \, {\cal T}_{\rm GW} \, \tilde \Omega_{\rm GW} \, K_{\rm int, exp}^2
    \HH_\ast \RR_\ast \ S_{\rm exp} ({k_c} \RR_\ast)\,. \label{OmGW_final_exp}
\end{equation}

Taking into account that
the power-law decay in flat space-time
should be taken in conformal time
$K (\tilde \eta) = K_0 \, (\Delta \tilde \eta/\Delta \tilde \eta_0)^{-b}$ due to the conformal
invariance of the dynamics for a radiation-dominated fluid,
then we need to express the absolute times in a flat space-time as time intervals in conformal time, $\Delta \tilde \eta = \delta \tilde \eta + \Delta \tilde \eta_\ast$,
where $\delta \tilde \eta = \tilde \eta - 1$ and $\Delta \tilde \eta_\ast = \Delta \tilde \eta_0 - \delta \tilde \eta_0$.
The resulting integral for $2 b \neq 1$ can be expressed as
\begin{equation}
    K^2_{\rm int, exp} = K_0^2 \, \Delta \tilde \eta_0^{2b}
    \int_1^{\tilde \eta_{\rm fin}}
    \frac{\dd \tilde \eta}
    {\tilde \eta^2} (\tilde \eta + \Delta \tilde \eta_\ast - 1)^{-2b}
    = K_0^2 \, \Upsilon_b (\delta \tilde \eta_{\rm fin})\,,  \label{Kexp_fit}
\end{equation}
where we have defined a suppression factor
\begin{equation}
    \Upsilon_b (\delta \tilde \eta_{\rm fin}) = \frac{{\cal F}_b (\delta \tilde \eta_{\rm fin}) - {\cal F}_b (0)}{1 - 2b}\,, \label{eq:Upb}
\end{equation}
with
\begin{equation}
    {\cal F}_b (\delta \tilde \eta) = \biggl(\frac{\Delta \tilde \eta_\ast + \delta \tilde \eta}
    {\Delta \tilde \eta_0}\biggr)^{1 -2b}
    \frac{\Delta \tilde \eta_0}
    {(\Delta \tilde \eta_\ast - 1)^2}  \,_2 F_1 \biggl[2, 1 - 2b, 2 - 2b, \frac{\Delta \tilde \eta_\ast + \delta \tilde \eta}
    {\Delta \tilde \eta_\ast  - 1} \biggr]\,, \label{hyper}
\end{equation}
being $\,_2 F_1$ the hypergeometric function.
The function $\Upsilon_b$ reduces to the one found for stationary
sources when $b = 0$, i.e., $\Upsilon_0 (\delta \tilde \eta_{\rm fin}) \equiv \Upsilon (\delta \tilde \eta_{\rm fin}) = \delta \tilde \eta_{\rm fin}/(1 + \delta \tilde \eta_{\rm fin})$ \cite{Guo:2020grp,RoperPol:2023dzg}.
We highlight that the emergence of a hypergeometric function has no deep physical meaning, since
\Eq{hyper} arises from introducing the chosen fit $K(\tilde \eta)$
in \Eq{Kexp_general}.
The relevant physical quantity is the modification
$\Upsilon_b$ with respect to $\Upsilon$ (i.e., with no decay of the source)
obtained from the additional decaying
function in the integral in \Eq{Kexp_fit}.
The value of $\Delta \eta_0 \equiv \Delta \tilde \eta_0/\HH_\ast$
corresponds to the characteristic time interval
$\Delta t_0 \equiv \Delta \tilde t_0/\beta$
used in the fit of $K^2$ in flat space-time, introducing
the implicit dependence of the resulting GW spectrum on the
ratio $\beta/H_\ast$ through $\Delta \tilde \eta_0 =
\Delta \tilde t_0 \, H_\ast/\beta$.
We note that, in principle, using $\Delta \tilde \eta_\ast = \tilde \eta_\ast - \tilde \eta_0 + \Delta \tilde \eta_0$ in the integrand of \Eq{Kexp_fit} allows to
compute the GW spectrum starting at any conformal time $\tilde 
\eta_\ast>\tilde \eta_0$.
We also find that for any value of $b$,
the functions $\Upsilon_b (\delta \tilde \eta_{\rm fin})$ always
reduce to the linear dependence with the source duration $\delta \tilde \eta_{\rm fin}$ for short duration,
$\delta \tilde \eta_{\rm fin} \ll 1$.
Then, the final GW spectrum becomes
\begin{equation}
    \Omega_{\rm GW, exp} (k) = 3 \, {\cal T}_{\rm GW} \, \tilde \Omega_{\rm GW} \, K_0^2 \
    \Upsilon_b(\delta \tilde \eta_{\rm fin})
    \, \HH_\ast \RR_\ast \,S_{\rm exp} ({k_c} \RR_\ast)\,,
    \label{OmGW_locally_stat_exp}
\end{equation}
where $\Upsilon_b$ is obtained from the integrated $K_{\rm int}^2$ and in particular it is given
by \Eqs{eq:Upb}{hyper} when the fit $K^2 = K_0^2 \, (\Delta \tilde \eta/\Delta \tilde \eta_0)^{-2b}$ holds at all times of the GW production.
We emphasize that the suppression factor
$\Upsilon_b$ works as a proxy
to estimate
the effect of the Hubble expansion,
which has not been accounted for in our simulations performed in flat space-time.

As mentioned above, in \Sec{decay_K2} we use the numerical
results of the simulations to find the values of the fit
parameters $b$
and $K_0$ for different PTs.
Then, we validate the assumption that \Eq{OmGW_general} applies
within the duration of our simulations
in \Sec{GW_spec_time},
and provide an estimate of the
GW amplitude as a function of the source duration $\delta \tilde \eta_{\rm fin}$,
for an expanding Universe using \Eq{OmGW_locally_stat_exp}.
We will further assume that the GW production starts
when all the simulation domain is in the broken phase,
i.e., $\tilde \eta_\ast = \tilde \eta_0$.

\section{Numerical setup}
\label{sec:numerical_setup}

In this section, we focus on describing the numerical setup of the Higgsless simulations: in \Sec{subsec:Updates to the simulation}, we comment on the updates in the numerical scheme with respect to Ref.~\cite{Jinno:2022mie}, and in \Sec{sec:params}, we describe the simulation suite considered for this work.

%%%%%%%%%%%%%%%%%%%%%%%%%%%%%%%%%%%%%%%%%%%%%%%%%%%%%%%%%%%%%%%%%%%%%%
\subsection{Updates to the numerical setup}
\label{subsec:Updates to the simulation}
%%%%%%%%%%%%%%%%%%%%%%%%%%%%%%%%%%%%%%%%%%%%%%%%%%%%%%%%%%%%%%%%%%%%%%

In this section, we highlight three updates to the
Higgsless simulations with respect to Ref.~\cite{Jinno:2022mie}
aimed at improving: {\em (1)} the time integration scheme,
{\em (2)} the mapping
between the discrete and the continuum momenta,
and {\em (3)} the criterion for numerical stability in simulations of strong first-order PTs ($\alpha = 0.5$). For a complete description of the Kurganov-Tadmor (KT) numerical scheme \cite{KURGANOV2000241} used for the Higgsless simulations, we refer to Refs.~\cite{Jinno:2020eqg,Jinno:2022mie}.

Commencing with {\em (1)}, in practice, the
integral in \Eq{eq:Fourier transform}
must be computed numerically on the grid of space and time. For the space grid, this is accomplished through a fast Fourier transform routine~\cite{Frigo:2005zln}. For the time grid, in order to overcome the practical limitation of memory (i.e., storing a large number of 3D time slices), one needs to resort to another method. In the first iteration of the Higgsless simulation code, the discrete integral in time of \Eq{eq:Fourier transform} was approximated as
\begin{equation}
\tilde T_{ij} (\tilde t_{\rm init}, \tilde t, \tilde q, \tilde \kk) =
\sum_{m=1}^{N_t}
\delta{\tilde{t}_m}\, e^{i \tilde q \tilde t_m} \tilde T_{ij} \bigl(\tilde t_m, \tilde \kk \bigr) \, ,
\end{equation}
where the time coordinate is discretized in $N_t$ intervals
$[\tilde t_{m} - \tilde t_{m - 1}]$
of size $\delta \tilde t_m = \tilde t_m - \tilde t_{m - 1}$,
i.e.,~through its Riemann sum, by stacking past time slices weighted by a complex factor from
$\tilde t_{\rm init}$ until
$\tilde t\leq \tilde t_{\rm end} = \tilde t_{N_t}$
for each time step $\delta{\tilde{t}}_m$ over which the GWs are sourced.
We assume a fixed time step, such that $\delta \tilde t_m = \delta \tilde t, \forall m$.
In the current version, we improve upon this scheme by treating $\tilde T_{ij}$ as a piecewise linear function interpolating between the support points, using a similar scheme to the one proposed in Ref.~\cite{RoperPol:2018sap} for solving the GW equation.
Since the integrand involving an oscillating exponential as well as the linearized $\tilde T_{ij}$ is now analytically integrated, this modified routine
allows to capture better
the behavior at large $k$, alleviating the time-step $\delta{\tilde{t}}$ required
to find accurate spectra in this
regime (see discussion in Ref.~\cite{RoperPol:2018sap}).
However, no sizable discrepancies have been observed in the UV range of the
GW spectra through this change
for the dynamical range and choice of $\delta{\tilde{t}}$ used in
our simulations.

Continuing with {\em (2)}, we begin by noting that the first version of the Higgsless simulations employed a $\sin$-prescription for the mapping of discrete momenta on the grid to their
correspondents in the continuum.
Care must be taken that on the grid of the simulation with $N$ points per dimension, Fourier modes with momenta $-l_i$ and $N-l_i$ (in the $i$th direction) are equivalent and mapped to the same
momenta in the continuum.
At the same time, momenta of order $l_i \simeq N$ are equivalent to $l_i \simeq 0$ and should be considered soft.
Depending on whether the observable under consideration is sensitive to the sign of the momentum, this 
motivated the mapping
\be
\tilde k_i = \frac{2 - a}{\delta{\tilde{x}}} \sin \left(\frac{a \pi l_i}{N}\right) \,,
\ee
where $\delta{\tilde{x}} = \tilde L/N$, with
$a = 1$ when the sign is relevant and $a = 0$ when it is not.
In the current simulations, we generally use a {\em saw} description for the momenta
\begin{equation}
\label{eq:momenta mapping}
\tilde k_i = \begin{cases}
2 \pi l_i / (N \delta{\tilde{x}} )\,, & l_i < N / 2\,, \\ 
0\,, & N / 2\,, \\ 
2 \pi (l_i - N) / (N \delta{\tilde{x}})\,, & l_i>N / 2\,.
\end{cases}
\end{equation}
As such, the saw-prescription avoids different descriptions in different contexts (such as the space Fourier transforms for the GW estimate or for the numerical fluid evolution) and maintains a good map of momenta all the way to $l_i \simeq N/2$, while the previous method is only accurate in the linear regime of the sine function.
At the moment, we do not find substantial differences between
the two implementations, but we expect the new implementation to improve
the results when increasing the resolution of the Higgsless simulations.

The third point {\em (3)} concerns the choice of the maximal local velocity $a_{j+1 / 2}$ (on a staggered cell in direction $j$), appearing in Eq.~(3.7) of Ref.~\cite{Jinno:2022mie}.
In summary, this quantity enters the flux limiter used in the KT scheme to preserve the shock structures in the lattice
by setting a minimal numerical viscosity
that reduces spurious oscillations
and improves the stability of the numerical scheme.
In the limit of small fluid velocities, i.e., for weak and intermediate PTs,
$a_{j+1 / 2}=\cs=1/\sqrt{3}$ is a good choice.
In the case of strong PTs, however, fluid velocities often supersede $1/\sqrt{3}$ and approach 1. To improve the numerical stability of the simulation, we therefore choose $a_{j+1 / 2}=1$ for strong PTs. 
In the weak regime, the numerical changes due to this choice are negligible but for stronger 
PTs, it improves the stability of the code significantly. 
In rare occasions and close to shocks, the simulation can lead to unphysical fluid velocities (essentially $v>1$) as a numerical artifact.
In these cases, we opted to enforce the local fluid velocity to $1$. This only happened in isolated points and had
no measurable impact on the conservation of $T^{0\mu}$ or the GW spectra. 

In all other regards, the current version of the Higgsless implementation is identical to the first version in Ref.~\cite{Jinno:2022mie}.

%%%%%%%%%%%%%%%%%%%%%%%%%%%%%%%%%%%%%%%%%%%%%%%%%%%%%%%%%%%%%%%%%%%%%%
\subsection{Simulations and parameter choices\label{sec:params}}
%%%%%%%%%%%%%%%%%%%%%%%%%%%%%%%%%%%%%%%%%%%%%%%%%%%%%%%%%%%%%%%%%%%%%%

\begin{table}[b!]
\begin{center}
\small
\scalebox{0.9}{
\begin{tabular}{|c||c|c|c|}
\hline
& \bf{reference}
& \bf{seeds} 
& \bf{single-bubble}\\
\hline
PT strength $\alpha$ 
& $\{0.0046,\,0.05,\,0.5\}$ 
& $\{0.0046,\,0.05,\,0.5\}$ 
& $\{0.0046,\,0.05,\,0.5\}$ \\
\hline
wall velocity $\vw$
& $\in [0.32, 0.8]$
& $\{0.32/0.36, 0.6, 0.8\}$
& $\in [0.32, 0.8]$ \\
\hline
box size $\tilde L/\vw \equiv L\beta/\vw$ 
& $\{20,\,40\}$ 
& $\{20,\,40\}$ 
& $\{20,\,40\}$ \\
\hline
sim. time $\tilde t_{\rm end} \equiv t_{\rm end} \beta$
& $32$
& $32$
& $\tilde L/[2\,\mathrm{max}(\vw, c_{\rm{s}})]$\\
\hline
grid size $N$ 
& $\{64,\,128,\,256,\,512\}$ 
& $\{64,\,128,\,256,\,512\}$ 
& $512$ \\
\hline
$\delta{\tilde{t}}/\delta{\tilde{x}}$ 
& $<1/4$ 
& $<1/4$ 
& $<1/4$ \\
\hline
\hline
count. 
& $304$ 
& $72\times 9\, {\rm seeds} = \,648$ 
& $76$ \\
\hline
\end{tabular}}
\caption{
Summary of simulation runs with physical and numerical parameter choices. \emph{Reference} indicates the simulations constructed from a single reference bubble nucleation history (for each box size $\tilde L/\vw$), thereby eliminating statistical differences among the sample of reference simulations.
\emph{Seeds} refers to simulations constructed from a set of 9 additional bubble nucleation histories, allowing to infer statistical sample variance for 3 selected wall velocities $\vw = 0.32$ (0.36 for strong PTs), 0.6, and 0.8,
which correspond to a deflagration, a hybrid, and
a detonation for weak and intermediate PTs, while $\vw = 0.8$ is still a hybrid for strong PTs.
\emph{Single-bubble} refers to simulations with only one single centrally nucleated bubble, allowing us to study the convergence of self-similar profiles. 
We take a range of $\vw \in [0.32, 0.8]$ in increments
of 0.04 besides for
strong transitions ($\alpha=0.5$) for which we take $\vw \in [0.36, 0.8]$.
A total of 1028 simulations have been performed.
\label{tab:simulation_summary}}
\end{center}
\end{table}

We list the parameters considered in this study in \Tab{tab:simulation_summary}.
We expand upon Ref.~\cite{Jinno:2022mie} by including in our parameter scan strong PTs with $\alpha = 0.5$.
We thus run simulations for $\alpha \in \{ 0.0046,\, 0.05,\, 0.5\}$ and wall velocities $v_{\rm{w}} \in \{0.32,\, 0.36,\, ..., 0.76,\, 0.8\},$
where the dots mean increases by $0.04$, 
except for strong PTs where $v_{\rm{w}}=0.32$ is excluded due to the non-existence of
deflagrations  for $\alpha \gtrsim \onethird (1 - \vw)^{-13/10}$
\cite{Espinosa:2010hh}, 
implying a total of $3\times 13 - 1 = 38$ PT parameter points. 
To extract our main results,
we run \emph{reference} simulations for each simulation box size, in which a single reference bubble nucleation history is used for all wall velocities, PT strengths, and grid sizes, thus eliminating the effect of sample variance when comparing the results for different parameter sets.

The bubble nucleation histories result in a number of bubbles
of the order of 
$N_b \simeq  \tilde L^3/(8 \pi \, \vw^3)$,
where $\tilde L \equiv L\beta$ is the simulation box size,
nucleated following a statistical distribution that is exponential in time and uniform in space, as described in \Sec{nucleation_hist},
and then removing bubbles that nucleate inside the future
causal cone of previous bubbles to take into account
the evolution of the broken-phase volume with time (see Ref.~\cite{Jinno:2022mie} for details).
In our simulations,
$\tilde L/\vw$
takes on values
of 20 and 40, yielding of the order of 300 and 2500 bubbles respectively.
Using the same numerical resolution, simulations with
$\tilde L/\vw = 40$ yield a reduction in the statistical variance by increasing the number
of bubbles and by offering an increased resolution of the measured quantities in the IR regime, while simulations with
$\tilde L/\vw = 20$ cover a larger dynamical range in the UV regime.
For comparison, the number of bubbles for $\tilde L/\vw = 40$ ($N_b \simeq 2500$) in our work
and previous Higgsless simulations \cite{Jinno:2022mie} is in general larger
than most\footnote{We note that Ref.~\cite{Hindmarsh:2015qta} uses $N_b = 32558$ for
a weak PT with $\vw =0.44$, while it takes either $N_b = 988$, 125, or 37 for the rest of the PT parameter space.
In Ref.~\cite{Hindmarsh:2017gnf}, 5376 bubbles are used
for some weak PTs, while 11 and 84 are used for other
weak PTs, and for intermediate ones.
Reference~\cite{Cutting:2019zws} considers 8 bubbles
for all simulations.} of the previous numerical simulations of the fluid-scalar system
\cite{Hindmarsh:2015qta,Hindmarsh:2017gnf,Cutting:2019zws}, especially for intermediate PTs,
allowing
for a reduction of the statistical variance.
A potential issue of small box sizes is that for small wall velocities,
the shock in front of the wall of the first nucleated bubble might collide with its mirror images
(due to the use of periodic boundary conditions) before the
end of the PT.
To avoid this issue, we take the minimum value of $\vw$ to be 0.32
in our simulations, such that the
numerical domain is filled with the broken phase
 before the largest bubble reaches the edges
of the simulation box even for the smaller box $\tilde L/\vw = 20$.

For each of the 76 parameter points $\{v_{\rm w}, \alpha, \tilde L/\vw \}$,
we then run simulations with different number of grid points $N^3$ with
$N\in \{64,\, 128,\, 256,\, 512\}$, yielding
a total of $76 \times 4 = 304$ {\em reference} simulations.
Running simulations of different grid sizes allows us 
to test the degree of convergence of our numerical
results and
to estimate physical 
quantities in the continuum limit by extrapolation (see \Sec{sec:convergence}).
To ensure the stability of our simulations,
we choose the number of time steps $N_t =\tilde t_{\rm{end}}/\delta{\tilde{t}}$ to satisfy
the Courant-Friedrichs-Lewy (CFL) condition
$\delta{\tilde{t}}/\delta{\tilde{x}}<1/4$
with $\delta{\tilde{x}} = \tilde L/N$.
We have confirmed that even for strong transitions, increasing $N_t$ beyond this threshold does not significantly change the numerical results.
For each parameter point $\{\vw, \alpha, \tilde L/\vw\}$, we have also run {\em single-bubble} simulations to track the convergence of the self-similar fluid profiles, leading to 76 simulations.
These results are presented in \App{sec:kinetic_ed} and will be used
to improve the extrapolated predictions of the {\em reference} multiple-bubble simulations in \Sec{sec:convergence}.
We note that {\em single-bubble} simulations are only run until $\tilde t_{\rm end} = \tilde L/[2\max(\cs, \vw)]$, being roughly the time when the fluid shell reaches the edge of the simulation domain.

In addition to the {\em reference} simulations, we also run 
multiple-bubble simulations based on 9 additional distinct bubble nucleation histories for all strengths, resolutions, and box sizes, for $\vw
\in\{0.32/0.36,\, 0.6,\, 0.8\}$, where the lower $\vw = 0.32$ is used for weak and intermediate transitions, and $\vw=0.36$ for strong ones.
These velocities correspond to deflagrations, hybrids, and detonations, respectively, except for strong transitions for which also $\vw=0.8$ corresponds to a hybrid. This implies a total of $3 \times 3 \times 2 \times 4 \times 9 = 648$ \emph{seed} simulations from which
the statistical variance of the results can be estimated.
We will use these simulations to provide error bars in our measured quantities,
corresponding to the standard deviation from the 10 different bubble
nucleation histories in \Sec{sec:results}.

All {\em reference} and {\em seed} simulations are run between
$0 < \tilde t \equiv t \beta < 32$ and the GW spectrum
is extracted from
the time interval spanning from
$\tilde t_{\rm init} = 16$ to $\tilde t_{\rm end} = 32$.
We set the origin of time coordinates,
$\tilde t_\mathrm{ref} = 0$, at
a reference value such that the first bubble nucleates at $\tilde t_n = 0.5$, based on the invariance of our equations to time translations when the expansion of the Universe can be ignored.
Then, the results of the simulations in flat space-time accurately model the PTs in
an expanding Universe
when $\beta/H_\ast \gg \tilde t_{\rm end} = 32$.
We specifically cut out the early times up to
$\tilde t_{\rm init} = 16$ to extract the contribution from the fluid perturbations after the collisions of bubbles, and to reduce the realization-dependent effects on the GW production.
Consequently, we also neglect the
contribution to the GW spectrum of the initial collisions 
(see also the discussion in Ref.~\cite{Jinno:2020eqg}).
In this regime, we then compute ${\cal I}_{\rm sim} (\tilde t_{\rm init}, \tilde t_{\rm fin}, \tilde k)$ that allows us to robustly
test the scaling of \Eq{OmGW_general} and compute
the GW efficiency $\tilde \Omega_{\rm GW}$ and the spectral shape
$S(k R_\ast)$.
The time $\tilde t_{\rm init} = 16$ is shortly after
the time when the broken phase fills up
the whole volume of the simulation,
$\tilde t_0 \simeq 10$, for the reference nucleation history
with $\tilde L/\vw = 20$.
We will consider times $\tilde t > \tilde t_0$
to fit the time evolution
of the kinetic energy fraction $K(\tilde t) = K_0 \, (\tilde t/\tilde t_0)^{-b}$ in \Sec{decay_K2}.

In total, we have performed 1028 simulations, which we summarize in \Tab{tab:simulation_summary}, with an estimated time of $\,\sim 10^6$ CPU hours.
We note that
each large-resolution simulation ($N=512$) takes $\sim 10^3$ CPU hours, a quite modest value that indicates the
numerical efficiency
of the Higgsless approach.

%%%%%%%%%%%%%%%%%%%%%%%%%%%%%%%%%%%%%%%%%%%%%%%%%%%%%%%%%%%%%%%%%%%%%%
\section{Numerical results\label{sec:results}}
%%%%%%%%%%%%%%%%%%%%%%%%%%%%%%%%%%%%%%%%%%%%%%%%%%%%%%%%%%%%%%%%%%%%%%

Before we present a detailed account of our numerical results, we would like to put them in perspective.
Overall, our results can be summarized as follows:

\begin{figure*}[t]
    \centering
    \includegraphics[width=0.99\columnwidth]{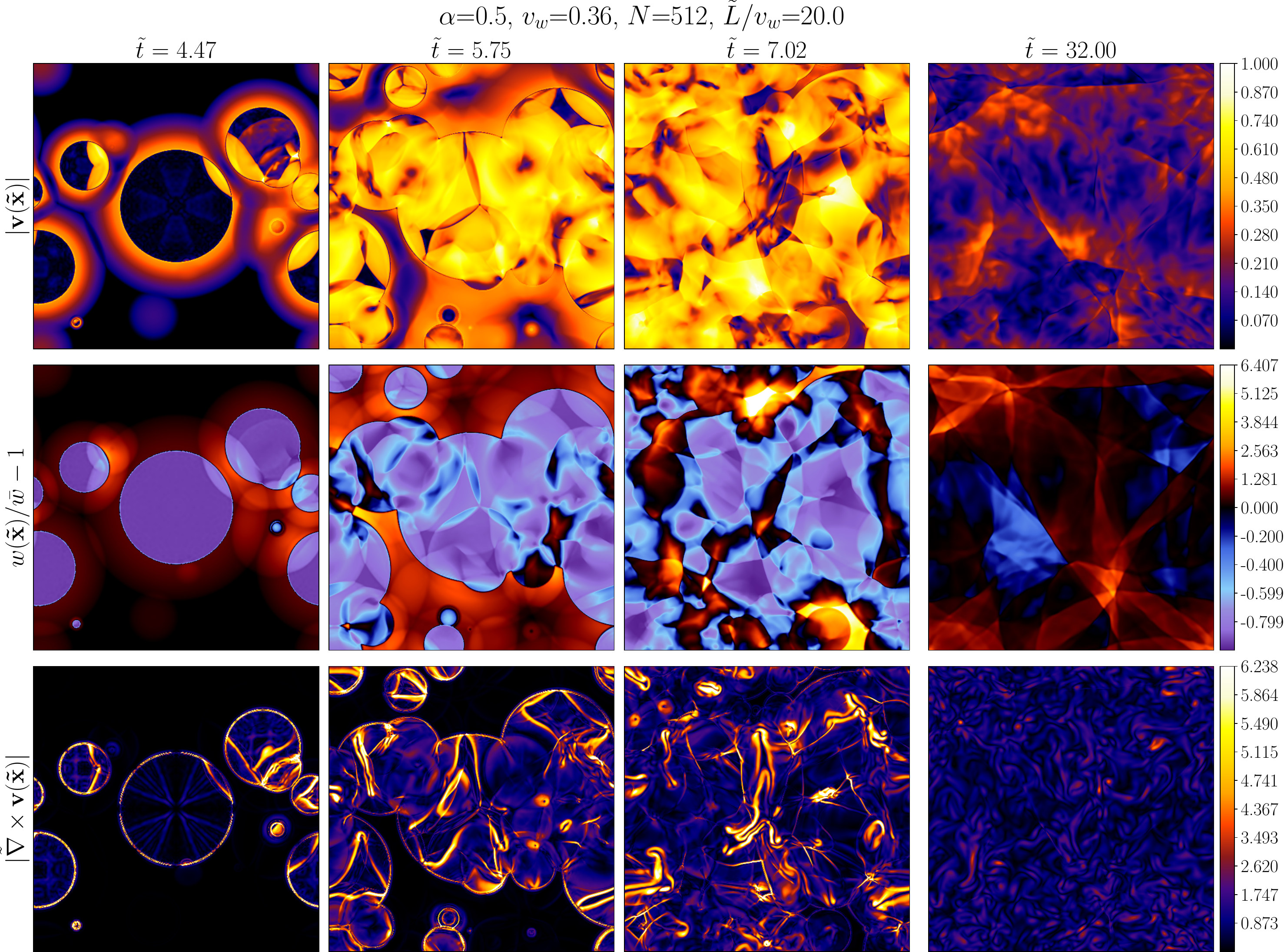}
    \caption{Velocity (upper panel), enthalpy fluctuations (middle panel),
    and vorticity (lower panel) in an $xy$-plane
    slice of the simulation volume at $z = 0$ and at
    different times $\tilde t$,
    for a strong PT with $\alpha = 0.5$ and wall velocity $\vw = 0.36$,
    which corresponds to a deflagration (see \Fig{fig:1d_profiles}). \label{fig:bubbles}
}
\end{figure*}

\begin{itemize}
\item {\em Simulations of strong first-order PTs with $\alpha = 0.5$}:
We present results of simulations of strong first-order PTs, covering a wide range of wall velocities and performing systematic checks of the numerical convergence of our results.
For the first time, we obtain GW spectra from fluid motion
for strong PTs.\footnote{Reference~\cite{Cutting:2019zws} also provides estimates of the kinetic energy and the integrated
GW spectra for $\alpha = 0.5$, but does not
present results about the spectral shape.}
Simulations of strong PTs are more challenging when it comes to numerical stability and proper resolution
of non-linearities.
At the same time, stronger PTs lead to a larger GW signal and therefore are
preferred for potential detection,
hence the importance of developing an accurate understanding of the resulting GW spectrum.
We provide in \Sec{sec:summary}
a template, based on the expected GW spectrum from 
compressional fluid perturbations (obtained within the assumption that the UETC of the
source is stationary in time), but
extended to intermediate and
strong PTs, for which the kinetic energy decays (see \Secs{GW_sw}{sw_extended}).
The resulting integrated GW amplitude is studied numerically in \Sec{GW_spec_time}
and the model of \Sec{sw_extended} is validated against the numerical results.
This allows us to incorporate
information from our simulations
to be used for 
phenomenological studies.
We show in \Fig{fig:bubbles} a simulation example
of a strong PT that corresponds to
a deflagration with $\vw = 0.36$.

\item {\em Development of non-linearities}:
For strong PTs, and some intermediate PTs with confined hybrids,
the simulations show
several phenomena that probably
stem from non-linear dynamics of the fluid.
First and foremost, we observe a decay in the kinetic energy density of
the fluid at late times (after the PT ends), which
could be compatible with the formation of non-linearities
leading to a cascading of energy from larger
to smaller scales in the fluid perturbations, making the
viscous dissipation at small scales more effective.
Potentially due to this decay,
we find that
the amplitude of the GW spectrum
deviates from the linear dependence with the source duration predicted by the sound-shell model for weak PTs, under the assumption of source stationarity. 
For intermediate and strong PTs we observe that, as the simulation proceeds, the amplitude of the GW spectrum grows slower than linearly with the source duration.
While we do not run the simulations sufficiently long to observe the ultimate saturation of the GW spectrum resulting from this behavior, we do indeed expect that the GW spectrum amplitude eventually saturates as the kinetic energy dampens and vorticity
dominates the fluid motion
\cite{RoperPol:2019wvy,RoperPol:2021xnd,RoperPol:2022iel}.
The cascading of kinetic
energy from large to small scales would
also impact the UV part of the GW spectrum,
from the
$k^{-3}$ found
for sound waves
\cite{Hindmarsh:2013xza,Hindmarsh:2015qta,Hindmarsh:2016lnk,Hindmarsh:2017gnf,Jinno:2020eqg,Jinno:2022mie,RoperPol:2023dzg,Sharma:2023mao} towards a shallower spectrum, for example 
the $k^{-8/3}$ inferred from the assumption of Kolmogorov spectrum
in vortical turbulence
\cite{Kosowsky:2001xp,Caprini:2006jb,Gogoberidze:2007an,Caprini:2009yp,
Niksa:2018ofa,RoperPol:2019wvy,RoperPol:2022iel}.
We indeed observe a shallower spectrum in the UV regime in simulations of intermediate and strong PTs.
While these phenomena are all consistent with what would be expected from the development of non-linearities in the bulk fluid motion, a solid physical interpretation of the simulations would still require more evidence and more checking, as numerical resolution can become an issue in simulating intermediate and strong PTs.
We test the numerical robustness of our results in the following of this section, and comment on future studies that would be 
required to confirm some of our findings.
In particular, we present in \App{sec:vort} a preliminary study
of the development of vorticity in our simulations, and also show the vorticity found in one 
of our simulations  in 
the lower panel of \Fig{fig:bubbles}.

\item
{\em Template parameterizations}:
We express all our numerical findings
in terms of 
a few physical quantities, to facilitate their use in phenomenological studies (numerical fits are sometimes still necessary, e.g., for the time-decay parameters).
All quantities evaluated in the simulations are dimensionless, such that $\beta/H_\ast$ does not
enter in the
numerical results, and only appears when we recover the
physical quantities, as indicated in \Eq{OmGW_Ik}.
This motivated the authors in Refs.~\cite{Jinno:2020eqg,Jinno:2022mie} to use the variable $Q'$
[see \Eq{Qprime_lin}] to interpret the numerical results.
In the present work, we instead characterize them with
$R_\ast$ and $K_{\rm int}^2$ [see \Eq{OmGW_general}]. 
This allows
to capture the essential results of GW generation by fluid motion
in a form as simple as possible, because $R_\ast$ and $K_{\rm int}^2$ are quantities determining the fluid dynamics, as opposed to quantities related to the PT such as $\beta$.
Indeed, normalizing to $R_\ast$ and $K_{\rm int}^2$, the GW efficiency $\tilde \Omega_{\rm GW}$ becomes
{almost} invariant over both $v_w$ and duration
of the source.
Furthermore, we introduce $K_{\rm int}^2$ to allow
for deviations with respect to
the linear dependence of the GW amplitude with the source duration, which is expected for
stationary sources, but is no longer valid for decaying ones
(see discussion in \Secs{GW_sw}{sw_extended}).
We also provide in \Sec{sec:expansion} a definition of $K_{\rm int, exp}^2$
that allows to incorporate {\em a posteriori}
the effect of the expansion of the Universe [see \Eqs{OmGW_final_exp}{Kexp_fit}],
taking into account the conformal invariance of
radiation-dominated fluid dynamics \cite{Brandenburg:1996fc,RoperPol:2025lgc}.

\end{itemize}

To test the validity of our numerical results,
we pay special attention to the following points,
addressed throughout this section.
In \Sec{sec:convergence}, we study the convergence 
of our results with respect to the grid spacing, $\delta{\tilde{x}}$.
In \Sec{decay_K2}, we study the time dependence of the fluid kinetic energy fraction $K$, and fit
the decaying power law presented in \Sec{sw_extended} to
the numerical results.
In \Sec{GW_spec_time}, we test the
scaling of the GW spectrum
with $K_{\rm int}^2$ and $R_\ast$ presented in \Secs{GW_sw}{sw_extended}, we analyze
the evolution of the integrated
GW amplitude with the source duration,
and compute the GW efficiency
$\tilde \Omega_{\rm GW}$, according to \Eq{OmGW_general}.
We also provide in \Sec{GW_spec_time} an estimate of the
expected GW amplitude in a flat Minkowski space-time, based on the numerical results of
\Secss{sec:convergence}{GW_spec_time},
and  in an expanding background, using the model presented
in \Sec{sec:expansion}.
Finally, in \Sec{sec:shape}, we study the spectral shape of the GW spectrum obtained in
the simulations. 
We pay special attention to the UV
regime, where we find deviations with respect to
the slope expected in the case of GW production by sound waves, i.e., $k^{-3}$.
These deviations are compatible with the shallower decay with $k$ that one would expect in the presence of a forward energy cascade due to non-linearities. 
However,
while we observe both
a decay of the kinetic energy and the distortions of the GW spectra,
to be able to confirm the presence of a forward
cascade in our simulations,
a detailed study of the dependence of the kinetic spectra on the numerical parameters
would be required.
This is beyond the scope of the present analysis and we defer it to future work.
Furthermore, 
the time evolution can also affect the spectral shape.
Indeed,
it is not in general expected that all
wave numbers evolve with the source duration in the same way, as shown in Ref.~\cite{RoperPol:2023dzg}.
Therefore, the assumption that all wave numbers follow the same evolution
as the integrated amplitude, which is studied in \Sec{GW_spec_time},
from the ending time of the simulations to the final time of GW sourcing can
affect the resulting spectral shape, especially in the IR,
where the assumption of infinitesimal compact support described in \Sec{sw_extended}
might not necessarily be satisfied.
The Universe expansion could also affect the resulting spectral shape, which is
only computed in flat space-time, for long source durations, as we discuss in \Sec{sec:expansion}.

%%%%%%%%%%%%%%%%%%%%%%%%%%%%%%%%%%%%%%%%%%%%%%%%%%%%%%%%%%%%%%%%%%%%%%
\subsection{Convergence analysis of the kinetic energy and GW amplitude} \label{sec:convergence}
%%%%%%%%%%%%%%%%%%%%%%%%%%%%%%%%%%%%%%%%%%%%%%%%%%%%%%%%%%%%%%%%%%%%%%

In the present study, for each parameter point $\{\alpha, v_{\rm{w}}\}$, we have run simulations
at four resolutions $N\in\{64,\,128,\,256,\,512\}$ and two box sizes $\tilde L/\vw \in \{20, 40\}$.
The Higgsless simulations use relatively sparse grids compared to simulations with scalar 
fields~\cite{Hindmarsh:2013xza, Hindmarsh:2015qta, Hindmarsh:2017gnf, Cutting:2019zws}.
Therefore, the resolution is not always high enough
to reproduce some of the 
self-similar fluid
profiles
surrounding the uncollided
bubbles at the initial stages of the simulations.
This occurs especially for parameter points with  $\vw \lesssim v_{\rm CJ}$,
where $v_{\rm CJ}$ is the Chapman-Jouguet speed, determining the transition between hybrids and
detonations. 
Indeed,
the fluid profiles
become very thin as $\vw$ approaches $v_{\rm CJ}$, and require a large number of lattice points to be properly resolved. 
Since the Chapman-Jouguet speed is $v_{\rm CJ} = \{0.63, 0.73, 0.89\}$ respectively for $\alpha = \{0.0046, 0.05, 0.5\}$,
for our choice of parameters (cf.~\Tab{tab:simulation_summary}),
very thin profiles develop when {\em (i)} $\vw =0.6$ and the PT is weak, {\em (ii)} $\vw = 0.72$ and the PT has intermediate
strength, and {\em (iii)} $\vw = 0.8$ and the PT is strong.
For reference, we show in \App{sec:kinetic_ed} (see \Figs{fig:1d_profiles}{fig:1d_profiles_with_sims}) the self-similar profiles
of the fluid perturbations
by uncollided expanding bubbles \cite{Espinosa:2010hh}, computed
by direct integration of the fluid equation across the wall using {\sc CosmoGW} \cite{cosmogw}.
The development of the thin hybrid solutions appears clearly.

The fluid profile is described in terms of the variable
$\xi \equiv r/(t - t_{n})$,
with $r$ being the radial distance to the nucleation location and $t_n$
the time of nucleation.
Therefore, the resolution in $\xi$ 
for a fixed $N$ is initially low and then improves as
time evolves. 
Despite, for the aforementioned cases with $\vw$ close to $v_{\rm CJ}$, this increase in resolution is
not enough to fully resolve the
self-similar profiles at the time
of bubble collision, which is the relevant one for us. 
Indeed, we are interested in the GW production by fluid motion after bubble collision, and not resolving the single bubble fluid profile at the collision time affects the total kinetic energy available for the GW production.
We study the rate of convergence of the kinetic energy in \App{sec:kinetic_ed}, comparing multiple-bubble {\em reference}
with {\em single-bubble} simulations. 
We confirm that the under-resolution of the fluid shell profiles for some set of parameters leads to an underestimation of the total kinetic energy available for the GW production, and we propose a method to partially correct for this under-estimation, also discussed at the end of this section.

In addition to the required resolution for thin self-similar profiles before collisions, as the fluid perturbations become non-linear, we expect large numerical resolution to be required also to
fully capture the dynamics during and after collisions.
In the following, we first analyze the convergence of the numerical
results for multiple-bubble {\em reference} runs, to provide estimates of the time-integrated squared kinetic
energy $K_{\rm int}^2 (\tilde t_{\rm init}, \tilde t_{\rm end})$, 
defined in \Eq{K2rms},
and of the integrated GW amplitude ${\cal I}_{\rm sim}^{\rm int} \equiv \int {\cal I}_{\rm sim} \dd \ln k$,
where ${\cal I}_{\rm sim} (\tilde t_{\rm init}, \tilde t_{\rm end}, \tilde k)$
is defined in \Eq{Weinberg}, for initial and final times $\tilde t_{\rm init} = 16$ and $\tilde t_{\rm end} = 32$,
corresponding to the time interval
over which the GW spectrum is computed in the simulations.
These
quantities will be used in \Secs{GW_spec_time}{sec:shape} to estimate
the
GW efficiency $\tilde \Omega_{\rm GW}$ and the spectral shape $S(kR_\ast)$.
We also provide estimates for
the kinetic energy fraction $K_0$ evaluated
at the time when the PT completes, $\tilde t_0 \simeq 10$.
We will then attempt to improve this estimate by including
the results of  {\em single-bubble} runs studied in \App{sec:kinetic_ed}, 
tracking the degree of convergence of each bubble at the
time when they collide, and leading to a new estimate, ${\cal K}_0$, defined in \Eq{eq:calK_0 definition} (see \App{sec:kinetic_ed} for details).
Note that we will use the improved estimates ${\cal K}_0$ 
in the GW
templates presented in \Sec{sec:summary}.

In order to take the effect of numerical resolution into account, we study 
the numerical results as a function of the number of grid points
$N$ and attempt to improve our estimates by extrapolating 
our results to $N\to \infty$, based on the underlying assumption that the extrapolation
method obtained for the computed values of $N$ also applies for $N > 512$.
On general grounds, it is possible to define a particular number of grid points $N_\ast$ such that, for $N \gg N_\ast$, a simulation has reached
a converged solution, meaning that the numerical results are unaffected,
within some acceptable tolerance, by changing $N$: hence,
we can assume that they accurately represent the continuum results.
Empirically, we find that, for insufficient resolution, 
the kinetic energy fraction $K$ is in general {\em underestimated}
when the grid resolution is insufficient (see also \App{sec:kinetic_ed}),
as the velocity profiles around
the peak are under-resolved.
This motivates us to use the following function when extrapolating the 
numerical values of the kinetic energy fraction:
\begin{equation}
\label{eq:E_kin_convergence}
K = \frac{K_\infty}{1+(N_*/N)^a}\,,
\end{equation} 
where $a$, $N_\ast$, and $K_\infty$ are found by fitting the numerical results as a function of $N$.
We also define the relative error in the kinetic energy, $\varepsilon_K$:  
\begin{equation}
    \varepsilon_K \equiv \frac{K_\infty - K}{K_\infty} = \left(\frac{\delta{\tilde{x}}}{\delta{\tilde{x}}_\ast}\right)^a + {\cal O} \left(\left(\frac{\delta{\tilde{x}}}{\delta{\tilde{x}}_\ast}\right)^{2a}\right)\,,
    \label{rel_errorK}
\end{equation}
where the last equality comes from expanding in $\delta{\tilde{x}}/\delta{\tilde{x}}_\ast\ll 1$ with 
$\delta{\tilde{x}}_\ast = \tilde L/N_\ast$,
which holds when $N\gg N_*$.
The last equality shows that
the value of $a$ in \Eq{eq:E_kin_convergence}
indicates
the degree of convergence of the numerical results.

\begin{figure*}
    \centering
    \includegraphics[width=.32\columnwidth]{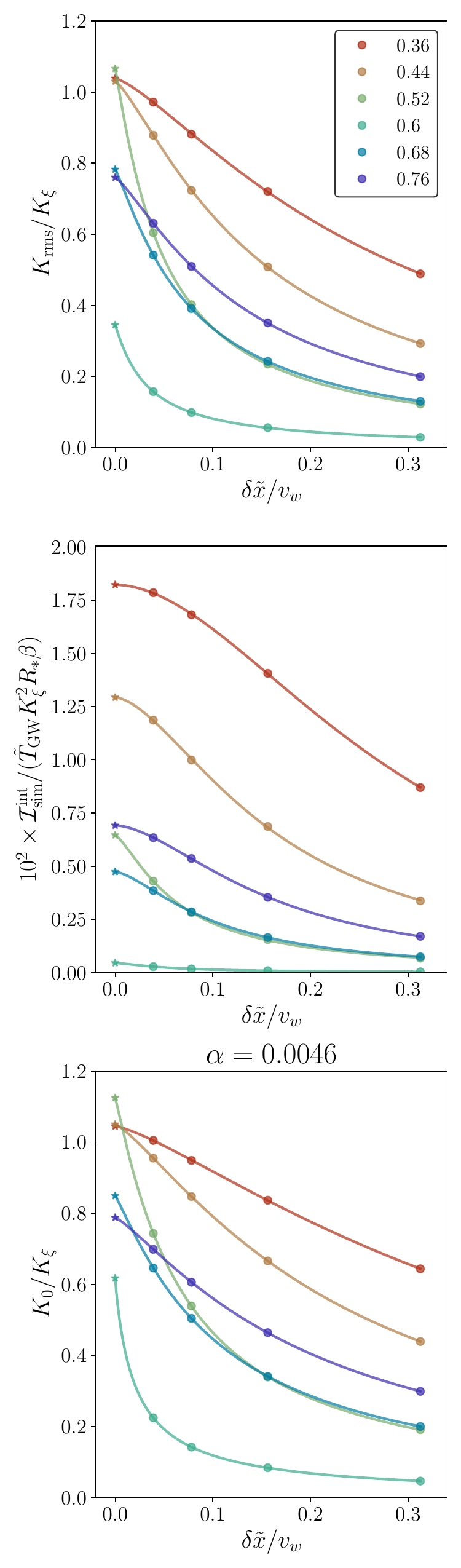}
    \includegraphics[width=.32\columnwidth]{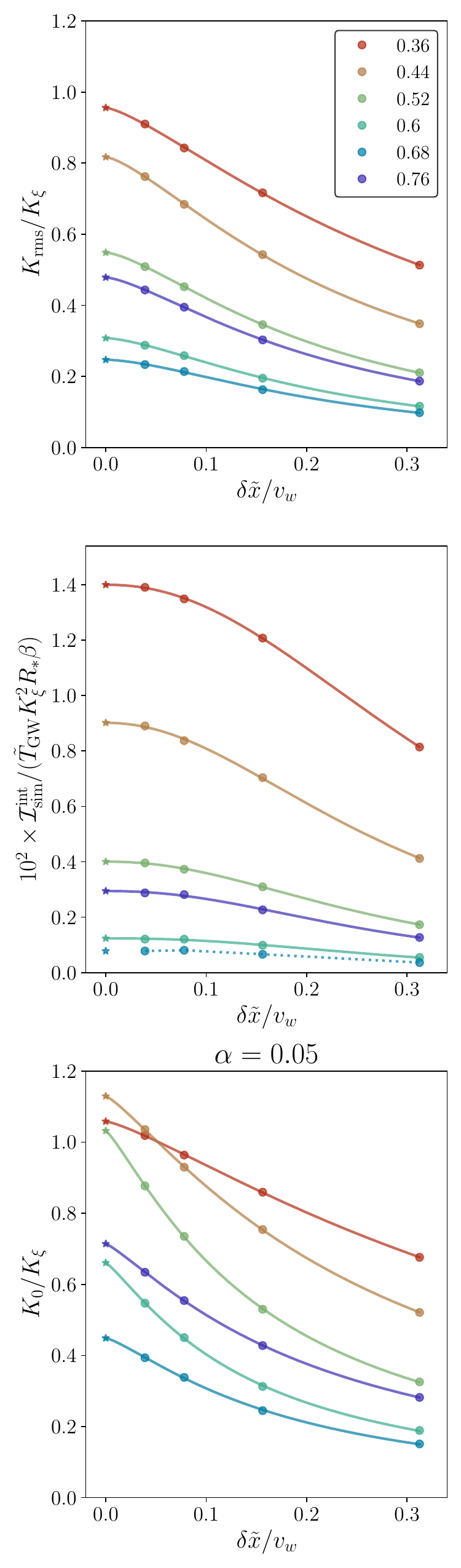}
    \includegraphics[width=.32\columnwidth]{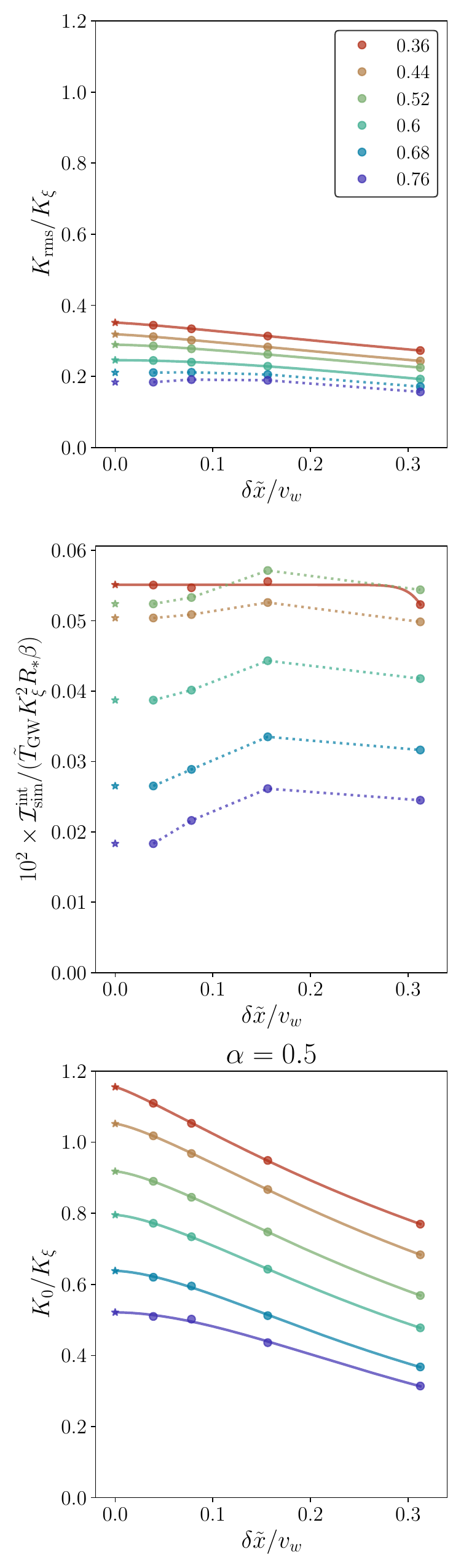}\\
    \caption{
    Plots showing the kinetic energy fraction $K$ and the integrated GW spectrum
    ${\cal I}_{\rm sim}^{\rm int}$
    as a function of grid spacing $\delta{\tilde{x}}/\vw = (\tilde L/\vw)/N$ for runs with $\tilde L/\vw = 20$.
    Upper and middle panels respectively show
    the rms value of the kinetic energy fraction
    $K_{\rm rms}\, \equiv K_{\rm int}/\tilde T_{\rm GW}^{1/2}$,
    normalized by the single-bubble kinetic energy fraction
    $K_\xi$ [see \Eq{Kxi} and \Tab{tab:kappas}], and the integrated GW spectrum ${\cal I}_{\rm sim}^{\rm int}$, normalized by a reference value
    $\tilde \Omega_{\rm GW} \sim 10^{-2}$ \cite{Hindmarsh:2013xza,Hindmarsh:2015qta,Hindmarsh:2017gnf}, and by $\tilde T_{\rm GW} K_\xi^2 R_\ast \beta$, based on the expected
    scaling of \Eq{OmGW_general}.
    Both $K_{\rm rms}$ and ${\cal I}_{\rm sim}^{\rm int}$ are computed for $\tilde t_{\rm init} = 16$ and $\tilde t_{\rm end} = 32$, with $\tilde T_{\rm GW} = 16$.
    The lower panels show the kinetic energy fraction
    at the time when
    all the simulation domain is in the broken phase around $\tilde t_0 \simeq 10$,
    $K_0$, also normalised by $K_\xi$. 
    Left, middle, and right columns are weak, intermediate, and strong PTs respectively.
    Solid lines show the least-squares fits of the extrapolation scheme given in \Eq{eq:E_kin_convergence}
    when the fit is valid, while dotted lines indicate the numerical trend when
    the fit is not valid (see discussion in the main text).
    Stars indicate the
    extrapolated values at $\delta{\tilde{x}} \to 0$.
    \label{fig:extrapolation}}
\end{figure*}

The result of applying the convergence analysis based on \Eq{eq:E_kin_convergence} to the {\em reference} runs
with multiple bubbles
is shown in \Fig{fig:extrapolation}
(for the full evolution of the kinetic energy, see \Sec{decay_K2} and \Fig{fig:E_kin_evolution}).
In the top row of panels in \Fig{fig:extrapolation}, 
we show the rms
kinetic energy fraction, computed from $K^2_{\rm int}(\tilde t_{\rm init},\tilde t_{\rm end})$ [see \Eq{K2rms}]: 
\begin{equation}
  K_{\rm rms}^2 \equiv \frac{1}{\tilde T_{\rm GW}}
   \int_{\tilde t_{\rm init}}^{\tilde t_{\rm end}} K^2 (\tilde t) \dd \tilde t =
   \frac{K_{\rm int}^2 (\tilde t_{\rm init}, \tilde t_{\rm end})}{\tilde T_{\rm GW}}
   \,, \label{Krms}
\end{equation}
where $\tilde T_{\rm GW} \equiv \tilde t_{\rm end} - \tilde t_{\rm init} = 16$. 
According to the model proposed in \Sec{sw_extended}, $K_{\rm int}^2 \equiv K_{\rm rms}^2 \, \tilde T_{\rm GW}$ is the relevant
quantity entering in the GW amplitude ${\cal I}_{\rm sim}^{\rm int}$.
The rms kinetic energy fraction is shown normalized by the single-bubble value $K_\xi$
(see \Tab{tab:kappas} for numerical values).

We focus on the simulations with $\tilde L/\vw = 20$, as these
cover smaller scales than those with $\tilde L/\vw = 40$
for a fixed
$N$, providing a better resolution of the kinetic 
energy density in the UV regime.
The dots in \Fig{fig:extrapolation} correspond to the four resolutions $N\in\{64,\,128,\,256,\,512\}$, given in terms of $\delta \tilde x/\vw=20/N$, while the continuous lines show the fit in \Eq{eq:E_kin_convergence}.
In \Tab{tab:conv_fit}, we provide the resulting
values of the fit
parameters of \Eq{eq:E_kin_convergence}:
$K_\infty^{\rm rms}$ normalized by the single-bubble
kinetic energy fraction
$K_\xi$ [see \Eq{Kxi}], the degree of convergence $a_K$,
and the relative errors $\varepsilon_K$ as defined in \Eq{rel_errorK},
for the set of PT parameters shown in \Fig{fig:extrapolation}.
In some simulations,
the numerical results have already converged for $N = 512$, as indicated by a small
relative error $\varepsilon_K \equiv |K - K_\infty|/K_\infty$
(see values in \Tab{tab:conv_fit}).
Moreover, the fit in \Eq{eq:E_kin_convergence} represents well most of the simulations, in particular whenever  
the kinetic energy fraction is underestimated at low resolution, and remains almost constant over the simulation time. 
However, there are a few cases with $\alpha = 0.5$ and large $\vw$, in which the trend is different (depicted as dashed lines in \Fig{fig:extrapolation}).
These exceptional cases correspond to strong PTs, in which non-linearities play a relevant role. As we will see in \Sec{decay_K2}, where we study the time evolution of $K(t)$ for different $N$, in strong PTs, increasing the resolution leads to a \emph{faster} decay of the kinetic energy density with time.
Therefore, $K_{\rm rms}$, which is time integrated, can become smaller when increasing $N$, inverting the trend with respect to \Eq{eq:E_kin_convergence}.
Whenever the fit of \Eq{eq:E_kin_convergence} is not valid,
we take $K^{\rm rms}_\infty$
to be the value computed in the simulations for $N = 512$, and the error $\varepsilon_{\rm K}$ is then estimated comparing this value to the one obtained for $N = 256$.

\begin{table}[b]
\centering
{\scriptsize
\begin{tabular}{|c|c|c|c|c|c|c|c|c|c|c|c|c|c|}
\hline
$\alpha$ & $\tilde{L}/\vw$ &  $v_w$ & $K^{\mathrm{rms}}_{\infty}/K_{\xi}$ & $a_{K}$ & $\varepsilon_K$ & ${\mathcal{I}_{\infty}^{\mathrm{int}}}$ & $a_{\mathcal{I}}$ & $\varepsilon_{\mathcal{I}}$ \\ \hline
\hline
  0.0046 &          20 &   0.36 &                                1.04 &              1.34 &                  $6.41\times 10^{-2}$ &                                          $9.97 \times 10^{-10}$ &                                  1.88 &                                        $2.04\times 10^{-2}$ \\
         &             &   0.44 &                                1.03 &              1.29 &                  $1.47\times 10^{-1}$ &                                          $2.11\times 10^{-9}$ &                                  1.64 &                                        $8.26\times 10^{-2}$ \\
         &             &   0.52 &                                1.07 &              1.11 &                  $4.33\times 10^{-1}$ &                                          $5.35\times 10^{-9}$ &                                  1.34 &                                        $3.35\times 10^{-1}$ \\
         &             &   0.60 &                                0.35 &              1.06 &                  $5.44\times 10^{-1}$ &                                          $5.36\times 10^{-9}$ &                                  1.36 &                                        $3.83\times 10^{-1}$ \\
         &             &   0.68 &                                0.78 &              1.16 &                  $3.08\times 10^{-1}$ &                                          $2.35\times 10^{-9}$ &                                  1.51 &                                        $1.87\times 10^{-1}$ \\
         &             &   0.76 &                                0.76 &              1.26 &                  $1.69\times 10^{-1}$ &                                          $1.19\times 10^{-9}$ &                                  1.69 &                                        $8.38\times 10^{-2}$ \\ \hline
    0.05 &          20 &   0.36 &                                0.96 &              1.35 &                  $4.80\times 10^{-2}$ &                                         $8.65\times 10^{-6}$ &                                  2.16 &                                        $6.70\times 10^{-3}$ \\
         &             &   0.44 &                                0.82 &              1.40 &                  $6.72\times 10^{-2}$ &                                          $1.24\times 10^{-5}$ &                                  2.04 &                                        $1.26\times 10^{-2}$ \\
         &             &   0.52 &                                0.55 &              1.46 &                  $7.21\times 10^{-2}$ &                                          $1.33\times 10^{-5}$ &                                  2.14 &                                        $1.20\times 10^{-2}$ \\
         &             &   0.60 &                                0.31 &              1.54 &                  $6.52\times 10^{-2}$ &                                          $1.04\times 10^{-5}$ &                                  2.43 &                                        $2.13\times 10^{-2}$ \\
         &             &   0.68 &                                0.25 &              1.61 &                  $5.52\times 10^{-2}$ &                                          $8.10\times 10^{-6}$ &                                 -- &                                       $3.49\times 10^{-2}$ \\
         &             &   0.76 &                                0.48 &              1.44 &                  $7.37\times 10^{-2}$ &                                          $6.98\times 10^{-6}$ &                                  2.21 &                                        $2.28\times 10^{-2}$ \\ \hline
     0.5 &          20 &   0.36 &                                0.35 &              1.25 &                  $2.06\times 10^{-2}$ &                                          $1.82\times 10^{-3}$ &                                 25.03 &                                        $7.14\times 10^{-4}$ \\
         &             &   0.44 &                                0.32 &              1.27 &                  $2.05\times 10^{-2}$ &                                          $2.18\times 10^{-3}$ &                                 -- &                                       $9.55\times 10^{-3}$ \\
         &             &   0.52 &                                0.29 &              1.44 &                  $1.26\times 10^{-2}$ &                                          $2.79\times 10^{-3}$ &                                 -- &                                       $1.74\times 10^{-2}$ \\ 
         &             &   0.60 &                                0.25 &              1.85 &                  $3.97\times 10^{-3}$ &                                          $2.56\times 10^{-3}$ &                                 -- &                                       $3.68\times 10^{-2}$ \\ 
         &             &   0.68 &                                0.21 &             -- &                 $3.97\times 10^{-3}$ &                                          $2.26\times 10^{-3}$ &                                 -- &                                       $8.88\times 10^{-2}$ \\
         &             &   0.76 &                                0.18 &             -- &                 $3.91\times 10^{-2}$ &                                          $1.80\times 10^{-3}$ &                                 -- &                                       $1.80\times 10^{-1}$ \\ \hline
\hline
\end{tabular}
}
\caption{
Numerical values of the fit parameters $a$, $K_\infty^{\rm rms}$,
and ${\cal I}_\infty^{\rm int}$ of \Eq{eq:E_kin_convergence} for the rms value of the
kinetic energy fraction $K_{\rm rms}/K_\xi$ and the integrated
GW amplitude ${\cal I}_{\rm sim}^{\rm int}$, respectively
shown in the top and middle panels of \Fig{fig:extrapolation}.
We present the relative errors $\varepsilon$
computed by comparing the extrapolated values
to those obtained in the largest resolution runs $N = 512$
when the fit works (see
discussion in the main text).
Otherwise (indicated with `--' in the values of $a$), the error is computed from the
relative difference between the two largest resolution
runs, $N = 256$ and $N = 512$.
}
\label{tab:conv_fit}
\end{table}

In the middle row of panels in \Fig{fig:extrapolation}, we show
the integrated GW spectrum obtained in the code, ${\cal I}_{\rm sim}^{\rm int}$,
as a function of $\delta{\tilde{x}}$, 
while the numerical values of the extrapolated
${\cal I}_\infty^{\rm int}$, the fit parameter $a_{\cal I}$,
and the relative error $\varepsilon_{\cal I}$ are given in \Tab{tab:conv_fit}.
For the integrated GW spectrum,
the fit of \Eq{eq:E_kin_convergence} is valid
for weak and intermediate PTs, but 
not for strong PTs, for most of the wall velocities. 
Indeed, the quantity plotted in \Fig{fig:extrapolation} is expected to be
[see \Eqs{OmGW_general}{Krms}]
\begin{equation}
\frac{10^2\times {\cal I}_{\rm int}^{\rm sim}}{\tilde T_{\rm GW} K^2_\xi\,R_*\beta }\simeq
\frac{K^2_{\rm rms}}{K^2_\xi}\,,
\end{equation}
and the presence of the square of the rms kinetic energy worsen the
effect already observed for $K_{\rm rms}$, due to the particular behaviour of the time evolution of the kinetic energy with increasing resolution.
In the cases when the fit is not valid, we take the extrapolated values in \Fig{fig:extrapolation} as those obtained for the
largest resolution runs with $N = 512$.

We find empirically that the exponent $a$ in \Eq{eq:E_kin_convergence} usually varies between one and two, 
indicating that the dynamics of the system reduces
the effective degree of convergence with respect to the
one expected from the numerical scheme, which corresponds to second order \cite{Jinno:2020eqg,Jinno:2022mie}.
We expect that further decreasing $\delta{\tilde{x}}$ significantly below $\delta x_\ast$
would be required to find an exact quadratic dependence of the error with $\delta{\tilde{x}}$.
In any case, we note that for most of the PTs (besides highly confined profiles with $\vw \lesssim v_{\rm CJ}$), we already find absolute errors below 10\%, as indicated in \Tab{tab:conv_fit}.
For confined profiles,
the relative error is large, and we need to take into account that the
extrapolated result $K_\infty$ presents a larger degree of uncertainty.
In these simulations, we expect
the lack of convergence to also become visible in the GW spectra:  For example, we observe that the expected UV behavior, $S(k) \sim k^{-3}$, found in the sound-shell model \cite{Hindmarsh:2016lnk,Hindmarsh:2019phv,RoperPol:2023dzg,Sharma:2023mao} and in PT simulations
\cite{Hindmarsh:2013xza,Hindmarsh:2015qta,Hindmarsh:2017gnf} 
is in these
cases obscured by an exponential decay (see the discussion in \Sec{sec:shape} and the fit used in Ref.~\cite{Jinno:2022mie}).

\begin{figure*}
    \centering
    \includegraphics[width=.5\columnwidth]{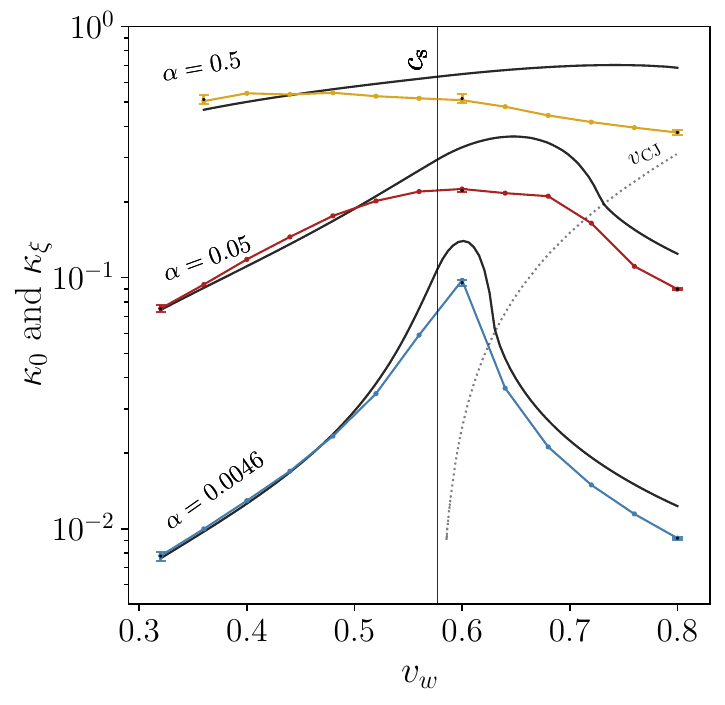}
    \includegraphics[width=1\columnwidth]{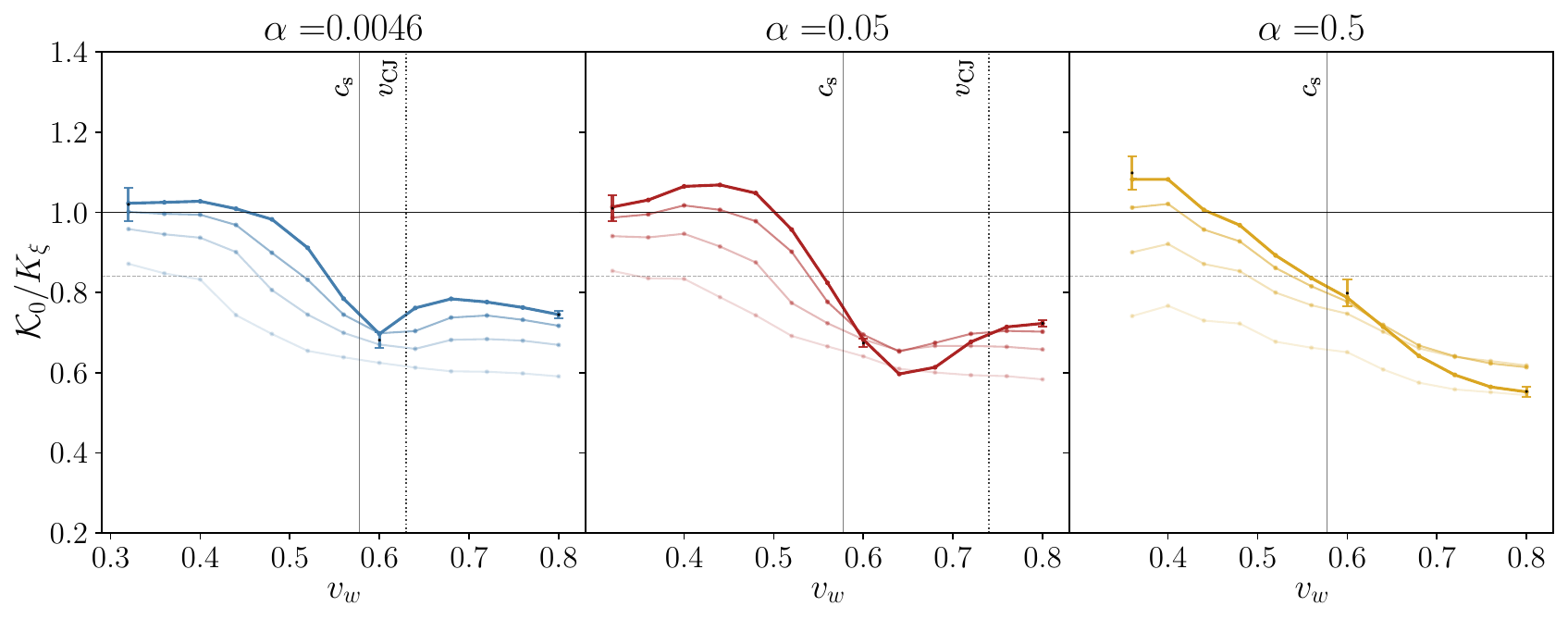}
    \caption{
   \emph{Upper panel:} Kinetic energy efficiency $\kappa_0 \equiv {\cal K}_0 (1 + \alpha)/\alpha$ obtained from
   correcting the numerical $K_0$ when $N = 512$
   and $\tilde L/\vw = 20$
   (see \App{sec:kinetic_ed}) for weak (blue), intermediate (red), and strong (orange) PTs, compared to $\kappa_\xi$ (black) for self-similar solutions [see \Eq{Kxi}].
   Vertical line corresponds to $\cs$, and $v_{\rm CJ}$ is indicated by the dotted gray line.
    \emph{Lower panel:}
    the single-bubble-corrected kinetic energy fraction at the time when the PT completes, ${\cal K}_0$, normalized to
    $K_\xi$ for self-similar profiles [see \Eq{Kxi}], for weak (left panel), intermediate (middle panel), and strong (right panel) PTs, as a function of
    $\vw$.
    Lines in increasing opacity correspond to increasing numerical resolution $N \in \{64, 128, 256, 512\}$.
    The horizontal line indicates the average value
    over $\vw$ and $\alpha$, 
    ${\cal K}_0/K_\xi \simeq 0.84$ [see \Eq{K0_fit}].
    The vertical solid gray line indicates the sound speed, $\cs$, while the dashed lines indicate the Chapman-Jouguet velocity, $v_{\rm CJ}$.
    Error bars show the standard deviation from 10 different bubble nucleation histories.
    \label{fig:kappa_eff_kappa}
}
\end{figure*}

Finally, we also display in \Fig{fig:extrapolation} the kinetic energy fraction $K_0$ at the
time when the PT ends, $\tilde t_0 \simeq 10$,
computed after getting rid of oscillations over time as obtained from the fit in \Eq{decay}
(see discussion in \Sec{decay_K2} and \Fig{fig:E_kin_evolution}).
As we will see later, in the parameterization we choose,
$K_0$  is essential to determine the amplitude of the GW spectrum, as apparent from \Eqs{OmGW_general}{Kint_decay}.
Therefore, to correctly capture the GW amplitude we need to accurately
reproduce $K_0$.
A first option can be to directly take the values $K_\infty^0$ extrapolated from
the fit of \Eq{eq:E_kin_convergence} (corresponding to the stars in the lowest panels of \Fig{fig:extrapolation}).
However,
the analysis performed in this section does not take into account 
that the self-similar profiles
have not reached convergence at the time when the bubbles collide, but only models the overall effect of the lattice resolution. We have seen that this is good enough in most of the cases for time-integrated quantities like $K_{\rm rms}$ and $\mathcal{I}_{\rm sim}^{\rm int}$. However, $K_0$ is the kinetic energy shortly after collisions, and it is therefore more sensitive to what happens in the fluid self-similar profiles.
In \App{sec:kinetic_ed},
we present a methodology to estimate the required correction
from the under-resolution of the fluid profiles.
The corrected values ${\cal K}_0/K_\xi$ are presented in \Fig{fig:kappa_eff_kappa}, together with
a modified ``efficiency'' $\kappa_0$, such that
\begin{equation} \label{eq:curlyK0}
    {\cal K}_0 \equiv \frac{\kappa_0\, \alpha}{1 + \alpha}\,,
\end{equation}
in analogy to \Eq{Kxi}.
We find a general trend that ${\cal K}_0/K_\xi \gtrsim 1$ when $\vw < \cs$,
while ${\cal K}_0/K_\xi \lesssim 1$ when $\vw > \cs$.
If we take an average value of ${\cal K}_0$ over the PT parameters
$\vw$ and $\alpha$, we find
\begin{equation}
    {\cal K}_0 = 0.84^{+0.24}_{-0.29} \, K_\xi\,, \label{K0_fit}
\end{equation}
where the super and subscripts are the maximum and minimum values found over all
wall velocities,
indicating that the typical use of $K_\xi$ for the kinetic energy would overestimate the GW production by a factor $ (K_\xi/{\cal K}_0)^2$, which can be as large as $0.55^{-2} \sim 3.3$, for example when $\alpha = 0.5$ and $\vw = 0.8$ (see \Fig{fig:extrapolation}).
For different PT parameters ($\alpha$ and $\vw$), one could take the values of \Fig{fig:kappa_eff_kappa}
to predict the correction to the resulting GW amplitude.
According to the locally stationary UETC presented in \Sec{sw_extended}, we find that $K_{\rm int}^2 \equiv K_{\rm rms}^2 \tilde T_{\rm GW}$ 
is
the relevant quantity determining the GW amplitude.
Therefore, based on the results of Ref.~\cite{Jinno:2022mie},
Ref.~\cite{Caprini:2024hue} used the estimated value $K_{\rm rms} \simeq 0.6 \, K_\xi$.
However, we note that using the power-law fit presented in \Sec{sw_extended} and validated in \Sec{decay_K2},
we can relate $K_{\rm int}^2$
to the corrected values ${\cal K}_0$ and the decay rate $b$
(as we will do in \Sec{GW_spec_time}).

\subsection{Time evolution of the kinetic energy}
\label{decay_K2}

\begin{figure*}[t]
    \centering
    \includegraphics[width=1\columnwidth]{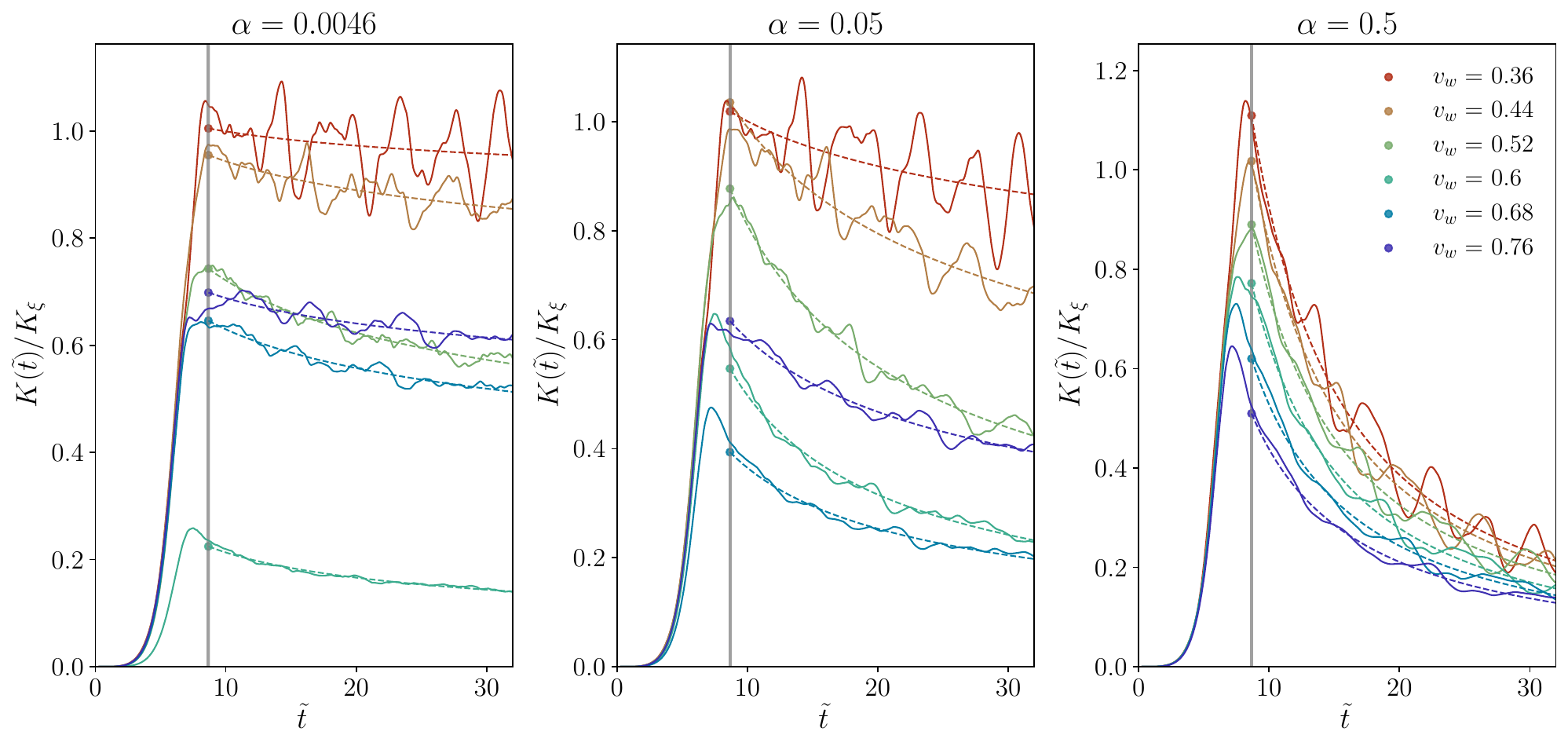}
    \includegraphics[width=1\columnwidth]{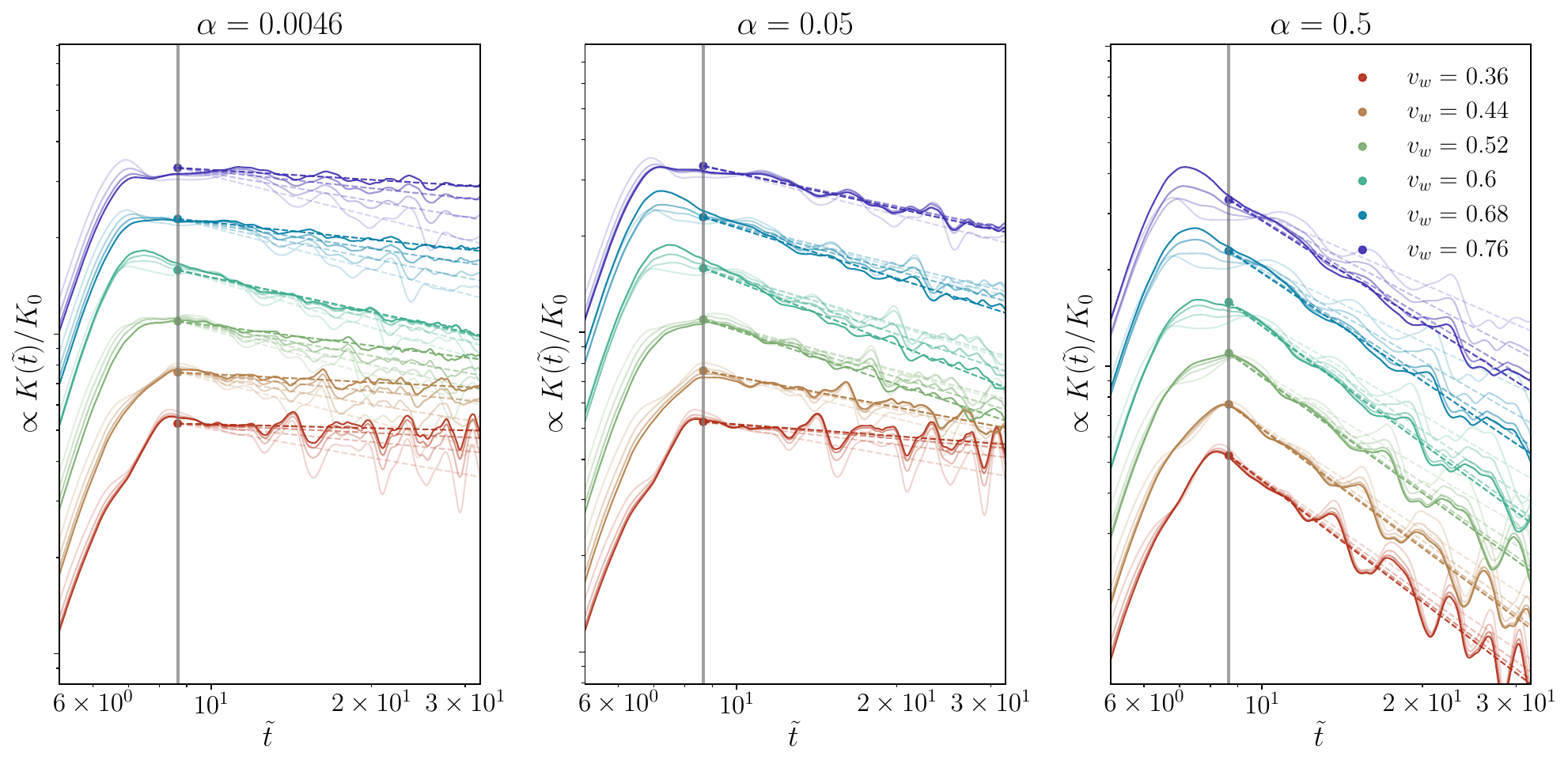}
    \caption{
    Evolution of the measured kinetic energy fraction $K(\tilde t)$ normalized to the 
    single-bubble values $K_\xi$ [see \Eq{Kxi} and \Tab{tab:kappas}] for weak (left columns), intermediate (middle columns), and strong (right columns) PTs, for $N=512$ (solid lines) and $\tilde L/\vw =20$, and the same wall velocities as those in \Fig{fig:extrapolation}.
    Dashed lines indicate the fits to the power-law decay of \Eq{decay} at times $\tilde t > \tilde t_0$.
    Values corresponding to $K_0/K_\xi$ are marked with circles.
    In the lower panel, the
    kinetic energy fraction is shown
    for different numerical discretizations
    $N = \{64, 128, 256, 512\}$ (solid lines with increasing opacity), normalized to
    the corresponding values of the fit $K_0$ at each resolution $N$. 
    The results for each $\vw$ are shifted by a constant to distinguish between wall velocities.
    This presentation in the lower panel
    is chosen to emphasize the dependence of the time
    decay on resolution.   \label{fig:E_kin_evolution}
}
\end{figure*}

In this section, we evaluate the time evolution of the
kinetic energy fraction $K$ for different numerical resolutions
$N$.
We show the results of $K(\tilde t)/K_\xi$ in \Fig{fig:E_kin_evolution} for the largest
resolution runs $N = 512$ (upper panel), and for a range of $N = \{64, 128, 256, 512\}$ (lower panel).
From the upper panels of \Fig{fig:E_kin_evolution} one can appreciate that, at late times $\tilde t \gtrsim \tilde t_0 \approx 10$, when the broken phase has filled the whole simulation volume, the kinetic energy decays with time. 
By analyzing how this decay depends on the numerical resolution, in the following we argue that the origin of this decay is most probably numerical for weak and most of intermediate
PTs, while it is probably physical for intermediate PTs with thin hybrid profiles ($\vw \lesssim v_{\rm CJ}$)
and strong ones.

In \Sec{sec:convergence} and \App{sec:kinetic_ed}, we demonstrated that, at early times, while the self-similar profiles develop in each nucleated bubble before it collides,
the kinetic energy is underestimated for low resolution
(see \Figs{fig:KoK}{fig:1d_profiles_with_sims}).
To study the dependence of the kinetic energy decay rate with resolution, we show in the lower panels of \Fig{fig:E_kin_evolution} the
evolution of $K (\tilde t)$ normalized by the corresponding values of $K_0$.
For weak and most of intermediate PTs,
one can appreciate that the decay of the kinetic energy with time
becomes less pronounced as we 
increase the resolution $N$,
which is consistent with the growth of both $K_{\rm rms}$ and $K_0$ with $N$ observed in \Fig{fig:extrapolation}.
Since the kinetic energy is typically damped by numerical viscosity,\footnote{In the Kurganov-Tadmor scheme used in our simulations \cite{Jinno:2022mie}, the numerical viscosity is expected to scale proportional to $(\delta{\tilde{x}})^3$ \cite{KURGANOV2000241} \label{num_visc}.}
it is in general expected that the
decay is less pronounced when the grid spacing is reduced:
the weak and most of intermediate PTs conform with this expectation.

However, the opposite trend is found
for intermediate PTs with thin hybrid profiles ($\vw \lesssim v_{\rm CJ}$) and strong ones:
in the decaying phase of the kinetic energy, 
the decay becomes
steeper with smaller grid spacing (see~the middle and right low panels of \Fig{fig:E_kin_evolution}).
In these cases,
the enhancement of the decay with resolution
would be consistent with the fact
that, as the
fluid shells carry larger kinetic energies at the time
of collisions, non-linearities are enhanced. 
From energy conservation,
we then expect that, as non-linearities develop, kinetic energy transfers from
larger
to smaller scales, where it can be converted to thermal energy at the scale 
determined by numerical viscosity.
If this physical process damps kinetic energy more efficiently than the numerical dissipation, which might indeed happen in the case of strong PTs, we expect the decay to increase with resolution, as higher resolution means a better modeling of the physical damping process.

In addition to the time decay, $K$ also oscillates in time. This occurs in particular for weak PTs, and is more pronounced for small $\vw$, away from hybrid solutions. 
In these cases, the oscillations
can be associated to the sound-wave regime, where an
oscillatory conversion between kinetic and thermal
energies is expected, and confirmed by the fact that we
conserve $T^{00}$ to machine precision (see results in App.~C of Ref.~\cite{Jinno:2022mie}).

To better quantify the decay of the kinetic energy with time, we use a power law, effectively getting rid of the oscillations over time.
We fit the numerical results at times $\tilde t > \tilde t_0$
when the PT is complete, using
the following
power-law decay with time,
\begin{equation}
    K (\tilde t > \tilde t_0) = K_0\, \biggl(\frac{\tilde t}
    {\tilde t_0}\biggr)^{-b}\,, \label{decay}
\end{equation}
where $b$ indicates the power-law decay rate of $K$.
This
prescription has already been discussed in \Sec{sw_extended} (note that here we set the initial simulation time $\tilde t_{\rm ref}=0$), and
accurately fits the numerical data as shown by  \Fig{fig:E_kin_evolution}.
We have checked that it remains accurate
up to $\tilde t_{\rm end} = 64$
for an example strong PT with $\alpha = 0.5$ and $\vw = 0.8$.
We define the half-life of the kinetic energy as the
time when $K (\tilde t_0 + \tilde \tau_{1/2}) = \half K_0$, i.e.,
\begin{equation}
    \tilde \tau_{1/2} = \Bigl(2^{\frac{1}{b}}-1\Bigr)   \, \tilde t_0 \,.
\end{equation}

\begin{figure*}
    \centering
    \includegraphics[width=1\columnwidth]{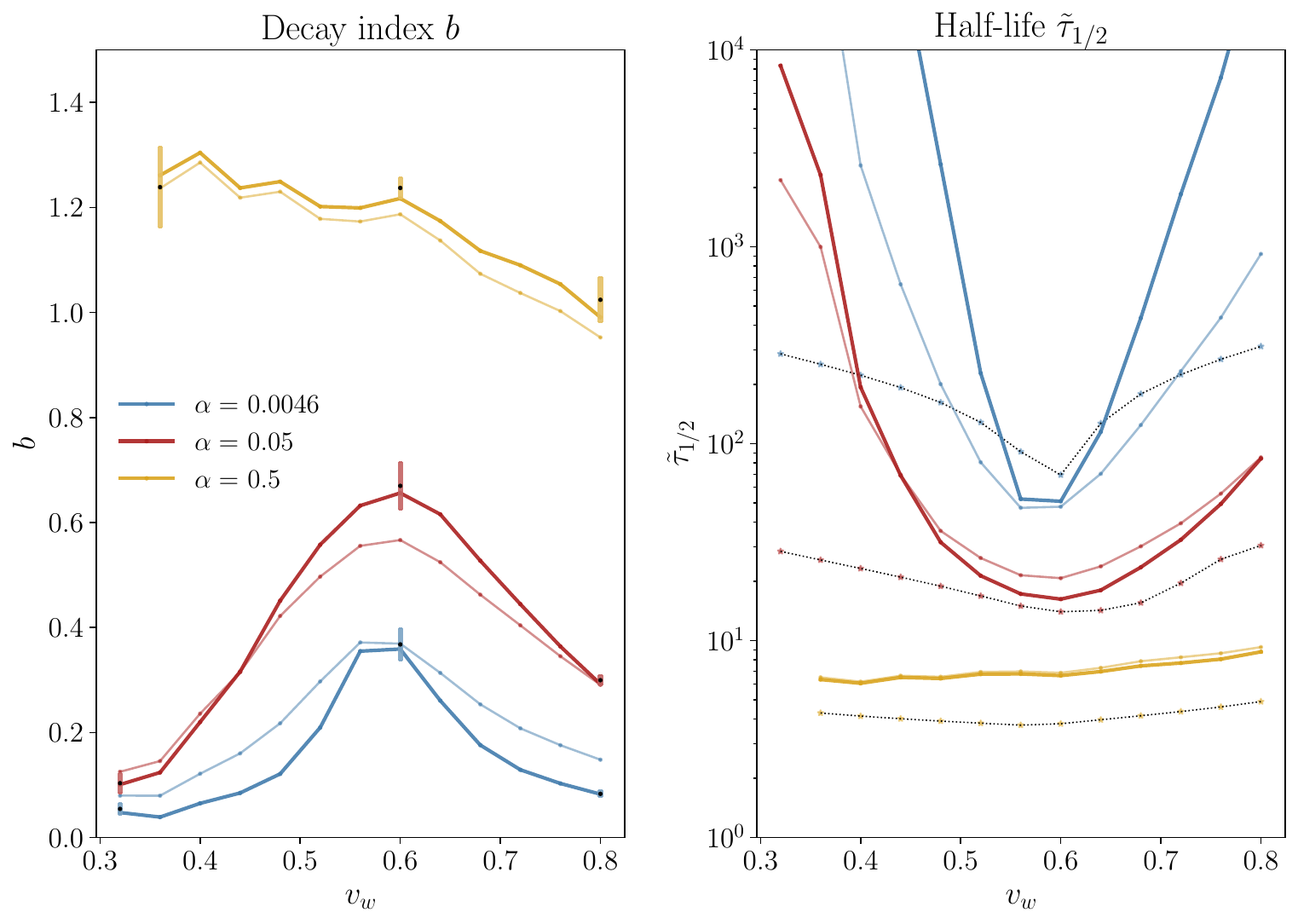}
    \caption{
    Decay index $b$ (left panel) and half-life of the kinetic energy fraction $\tilde \tau_{1/2}$ (right panel) as a function of $v_{\rm{w}}$ for $N = \{256, 512\}$ in increasing opacity for weak (blue lines), intermediate (red lines), and strong
    (orange lines) PTs.
    Error bars in the left panel show the standard deviation from 10 different bubble nucleation histories for $N = 512$.
    Dashed black lines with colored stars in the right panel correspond to the shock time
    $\tilde \tau_{\rm sh} = \beta R_\ast/\sqrt{K_\xi}$ (see values in \Tab{tab:kappas}),
    which we compare with $\tilde \tau_{1/2}$
    as we expect both time scales to be inversely proportional to $K_\xi$.
    \label{fig:decay_2}
    }
\end{figure*}

We display in \Fig{fig:decay_2} the fit of the decay index $b$ (left) as well 
as the half-life $\tilde \tau_{1/2}$ (right)
as a function of $v_{\rm{w}}$ for weak, intermediate, and strong PTs.\footnote{This time, we refrain from extrapolating to infinite resolution due to the complex behavior of the 
index.
We only plot the results obtained with the two largest resolutions $N=\{256,512\}$, and with the best resolution in the UV regime, $\tilde L/\vw = 20$.}
In the right panel of \Fig{fig:decay_2}, we also plot the shock time,
which corresponds to the time scale of
shock formation in the plasma and it is expected to determine
the time of decay into non-linear turbulent motion. 
Comparing the numerical $\tilde \tau_{1/2}$ to $\tilde \tau_{\rm sh}$
is therefore useful to interpret the development of non-linearities in the simulations. In order to avoid the uncertainty due to the underestimation of the kinetic energy in simulations, we evaluate $\tilde \tau_{\rm sh}$ with
the kinetic energy ratio expected for uncollided bubbles, 
$\tilde \tau_{\rm sh} = \beta R_\ast/\sqrt{K_\xi}$
(see \Tab{tab:kappas} for numerical values).
The shock time becomes $\tilde \tau_{\rm sh} \simeq 4$ for strong
PTs, $\tilde \tau_{\rm sh} \sim {\cal O} (10)$
for intermediate PTs, and $\tilde \tau_{\rm sh} \sim {\cal O} (100)$
for weak PTs.
Therefore,
the time scale for non-linearities to develop can be
reached within the simulation for strong
and some intermediate PTs, while
weak PTs should remain in the linear regime for the entire duration of the simulations,
and they would develop non-linearities at later times.
In some
intermediate PTs, the shock time
is reached towards the end of our simulations.
These expectations are in agreement with the simulation results.
We briefly discuss in \App{sec:vort}
the presence of vorticity in our simulations, and present some preliminary results.
To properly evaluate the development of vortical motion more detailed work is needed.

For weak transitions, the rate of kinetic energy damping is
reduced when increasing the resolution, as already discussed above in the context of the low panels of \Fig{fig:E_kin_evolution}. 
We interpret this trend as due to the reduction of the numerical viscosity 
(see footnote \ref{num_visc}).
This is indeed confirmed by the behaviour of both $b$ and $\tilde \tau_{1/2}$ with resolution, shown in \Fig{fig:decay_2}, blue lines: the decay exponent diminishes with increasing resolution, while the half-life of the kinetic energy grows, and bears no relation with the shock time.
This
means that for weak transitions, the decay is always dominated by numerical viscosity.
Only for the hybrid solution with $v_{\rm{w}} = 0.6 \lesssim v_{\rm CJ}$, when larger fluid velocities can be achieved (see self-similar profiles in \Fig{fig:1d_profiles}),
does $b$ (and hence $\tilde \tau_{1/2}$) appear to stagnate with increasing
resolution, pointing towards the onset of resolving the physics
responsible for the damping.
However, in this extreme case, the fluid profile is highly confined and
the simulations are far from reaching the converged profiles (see \Figs{fig:KoK}{fig:1d_profiles_with_sims}),
so it is not completely clear whether the obtained
decay rate $b$ is physical.

The results are more interesting in the case of intermediate transitions.
For small $v_{\rm{w}}$, corresponding to subsonic deflagrations, $b$ decreases with increasing resolution, and $\tilde \tau_{1/2}$ increases correspondingly.
However, for a large range of intermediate velocities $\vw \in \{0.52, 0.6, 0.68\} \lesssim v_{\rm CJ}$, the trend is reversed.
We interpret this point of reversal as a transition from a decay of the kinetic energy 
dominated by numerical viscosity to a decay determined by the development of non-linearities.
We note that in this case, some of the confined hybrids are still under-resolved, but this is no longer the case for the subsonic
deflagration with $\vw = 0.52$, indicating that the decay rate seems to
be physical (see \Figs{fig:E_kin_evolution}{fig:1d_profiles_with_sims}).
Furthermore, a similar decay of the kinetic energy was
already found for intermediate PT simulations
of the scalar-fluid system \cite{Hindmarsh:2017gnf,Cutting:2019zws}.

For strong transitions, we are universally in the regime where increasing the numerical resolution $N$ leads to
a larger decay rate, indicating that the physical
non-linear decay dominates over the numerical viscosity.
As discussed above, this is expected to be the case, as
the expected time scale for non-linearities to develop, i.e., the
shock formation time, is around
$\tilde \tau_{\rm sh} \simeq 5$,
occurring during the duration of our numerical simulations.
The fact that both  $b$ and $\tilde \tau_{1/2}$ are virtually independent on the value of $\vw$ indicates that the possible under-resolution of the initial fluid profiles is no longer an issue as far as the fluid dynamics is concerned.
However, this could also be a consequence of the low variability of $K_\xi$ with $\vw$ for strong
PTs (see values in \Tab{tab:kappas}).

\subsection{Dependence of the integrated GW amplitude with the source duration and GW efficiency}
\label{GW_spec_time}

In this section, we analyze how the GW energy density depends on the source
duration, with the aim of consolidating,
with the result of simulations, both the model described in \Sec{GW_sw}, and its generalization to decaying sources proposed in \Sec{sw_extended}. 
For this purpose, we study two quantities as a function of the simulation time,
which corresponds to the source duration: the
integrated GW amplitude ${\cal I}_{\rm sim}^{\rm int} \equiv \int {\cal I}_{\rm sim} \dd \ln k$, and the GW efficiency $\tilde \Omega_{\rm GW}$.

\begin{figure*}[t]
    \centering
    \includegraphics[width=0.32\columnwidth]{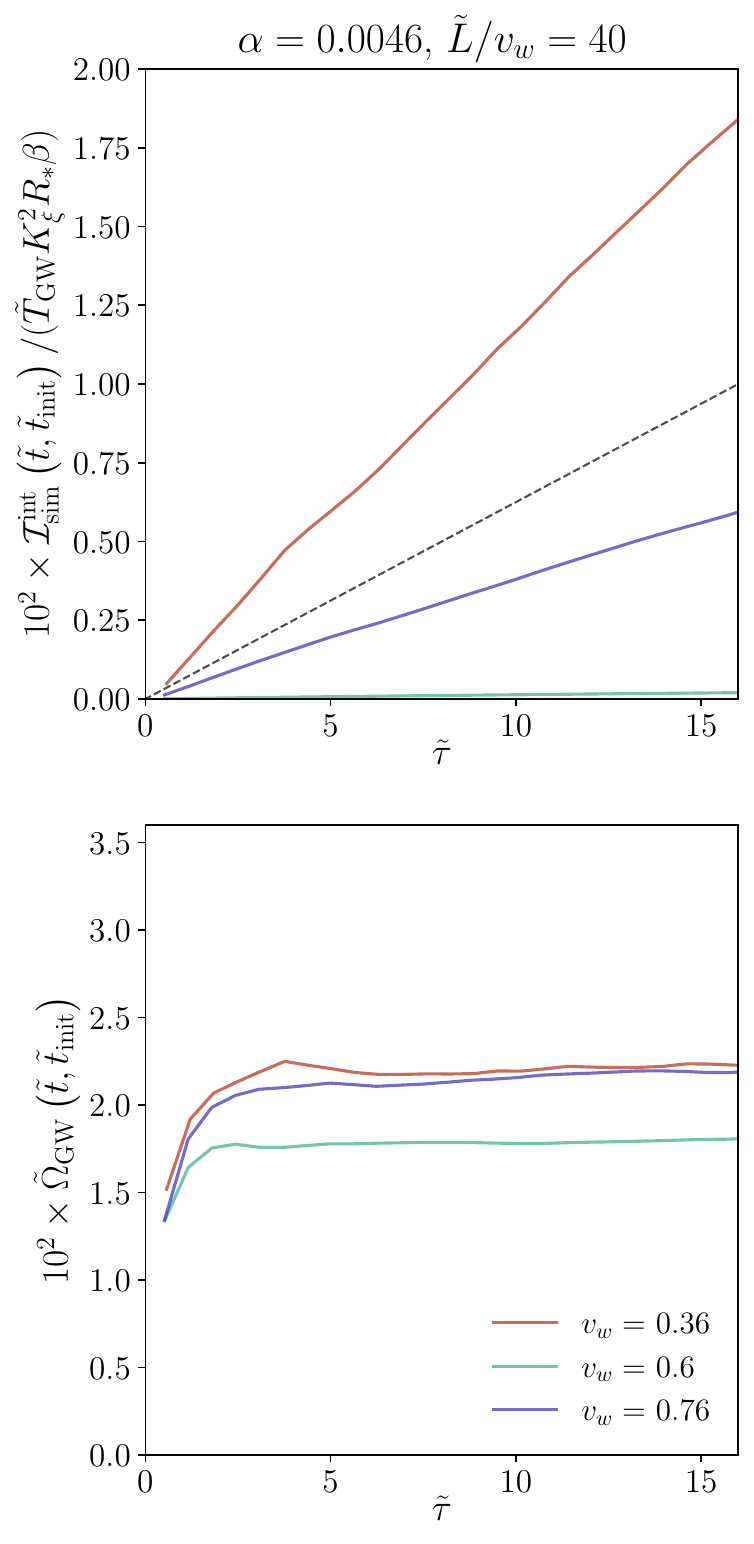}
    \includegraphics[width=0.32\columnwidth]{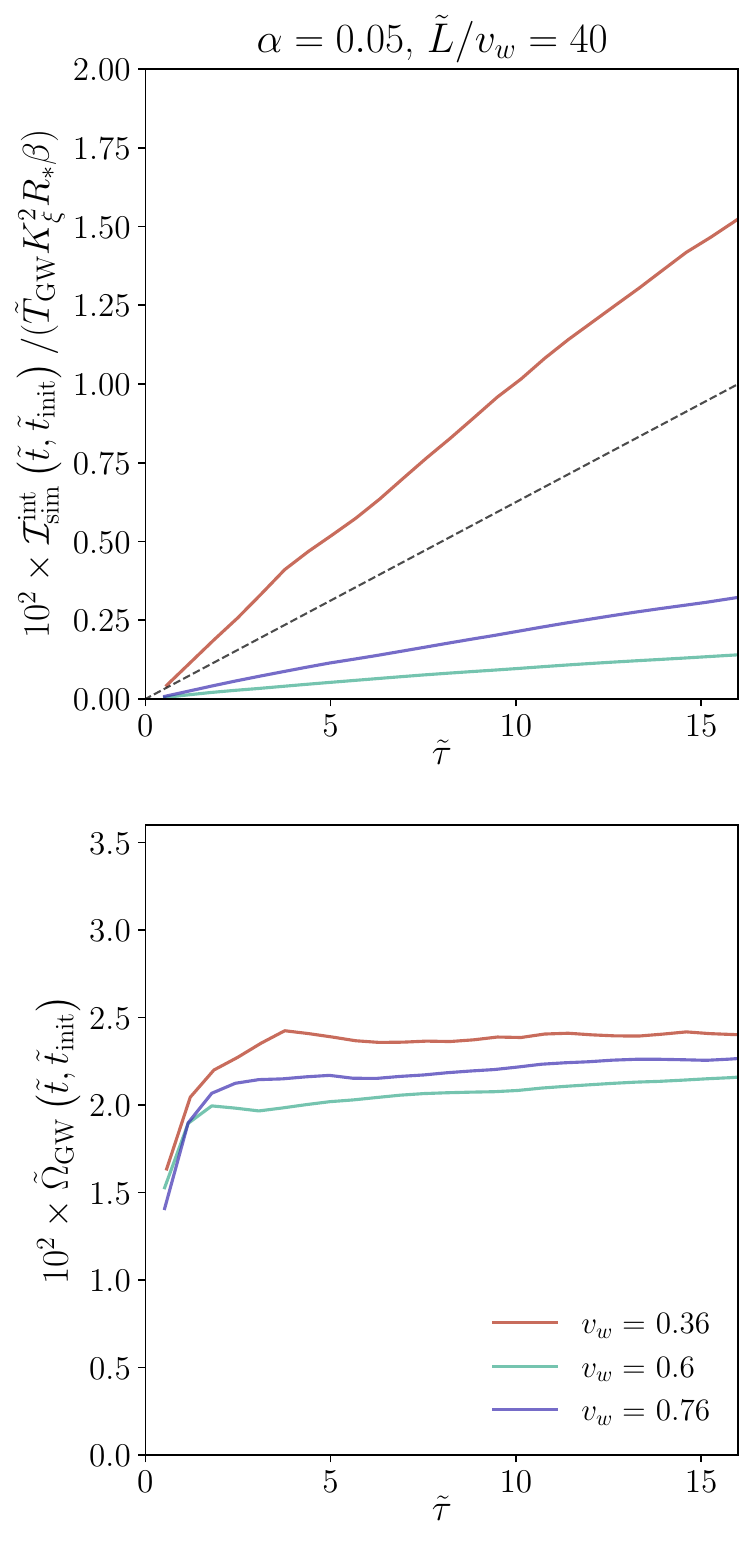}
    \includegraphics[width=0.32\columnwidth]{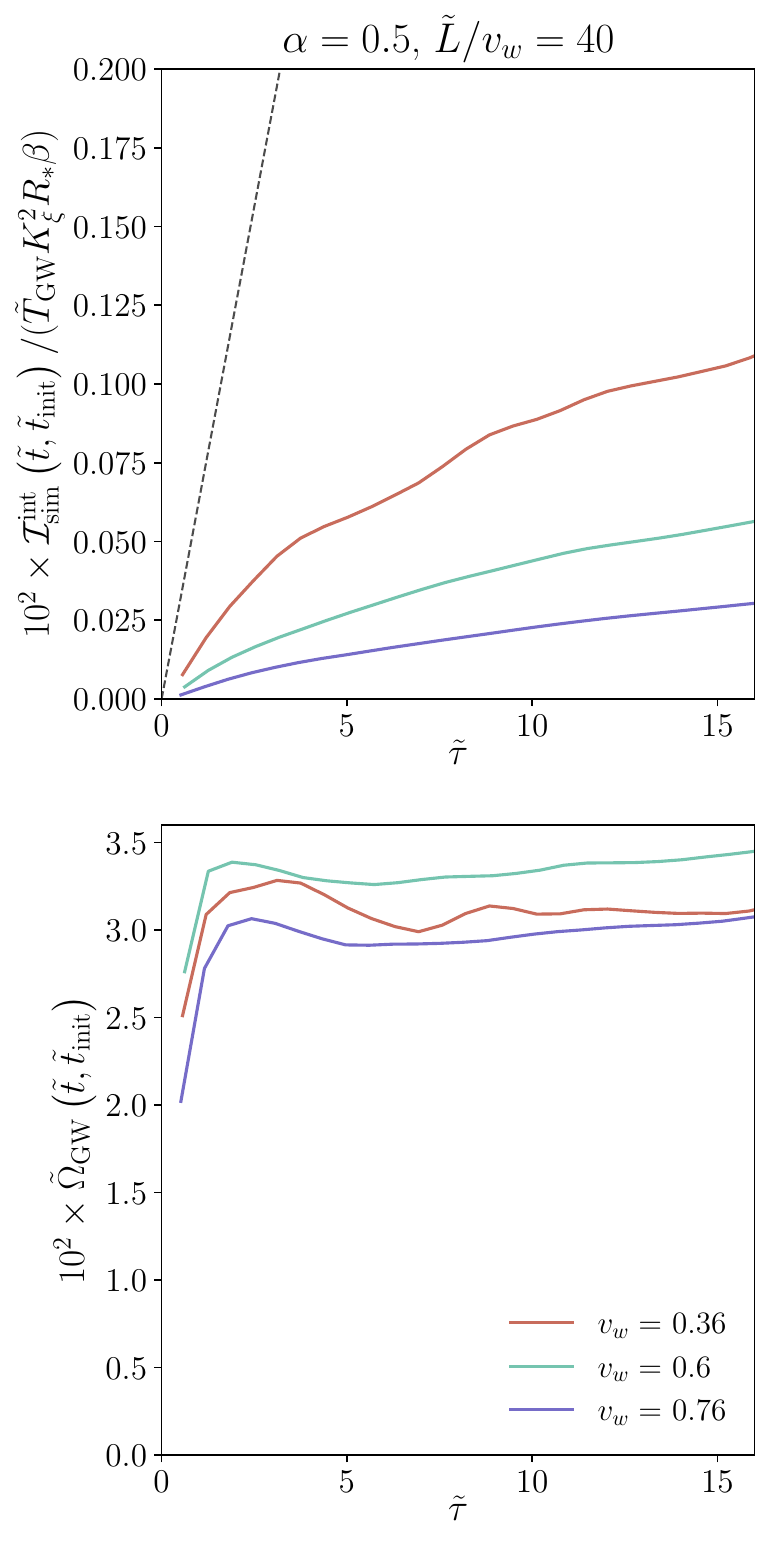}
    \caption{
    {\em Upper panel}: the numerical integrated
    GW amplitude found in the simulations with $\tilde L/\vw = 40$ and $N = 512$, as a function of the
    source duration $\tilde \tau \equiv \tilde t - \tilde t_{\rm init}$ for weak (left columns), intermediate (middle columns), and strong (right columns) PTs.
    The integrated GW amplitude is normalized as in the middle panel of \Fig{fig:extrapolation}. 
    Dashed lines 
    correspond to the linear growth with
    $K_{\rm rms}^2 = K_\xi^2$
    and $\tilde \Omega_{\rm GW} = 10^{-2}$.
    {\em Lower panel}: the GW production efficiency $\tilde \Omega_{\rm GW}$ also as a function of time, computed using \Eq{OmGW_numerical}.
    } \label{fig:GWgrowth}
\end{figure*}

The first quantity is shown in the upper panels of \Fig{fig:GWgrowth},
where ${\cal I}_{\rm sim} (\tilde t_{\rm init}, \tilde t, k)$ is evaluated at $\tilde t_{\rm init} = 16$, while $\tilde t$ can vary from $\tilde t_{\rm init}$ to $\tilde t_{\rm end} = 32$ [see \Eq{Weinberg}].
For weak and some intermediate PTs, the usual stationary
assumption of the sound-wave regime presented in \Sec{GW_sw} should apply, as illustrated in previous numerical work
\cite{Hindmarsh:2013xza,Hindmarsh:2015qta,Hindmarsh:2017gnf,
Jinno:2020eqg,Jinno:2022mie}.
Indeed, we confirm that in these cases, $K$ does not significantly evolve with
time within the simulations (see \Fig{fig:E_kin_evolution}).
Therefore, according to the model of \Sec{GW_sw}, we  expect
${\cal I}_{\rm sim}^{\rm int} (\tilde t_{\rm init}, \tilde t)$ to grow linearly
in time with the source duration $\tilde \tau \equiv
\tilde t - \tilde t_{\rm init}$, cf.~\Eq{OmGW_sshell}:
\begin{equation}
    \frac{{\cal I}_{\rm sim}^{\rm int} (\tilde t_{\rm init}, \tilde t)}{\tilde T_{\rm GW}\,K^2_\xi\,\beta R_*}=\tilde \Omega_{\rm GW}\left(\frac{K}{K_\xi}\right)^2
    \frac{\tilde \tau}{\tilde T_{\rm GW}},
    \label{eq:Inumlinearintime}
\end{equation}
where $\tilde \tau_{\rm fin} = \tilde t_{\rm fin} - \tilde t_\ast$ in \Eq{OmGW_sshell}
is now replaced by
$\tilde \tau$, and we have normalized the left hand side as in the upper panels of \Fig{fig:GWgrowth} 
(and middle panels of \Fig{fig:extrapolation}).
From this figure, one can appreciate that the linear dependence with the source duration of the GW amplitude, predicted by the sound-shell model, is indeed in good agreement with the simulation results for weak and some intermediate PTs. 
From \Eq{eq:Inumlinearintime}, it appears that the slope of the linear growth, as plotted in \Fig{fig:GWgrowth}, depends on the ratio $(K/K_\xi)^2/\tilde T_{\rm GW}$.
For confined hybrids, e.g., for 
$\alpha = 0.0046$ and $\vw = 0.6$,
the underestimation of the kinetic energy $K$ with respect to the single bubble one $K_\xi$, due to poor numerical resolution (see \Fig{fig:extrapolation} and discussion in \App{sec:kinetic_ed}), causes the slope to be small, and the GW amplitude to increase slowly.

On the other hand, for strong PTs with $\alpha = 0.5$, and also for $\alpha = 0.05$
with intermediate wall velocities
$\vw \in \{0.6, 0.68\} \lesssim v_{\rm CJ}$,
the decay of $K$ is significant within the time
of our simulations, and we observe deviations
with respect to the linear dependence with $\tilde \tau$.
In these cases, the effect of the time evolution of $K$ needs to be incorporated, as put forward in the context of the 
locally stationary UETC presented in \Sec{sw_extended}, 
which proposes to generalize the linear
growth to a growth proportional to $K^2_{\rm int} (\tilde t_{\rm init}, \tilde t)$, see \Eq{OmGW_general}. 
The integrated GW amplitude plotted in \Fig{fig:GWgrowth} becomes then
\begin{eqnarray}
    \frac{{\cal I}_{\rm sim}^{\rm int} (\tilde t_{\rm init}, \tilde t)}{\tilde T_{\rm GW}\,K^2_\xi\,\beta R_*}
    &=&
    \tilde \Omega_{\rm GW} \frac{K_{\rm rms}^2}{K_\xi^2} = 
    \tilde \Omega_{\rm GW}\left(\frac{K_0}{K_\xi}\right)^2
    \frac{\tilde t_{\rm init}}{\tilde T_{\rm GW}} \, \left(\frac{\tilde t_0}{\tilde t_{\rm init}}\right)^{2b} \, \frac{\left(\tilde t/\tilde t_{\rm init}\right)^{1 - 2 b} - 1}{1-2b} \label{eq:Inumbintime}\,,
\end{eqnarray}
where the second equality corresponds to \Eq{Kint_decay}
(with $\tilde t_{\rm ref}=0$), and holds
as long as the fit $K(\tilde t') = K_0 \, (\tilde t'/\tilde t_0)^{-b}$ accurately represents
the numerical results 
at times $\tilde t' \in [\tilde t_{\rm init}, \tilde t]$ (see \Fig{fig:E_kin_evolution}).
From \Fig{fig:decay_2}, we see that $b>1/2$ for strong PTs and intermediate ones with velocities close to $v_{\rm CJ}$. 
\EEq{eq:Inumbintime} then implies that the integrated GW amplitude grows slower than linearly
for $\tilde t> \tilde t_{\rm init}$, eventually saturating to a constant value if the duration is long enough [cf.~also \Eq{growthrate_K}]. Indeed, the simulations confirm this behaviour, as shown in \Fig{fig:GWgrowth}, supporting the model proposed in \Sec{sw_extended}.

Another way to test the validity of the models put forward in \Secs{GW_sw}{sw_extended} is to study the GW efficiency $\tilde \Omega_{\rm GW}$ defined in  \Eq{eq:OmGWtilde}, which, according to them, should be constant with the source duration
[cf.~\Eq{OmGW_sshell} and its generalization \eqref{OmGW_general}].
Therefore,
we plot in the lower panel of \Fig{fig:GWgrowth} the
following ratio
\begin{equation}
    \tilde \Omega_{\rm GW} (\tilde t) = \frac{{\cal I}_{\rm sim}^{\rm int} (\tilde t_{\rm init}, \tilde t)}
    {K^2_{\rm int} (\tilde t_{\rm init}, \tilde t) (\beta R_\ast)}\,,
    \label{OmGW_numerical}
\end{equation}
with the objective of estimating the
GW efficiency as a function of the source duration, while also including
the effect of the decay of $K$.
We note that when $K$ does not significantly decay with time,
$K_{\rm int}^2 \to K^2 \, \tilde T_{\rm GW}$ and \Eq{OmGW_numerical} goes back to the usual definition \Eq{eq:OmGWtilde}, valid within the stationary assumption of the sound-shell model.
After an initial steep increase with time, 
which is a just numerical artifact from abruptly
starting the GW computation at $\tilde t_{\rm init}$, 
$\tilde \Omega_{\rm GW}(\tilde t)$ stabilises to virtually constant values across all PT strengths and all values of $\vw$, again
validating the generalization of the linear
growth to a growth proportional to $K^2_{\rm int} (\tilde t_{\rm init},\tilde t)$
as found within the locally stationary UETC proposed in \Sec{sw_extended}.

\begin{figure*}
    \centering
    \includegraphics[width=1.0\textwidth]{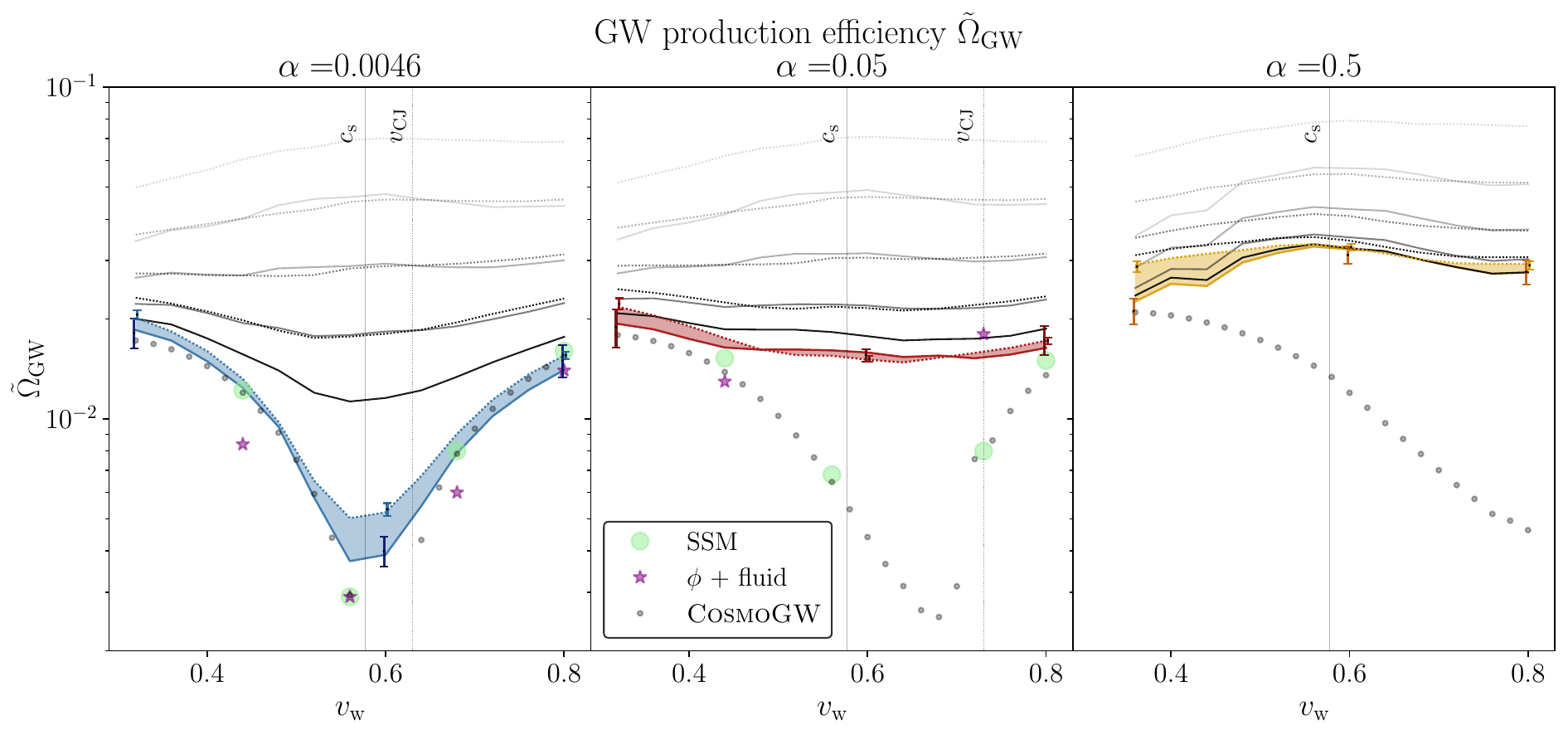}    
    \caption{
    Gravitational wave production efficiency $\tilde{\Omega}_{\GW}$ for weak (left), intermediate (middle), and strong (right) first-order PTs. Solid (dotted) lines correspond to $\tilde L/\vw = 20$ (40). Black lines with increasing opacity correspond to increasing resolutions $N\in \{64, 128, 256, 512\}$, while colored lines are the values extrapolated to the continuum limit $\delta{\tilde{x}} \to 0$.
    Green dots and violet stars mark $\tilde \Omega_{\rm GW}$ as presented in Tabs.~2 and 3 of Ref.~\cite{Hindmarsh:2019phv} corresponding to predictions from the sound-shell model (SSM) for exponential nucleation of bubbles~\cite{Hindmarsh:2019phv} and scalar field-hydrodynamical simulations for simultaneous nucleation~\cite{Hindmarsh:2017gnf}, respectively.
    Gray dots correspond to sound-shell model values found using
    the assumption described in \Sec{GW_sw}, following App.~B of Ref.~\cite{RoperPol:2023dzg}, and computed using {\sc CosmoGW} \cite{cosmogw}.
    Error bars indicate the standard deviation from 10 different bubble nucleation histories for $\tilde L/\vw = 20$ (darker) and $40$ (lighter).}
    \label{fig:GW_efficiency}
\end{figure*}

Since we find that $\tilde \Omega_{\rm GW} (\tilde t)$
is roughly constant with the source duration after incorporating $K_{\rm int}^2$
in the scaling of the GW amplitude (see \Eq{OmGW_numerical} and \Fig{fig:GWgrowth}), in the rest of the paper
we take its value at the end of the simulations $\tilde t_{\rm end}$.
The resulting GW efficiency
$\tilde \Omega_{\rm GW}(\tilde t_{\rm end})$, computed from \Eq{OmGW_numerical},
is shown in \Fig{fig:GW_efficiency}
for different numerical resolutions $N$ and box sizes
$\tilde L/\vw = 20$ and 40, and also
using the extrapolated values to $\delta{\tilde{x}} \to 0$ of ${\cal I}_{\rm sim}^{\rm int}$ and $K_{\rm int}$,
as described in \Sec{sec:convergence} (see also \Fig{fig:extrapolation}).
The error bars
show the standard deviation obtained from 10 different bubble nucleation histories, corresponding to the {\em seeds} set of simulations listed in \Tab{tab:simulation_summary}.
We also compare in \Fig{fig:GW_efficiency} the extrapolated
$\tilde \Omega_{\rm GW} (\tilde t_{\rm end})$
with the values of the efficiency found using
the sound-shell model \cite{Hindmarsh:2016lnk,Hindmarsh:2019phv} as described in
\Sec{GW_sw} (see also App.~B of Ref.~\cite{RoperPol:2023dzg}),
and with those obtained from numerical simulations of the full coupled scalar field-fluid system
\cite{Hindmarsh:2015qta,Hindmarsh:2017gnf}.
Note that the latter are found using
simultaneous bubble
nucleation, which in general leads to smaller
$\tilde \Omega_{\rm GW}$ compared to exponential
nucleation used in this work (see Tabs.~2 and 3 in Ref.~\cite{Hindmarsh:2019phv}).
We have also modified the values of $\tilde \Omega_{\rm GW}$
from Refs.~\cite{Hindmarsh:2015qta,Hindmarsh:2017gnf,Hindmarsh:2019phv} to take into account that they consider $\beta R_\ast = (8\pi)^{1/3} \, \vw$, instead of the corrected $\beta R_\ast = (8\pi)^{1/3} \max(\vw, \cs)$ that we use in \Eq{OmGW_general}.
Concerning the values of $\tilde \Omega_{\rm GW}$ found using
the sound-shell model, they
should also be modified to include the structure that
develops in the GW spectrum below the peak
for some values of the PT parameters, an effect pointed out for the first time in Ref.~\cite{RoperPol:2023dzg}.
However, we neglect this effect here, for two reasons: {\em (1)} 
as discussed in \Sec{GW_sw}, when the kinetic energy is small, the stationary assumption of the sound-shell model is valid. Therefore, for weak PTs, the integral in wave numbers necessary to evaluate $\tilde \Omega_{\rm GW}$ should not be strongly affected by including the correct spectral shape at low wave numbers.
{\em (2)}
the dynamical range in the IR available in our and previous simulations is usually not large enough to clearly reconstruct
the exact spectral shape described in Ref.~\cite{RoperPol:2023dzg}, see results in \Sec{sec:shape}.

For weak transitions (left panel of \Fig{fig:GW_efficiency} with $\alpha = 0.0046$), the extrapolated $\tilde \Omega_{\rm GW}$ obtained from the
 Higgsless simulations accurately reproduces 
both the results of the sound-shell model and of the coupled scalar field-hydrodynamical simulations
\cite{Hindmarsh:2015qta,Hindmarsh:2017gnf}.
The agreement between three independent approaches supports the conclusion that the dependence
of $\tilde \Omega_{\rm GW}$ with $\vw$ may then be physical.

However, for intermediate PTs, we begin to observe deviations from the sound-shell model, in particular for large $\vw$.
The middle panel of \Fig{fig:GW_efficiency} shows that the $\vw$ dependence present for weak PTs has flattened, and that the overall efficiency $\tilde \Omega_{\rm GW}$ is larger than the sound-shell model results.
We are reasonably confident that this result is physical, for two reasons. First of all,
the extrapolation method described in \Sec{sec:convergence} and presented in \Fig{fig:GW_efficiency} as solid and dotted lines 
seems to behave very well, delivering agreement between the numerical results from both 
simulation domains $\tilde L/\vw = 20$ and $40$.
Second, our findings are also consistent with the two available data points for scalar field-hydrodynamical
simulations from Ref.~\cite{Hindmarsh:2017gnf} (violet stars). 
As $\alpha$ grows, non-linearities become more relevant, and simulations are then necessary to push beyond the validity reach of the sound-shell model;
however, only a few points of reference data for $\tilde \Omega_{\rm GW}$ exist for intermediate PTs ($\alpha = 0.05$),
and so far none\footnote{Reference~\cite{Cutting:2019zws} presents results of
$\Omega_{\rm GW}/\Omega_{\rm GW, exp} = {\cal I}_{\rm sim}^{\rm int}/{\cal I}_{\rm exp}^{\rm int}$, where ${\cal I}_{\rm exp}^{\rm int}$
would correspond to the value found using \Eq{OmGW_sshell} with $K = K_\xi$ and $\tilde \Omega_{\rm GW} = 10^{-2}$.
The ratio that ref.~\cite{Cutting:2019zws} presents therefore corresponds to a combined
estimate of $\tilde \Omega_{\rm GW} K_{\rm rms}^2/K_\xi^2$ and extraction
of $\tilde \Omega_{\rm GW}$ for comparison is not straightforward.} for strong PTs ($\alpha=0.5$).
Still, the broad agreement between the results of the Higgsless and of
the scalar field-hydrodynamical simulations points towards
a consistent departure from linearity in the fluid perturbations and, hence,
from the sound-shell model, at least for the point at large $\vw$ (i.e., 0.76).
We note that the reference data points
$\tilde \Omega_{\rm GW}$ in Refs.~\cite{Hindmarsh:2019phv,Hindmarsh:2017gnf} are computed
assuming a linear growth with the source duration as in \Eq{OmGW_sshell}:
hence, incorporating $K_{\rm int}$ as in \Eq{OmGW_general} can modify the
value of $\tilde \Omega_{\rm GW}$ when the source decays.
Furthermore, small
discrepancies with the numerical results of Ref.~\cite{Hindmarsh:2017gnf} might also be due to the different nucleation histories
considered (simultaneous in Ref.~\cite{Hindmarsh:2017gnf} and exponential in our simulations). 

Finally, for strong PTs, we observe even larger efficiencies overall (see the right panel of \Fig{fig:GW_efficiency}). 
In this case as well, the impact of the non-linearities 
washes out the dependence of $\tilde \Omega_{\rm GW}$ on the wall velocity.
Again, extrapolation seems to work overall, as the extrapolated values agree reasonably well for the two simulation sizes $\tilde L/\vw = 20$ and 40.
We note that, for weak PTs, the 
relative difference between the extrapolated values and those
obtained in our largest resolution runs with $N = 512$ is rather large.
Hence, the values provided in \Fig{fig:GW_efficiency}
are still subject to
numerical errors, related to the values of $\varepsilon_{\cal I}$
listed in \Tab{tab:conv_fit}.
Indeed, we find these relative differences
to be larger for weak PTs (up to 50\%
for extremely thin profiles, and usually below 10\% otherwise),
where we can compare our extrapolated results to those found
within the sound-shell model, while for intermediate and strong PTs,
the relative difference of our extrapolated values
to the numerical results for the largest resolution $N = 512$
are below 10\% for all wall velocities.

It is very important to remark that the definition of $\tilde \Omega_{\rm GW}$ in terms of the integrated kinetic energy  $K^2_{\rm int}(\tilde t_{\rm init},\tilde t_{\rm end})$ reduces its variation with the wall velocity and with the strength of the PT significantly --- compared to normalizing 
it to a stationary kinetic energy ratio (e.g., to
$K_\xi$ or to $K_0$) multiplied by the source
duration, as done previously.
This is due to the decay
found in the kinetic energy that is not
captured by the stationary assumption for the UETC of the sound-shell model (see discussion
in \Sec{sw_extended} and numerical results in \Sec{decay_K2}).
Furthermore, the universality
of $\tilde \Omega_{\rm GW}$ is also due
to the use of \Eq{OmGW_general} instead of $Q'/K^2$ considered
in previous work \cite{Jinno:2020eqg,Jinno:2022mie}, as discussed in \Sec{GW_sw}.

The average of the extrapolated values of $\tilde \Omega_{\rm GW}$
over $\vw$, for each strength $\alpha$, from the simulations, are the
following:
\be \label{eq:Omegatilde}
10^2 \, \tilde \Omega_{\rm GW} = 
\begin{cases}
1.04^{+0.81}_{-0.67}\,,  & \quad \textrm{for} \quad \alpha = 0.0046\,; \\ 
1.64^{+0.29}_{-0.13}\,,  & \quad \textrm{for} \quad \alpha = 0.05\,; \\
3.11^{+0.25}_{-0.19}\,,  & \quad \textrm{for} \quad \alpha = 0.5\,, 
\end{cases}
\ee
where the super and subscripts refer to the maximum and
the minimum over $\vw$ of the extrapolated values.\footnote{Based on the parameterization of \Eq{OmGW_stat2}
and the results of Ref.~\cite{Jinno:2022mie}, Ref.~\cite{Caprini:2024hue} reported a value $A_{\rm sw} = 3 \, \tilde \Omega_{\rm GW} \simeq 0.11$,
slightly larger than the extrapolated values in \Eq{eq:Omegatilde} with our updated
numerical simulations and results.}
Note that \Eq{eq:Omegatilde} only takes into account the variation of $\tilde \Omega_{\rm GW}$ with $\vw$, but not from
numerical inaccuracies, since it provides the extrapolated values.

As discussed in \Sec{GW_production} and \App{sec:GW_prod}, in principle the GW signal evaluated from the simulations corresponds to the physical one detectable today only if the source has stopped operating by the end of the simulation $\tilde t_{\rm end}$. 
For weak and most of intermediate PTs, the decay of the kinetic energy is very mild, and the bulk fluid motion is still sourcing GWs at the end of the simulation: indeed, in these cases we
find a linear growth rate with the source duration
(see \Fig{fig:GWgrowth}) and, hence, 
the free-propagation regime of the GW
amplitude has not been reached across all wave numbers.
This result is in agreement with previous simulations \cite{Hindmarsh:2013xza,Hindmarsh:2015qta,Hindmarsh:2017gnf}, and with the sound-shell model \cite{Hindmarsh:2019phv}. 
In this case, in order to provide a proxy for the GW signal today,
the usual assumption in the literature is that the linear growth persists until the development of non-linearities at $\tilde t_* - \tilde t_{\rm init} \equiv \tilde \tau_{\rm fin}
\sim \tilde \tau_{\rm sh} \simeq \beta R_\ast/\sqrt{K}$ (see values in \Tab{tab:kappas}),  
and that after this time, the source is abruptly switched off \cite{Caprini:2015zlo,Caprini:2019egz}.
This leads to the linear dependence with $\tilde \tau_{\rm fin}$ of \Eqs{OmGW_sshell}{OmGW_stat2}.
For PTs where the non-linearities timescale has been reached
within the duration of the simulation, i.e., strong and some intermediate
PTs (see discussion in \Sec{decay_K2} and the right panel of \Fig{fig:decay_2}), the decay of the kinetic energy reduces the GW sourcing during the simulation.
Nevertheless,
we find that,
even though the GW amplitude
is growing with the source duration much slower than linearly, it
is still growing after $\tilde \tau_{\rm sh}$, and actually until
the final time of our simulations (cf.~the upper panels of \Fig{fig:GWgrowth} for $\alpha=0.05$ and 0.5). 
This means that, also in these cases, the simulations have not run long enough
to reach the end of the GW sourcing.
More importantly, this shows that the growth of the GW amplitude with the source
duration persists after the development of non-linearities for an additional duration that cannot
be predicted with current simulations.

Indeed, the power-law decay model of the kinetic energy $K^2 (\tilde t) \propto \tilde t^{-2b}$
inferred from the simulations (see \Sec{decay_K2}),
allows us to extrapolate the
resulting GW amplitude by extending $K^2_{\rm int}$ to 
times beyond the final time of the simulation, using \Eq{OmGW_general}.
According to this model, the GW amplitude \Eq{eq:Inumbintime} 
grows asymptotically for $\tilde t\gg \tilde t_*$
proportional to 
${\cal I}_{\rm sim}^{\rm int} \sim \tilde t^{\,1 - 2b}$ for $b < \half$,
${\cal I}_{\rm sim}^{\rm int} \sim \ln \tilde t$ for $b = \half$, and ${\cal I}_{\rm sim}^{\rm int} \sim \tilde t^{\,0}$ for $b > \half$ [see also \Eq{growthrate_K}].
For the weak and intermediate PTs that are featuring slow decay, with $0 \leq b<\half$, the GW
amplitude keeps growing unbounded, 
and the linear growth predicted by the sound-shell model gets generalized to ${\cal I}_{\rm sim}^{\rm int} \sim \tilde t^{\,1 - 2b}$.
On the other hand, as discussed below \Eq{eq:Inumbintime}, $b > \half$ for strong and some intermediate PTs: in these cases, the GW amplitude then saturates to a constant value
in the limit $\tilde t\gg \tilde t_*$.
We emphasize that these results seem to indicate that, after
the fluid perturbations enter the non-linear regime, the GW
amplitude still takes some time to saturate to its free-propagation
value. Hence, the usual procedure that consists in
assuming a linear growth cut
at $\tilde \tau_{\rm fin} = \tilde \tau_{\rm sh}$
most probably underestimates the GW amplitude.
For strong PTs we find that, because of the large decay of the source, the GW amplitude at the end of the simulation has almost reached its saturated value.
Hence,
the uncertainty on the exact saturation time
has a much smaller impact on the amplitude of the GW spectrum.

These considerations hold as long as the UETC
assumed in \Sec{GW_sw} describes the source dynamics.
However, we expect that the UETC deviates from our model as vortical motion and turbulence
dominates
\cite{Caprini:2009fx,Niksa:2018ofa,RoperPol:2022iel,Caprini:2024ofd}.
Further analyses and simulations are necessary to fully understand this regime. 
In the present work, we have characterized the initial phase in which non-linearities
and vortical motion start to develop for intermediate and strong PTs,
during which we find that the proposed model
describing the source UETC as locally stationary works well for the duration
of the simulations.
We therefore propose
effectively stopping
the GW sourcing 
after an appropriate choice of
the source duration $\tilde \tau_{\rm fin}$, which we leave as a free parameter in our current
estimates.

\begin{figure*}[t]
    \centering
    \includegraphics[width=0.99\columnwidth]{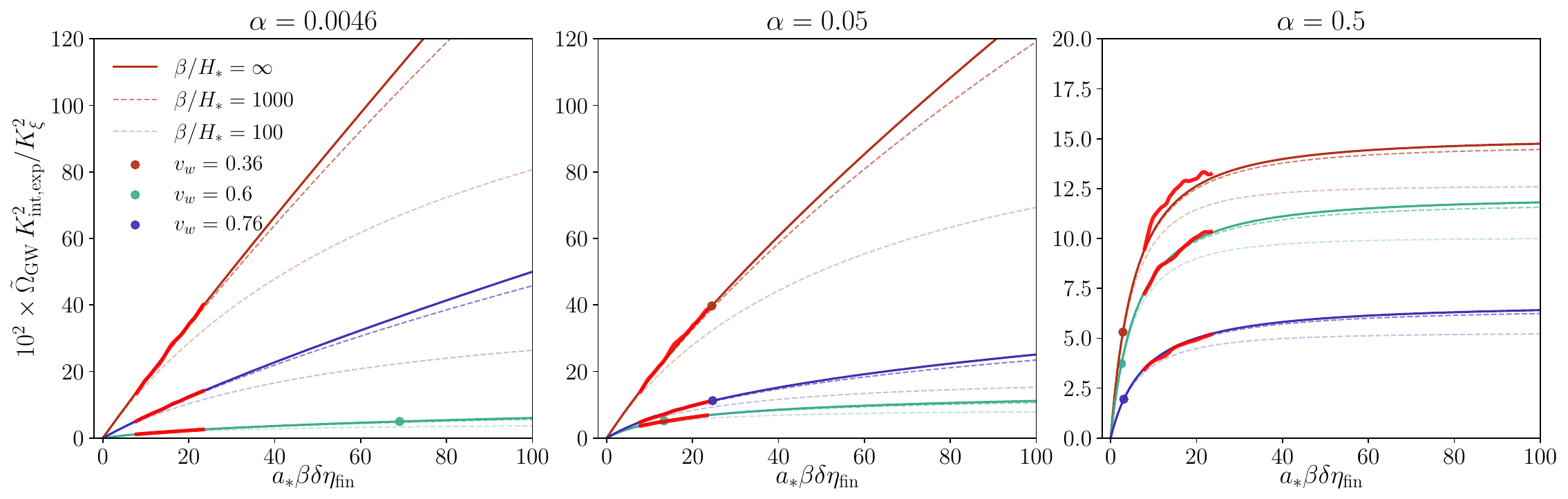}
    \caption{
    Dependence of the
    GW integrated amplitude 
    with the conformal source
    duration
    $a_\ast \beta \delta \eta_{\rm fin} 
    = a_\ast \beta \eta_{\rm fin} 
    - a_\ast \beta \eta_\ast = a_\ast \beta \eta_{\rm fin}
    - \beta/H_\ast$,
    assuming that the GW production
    starts at
    $\eta_\ast = \eta_0$,
    normalized by the reference value $\tilde \Omega_{\rm GW} \simeq 10^{-2}$ \cite{Hindmarsh:2017gnf} and by $K_\xi^2 R_\ast \beta$ [cf.~\Eq{eq:Inumbintime}], for
    wall velocities $\vw = 0.4$, $0.56$, and 0.8.
    Red segments indicate simulation results,
    already plotted in
    \FFig{fig:GWgrowth} but here re-scaled by the
    extrapolated values of the kinetic energy fraction
    at $\tilde t_0$, ${\cal K}_0^2$,
    found in \Sec{sec:convergence} (see \Fig{fig:kappa_eff_kappa}),
    and adding the expected contribution to the GW production
    from the interval of time
    $\eta \in [\eta_0, \eta_{\rm init}]$.
    We remind that in the simulations $\Delta \eta_0\simeq 10/(\beta a_*)$ and $\Delta \eta_{\rm init}=16/(\beta a_*)$.
    Solid lines indicate the expected
    amplitudes extrapolated to times after the end of our
    simulations according to the model of \Sec{sw_extended}, neglecting
    the Universe expansion 
    ($\beta/H_\ast \to \infty$), while dashed
    lines correspond to the expected amplitudes after including expansion according to the model of \Sec{sec:expansion}, for $\beta/H_* = 1000$ and $100$ in decreasing
    opacity.
    Dots indicate the shock formation time
    $a_* \beta \,\delta \eta_{\rm sh} = \beta R_\ast/\sqrt{K_\xi}$ (see values in \Tab{tab:kappas}),
    which determines the
    expected scale for non-linearities to develop (they do not appear in the plot for weak PTs with $\vw = 0.4$ and $0.8$).
    } \label{fig:GWmodel}
\end{figure*}

In \Fig{fig:GWmodel}, we compare the integrated GW amplitude obtained in the simulations (also shown in the upper panel of \Fig{fig:GWgrowth}) with the prediction from our
locally stationary model for the UETC presented in \Sec{sw_extended}. 
Furthermore, we also show the effect of the expansion of the Universe, following the model presented in \Sec{sec:expansion}. 
Because of this, the integrated GW amplitude is plotted as a function of conformal time,
$a_\ast \beta \delta \eta_{\rm fin} 
\equiv a_\ast \beta \eta_{\rm fin} 
- \beta/H_\ast$ (we remind that $a_\ast \eta_\ast\equiv a_\ast \HH_\ast^{-1} = H_\ast^{-1}$ and that $\delta \eta_{\rm fin}$ denotes the source duration in conformal time).
The analytical, integrated GW amplitude is obtained from the scaling of \Eq{OmGW_general}, leading to \Eq{eq:Inumbintime}, adopting the power-law fit
$K(\eta) = K_0 \, (\Delta \eta/\Delta \eta_0)^{-b}$,
assuming that the GW production starts at the time
when the PT is completed 
$a_\ast \beta \Delta \eta_0 \simeq 10$,
and using the decay rate $b$ provided in \Fig{fig:decay_2} for each parameter set. 
This agrees extremely well with the integrated GW amplitude obtained from the simulations 
${\cal I}_{\rm sim}^{\rm int}/(K_\xi^2\, \beta R_*)$, provided we
{\em (i)} correct for the initial time of GW production from
$a_\ast \beta \Delta \eta_{0} \simeq 10$ to
$a_\ast \beta \Delta \eta_{\rm init} = 16$ by adding the integrated
$K^2_{\rm int}(\eta_0,\eta_{\rm init})$ to the numerical integrated GW amplitude plotted in \Fig{fig:GWmodel}, and {\em (ii)} we use the extrapolated values ${\cal K}_0$ presented in \Sec{sec:convergence} to account for the corrections
due to the under-resolution of the self-similar
profiles.
Note that we are extending the validity of the scaling of \Eq{OmGW_general}, and of the power law in \Eq{eq:Ktildegeneral}, outside the regime in which they have been tested by simulations: namely, in the interval $\eta \in [\eta_0, \eta_{\rm init}]$, and to times after the end of the
simulations, $\eta \in [\eta_{\rm end}, \eta_{\rm fin}]$.
The effect of the expansion of the Universe is accounted for according to \Eq{Kexp_general}, leading to \Eq{Kexp_fit}; we further choose as examples the values $\beta/H_\ast = 100$ and $1000$. 

It is important to remark that in general GWs start to be sourced from the time of the first bubble collision. 
However, in this work we are neglecting 
the GW production before $\tilde t_0$, the time at which the full simulation volume is in the broken phase: this implies that the GW amplitude extrapolated from the simulations is underestimated. 
Indeed, by the time $\tilde t_0$, $K$ has already significantly decayed in strong PTs, as can be appreciated by \Fig{fig:E_kin_evolution}.
However, in the collision regime the results are expected
to strongly depend on the nucleation history; furthermore, the stationary and locally stationary assumptions might not be valid, neither would be the power-law decay description of the kinetic energy evolution. 
The GW production near the collision regime is therefore beyond the scope of the present analysis, and deserves further study.

%%%%%%%%%%%%%%%%%%%%%%%%%%%%%%%%%%%%%%%%%%%%%%%%%%%%%%%%%%%%%%%%%%%%%%
\subsection{Gravitational wave spectral shape\label{sec:shape}}

In this section, we present the numerical results concerning the spectral shape for weak, intermediate, and strong transitions
and a range of wall velocities.
We present fits to the data and extract spectral features.
Results for weak and intermediate transitions were previously obtained in hybrid simulations in Ref.~\cite{Jinno:2020eqg} and Higgsless simulations in Ref.~\cite{Jinno:2022mie}.
Utilizing the improved Higgsless code, we update the results of Ref.~\cite{Jinno:2022mie} and present new results for strong transitions.
In addition to updating the results, we present scaling relations derived from normalizing to $R_*$ rather than to $\beta$, evidently revealing a better scaling behavior of the knee position in the spectrum associated with the typical bubble size.

The findings in Ref.~\cite{Jinno:2022mie} indicate that the GW spectrum $\OmGW(k)$ is characterized by a double broken power law:  
at small $k$, a $\OmGW(k)\propto k^3$ scaling was observed, which is also expected from causality \cite{Caprini:2009fx}.
At large $k$, the spectrum decays as $\OmGW(k)\propto k^{-3}$.
At intermediate scales, a linear scaling regime $\OmGW(k)\propto  k$ was observed.
Due to limited resolution, the spectrum appears to exponentially decay
beyond a damping scale $k_e$ as a result of numerical viscosity.
At scales around or beyond the Nyquist wave number, $\OmGW(k)$ behaves erratically and is always neglected for the purpose of analysis and parameter extraction. 

To capture the behavior of the spectral
shape $S(\tilde k) \equiv {\cal I}_{\rm sim} (\tilde k)/{\cal I}_{\rm sim}^{\rm int}$, we use the following double broken
power law function,
\begin{equation}
\label{eq:shape function}
S(k,\,k_1,\,k_2,\,k_\damp)=S_0\, \left(\frac{k}{k_1}\right)^{n_1}\left[1+\left(\frac{k}{k_1}\right)^{a_1}
\right]^{\frac{-n_1+n_2}{a_1}}\left[1+\left(\frac{k}{k_2}\right)^{a_2}\right]^{\frac{-n_2+n_3}{a_2}}
\, e^{-\left(k / k_e\right)^2},
\end{equation}
which corresponds to the shape function used, e.g.,
in Refs.~\cite{RoperPol:2023bqa,Caprini:2024hue}, with an additional exponential damping factor
effective above the damping scale $k>k_e$.
We expect that the exponential damping found in the simulated spectra is purely due to numerical viscosity so we disregard the parts of the spectra where the exponential
damping is relevant.
Assuming $k_1 < k_2$ and $k < k_e$,
the fitting parameters correspond to the slopes $n_1$, $n_2$, and $n_3$, such that $S(k) \sim k^{n_1}$ at small
wave numbers $k < k_1$, $S(k) \sim k^{n_2}$ at intermediate $k_1 < k < k_2$, and $S(k) \sim k^{n_3}$ at large $k > k_2$.
The parameters $a_1$ and $a_2$ allow to control the sharpness/smoothness of the spectral shape
around the knee and peak at $k_1$ and $k_2$.
$S_0$ is a normalization constant defined by the condition that $\int S \dd \ln k = 1$.
We note that the choice $a_1=2$, $a_2=4$, $n_1=3$, $n_2=1$, and $n_3=-3$ renders \Eq{eq:shape function} equivalent to
\begin{equation}
\label{eq:old_shape_function}
    S_f(k,\,k_1,\,k_2,\,k_e)=S_0 \times \frac{\left(k / k_1\right)^3}{1+\left(k / k_1\right)^2\left[1+\left(k / k_2\right)^4\right]} \times e^{-\left(k / k_\damp\right)^2},
\end{equation}
which was previously used in Ref.~\cite{Jinno:2020eqg}.
\EEq{eq:shape function}, however, allows for a more adaptable
recovery of the GW spectrum peak position and slopes
by adapting the sharpness/smoothness
of the spectral shape around the knee and the peak to the one
found in the numerical data.

We expect the characteristic knee and peak of the GW spectra to be determined by the scale of the fluid perturbations $R_\ast$.
Another important length scale is the fluid-shell thickness \cite{Hindmarsh:2016lnk,Hindmarsh:2019phv}
\begin{equation}
    \xi_{\text {shell }}:=\xi_{\text {front }}-\xi_{\text {rear }}\,,
    \label{eq:xi_shell}
\end{equation}
where   
\be
    \xi_{\text {front }}=
\begin{cases}
\xi_{\text {shock\,, }} & \text { for deflagrations and hybrids\,, } \\
\vw \,,
& \text { for detonations\,, }
\end{cases} 
\ee
and
\be
\xi_{\text {rear }}= 
\begin{cases}\vw\,,
& \text { for deflagrations\,, } \\
\cs\,,
& \text { for detonations and hybrids\,.}
\end{cases}
\ee
The scale $R_\ast \, \xi_{\rm shell}$ is
expected to determine the peak of the GW spectrum \cite{Hindmarsh:2013xza,Hindmarsh:2015qta,Hindmarsh:2017gnf,Hindmarsh:2019phv,Caprini:2019egz,Gowling:2021gcy,RoperPol:2023dzg}.

%%%%%%%%%%%%%%%%%%%%%%%%%%%%%%%%%%%%%%%%%%%%%%%%%%%%%%%%%%%%%%%%%%%%%%
\subsubsection*{Fitting to the numerical data}

\begin{figure}
    \centering
    \includegraphics[width=0.83\linewidth]{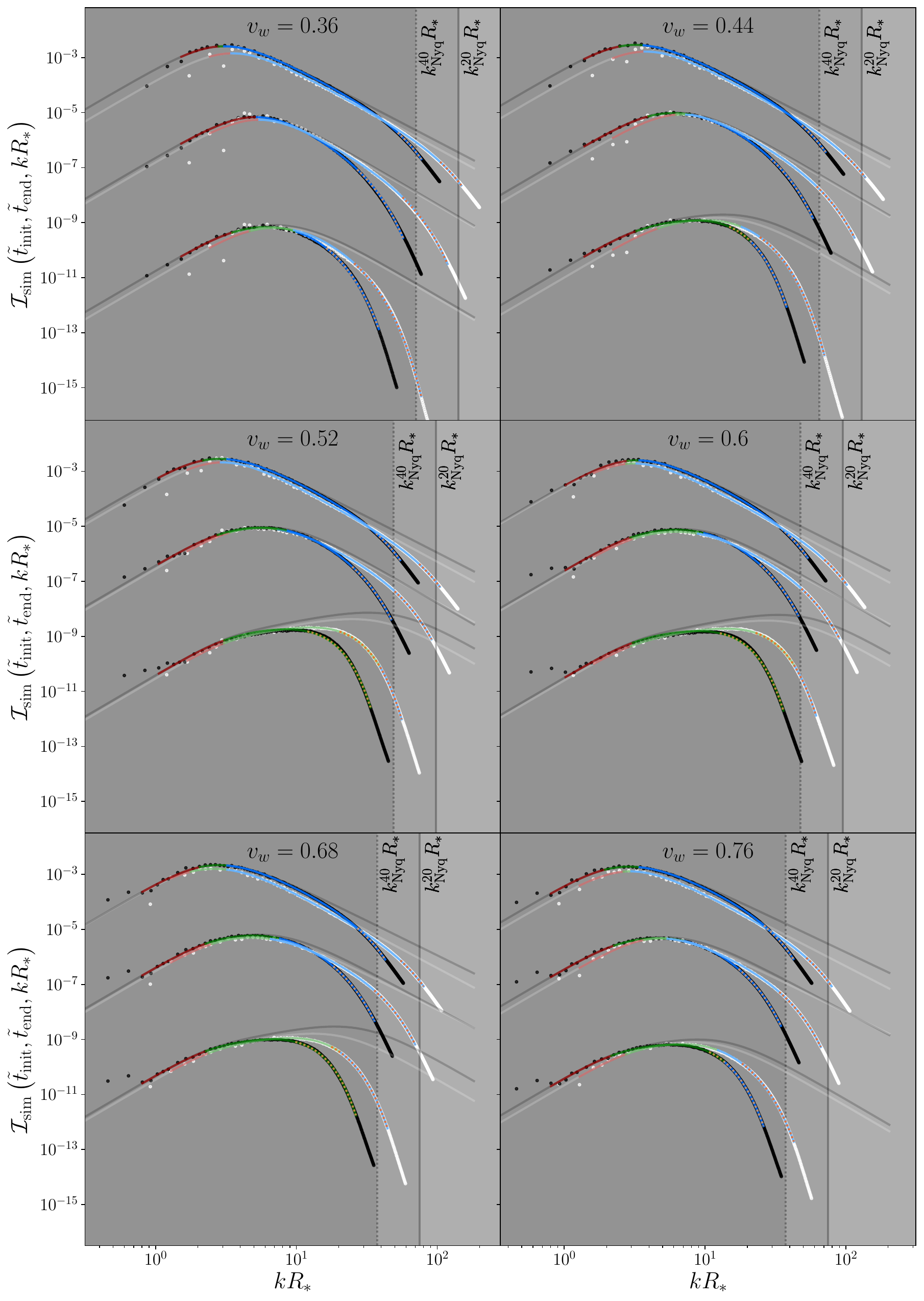}
    \caption{
    Fits of \Eq{eq:shape function} to the numerical results from weak, intermediate, and strong PTs (in each plot, amplitudes increase with larger $\alpha$)
    with $N = 512$ for a range of $v_{\rm{w}}$, and for $\tilde L/\vw = 20$ in brighter colors (white dots for the numerical
    data), and $\tilde L/\vw = 40$
    in darker colors (black dots for the numerical data).
    Red lines indicate wave numbers below the knee $k_1$, green
    indicates
    intermediate wave numbers
    $k_1 < k < k_2$, and blue corresponds to wave
    numbers above the peak $k_2$.
    The dotted orange lines indicate
    wave numbers $k > k_\damp$,
    where exponential damping dominates.
    The light and dark gray lines indicate the resulting fitted double broken power laws excluding the exponential damping.
    Vertical lines indicate the
    Nyquist wave numbers $k_{\rm{Nyq}} R_\ast = \beta R_\ast N/\tilde L $.
    \label{fig:fits}
}
\end{figure}

We fit \Eq{eq:shape function} to our numerically computed 
GW
spectra and thus extract spectral features from our data.
We show in \Fig{fig:fits} the numerical GW spectra ${\cal I}_{\rm sim} (\tilde t_{\rm init}, \tilde t_{\rm end}, \tilde k)$ 
found
in the simulations with numerical resolution $N = 512$ and box sizes $\tilde L/\vw = 20$ and $40$, for a range of wall velocities,
and for weak, intermediate, and strong PTs, together with the
analytical fits.
We use $\tilde t_{\rm init} = 16$ and $\tilde t_{\rm end} = 32$ to evaluate the GW spectra.
In the fitting procedure, we impose the constraint that $k_1 < k_2$. 
However, since $k_\damp$ does not represent a physical scale, we do not require that $k_2<k_\damp$, but allow $k_\damp$ to take on any value independently.
In the cases where $k_2>k_\damp$, the spectral peaks are not resolved properly and suffer from numerical viscosity.

\begin{figure}
    \centering
    \includegraphics[width=0.99\linewidth]{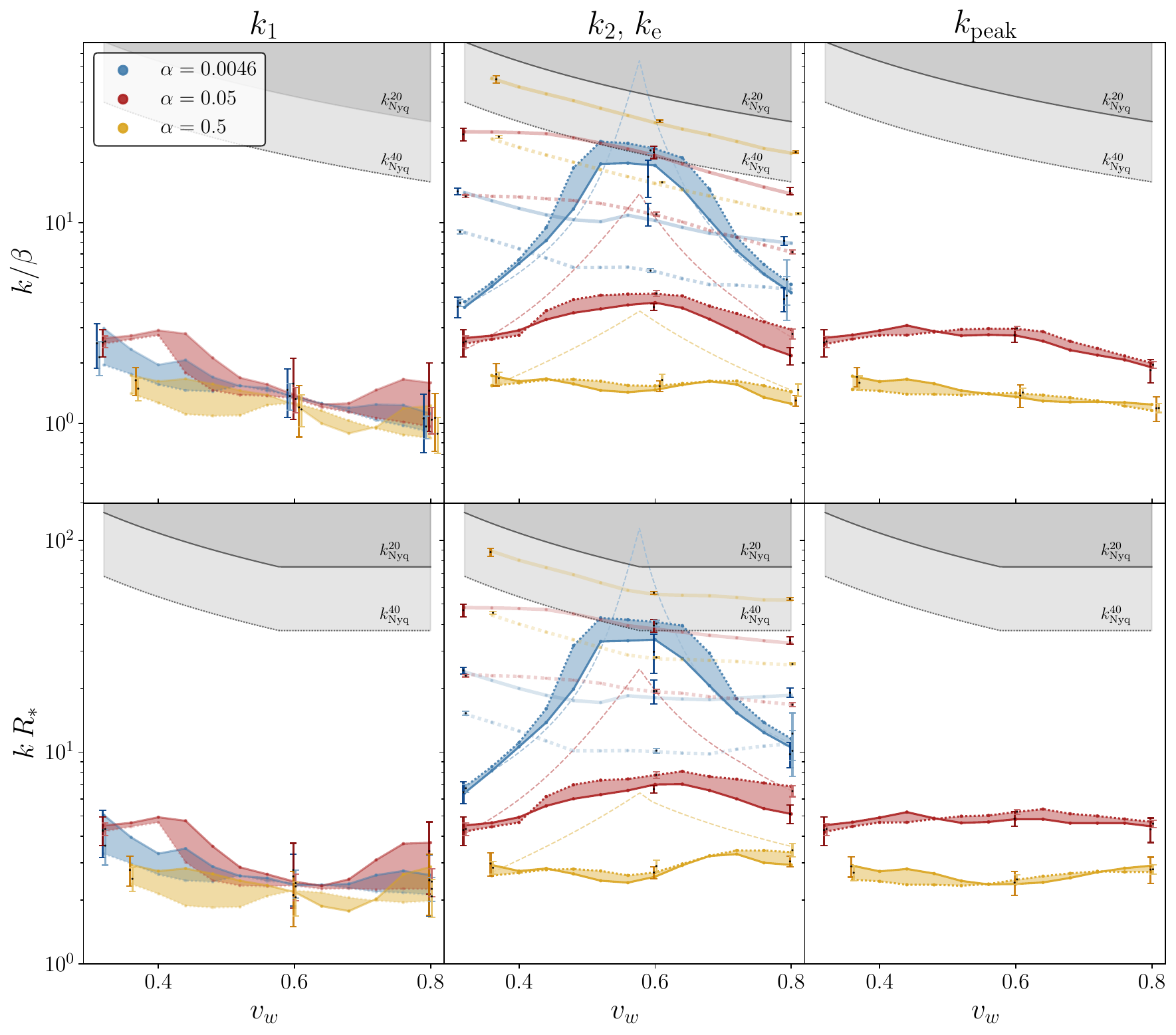}
    \caption{
    Fitted characteristic wave numbers $k_1$ (left column), $k_2$
    and $k_{\damp}$ (middle column), and $k_{\rm{peak}}$
    (right column) for weak (blue), intermediate (red), and strong (orange) PTs, using simulations with $N = 512$ and
    $\tilde L/\vw = 20$ (40) in solid (dotted) lines.
    Gray regions indicate the Nyquist frequency
    $k_{\rm{Nyq}} = N/\tilde L$.
    In the upper panel, wave numbers are normalized as $\tilde k \equiv k/\beta$, as presented in Ref.~\cite{Jinno:2020eqg}, while
    in the lower panel, they are normalized as $k\,R_\ast$.
    Thick colored lines of low opacity in middle column
    indicate $k_{\damp}$ for $\tilde L/\vw = 20$ (40)
    in solid (dotted) lines.
    In the upper middle panel, thin color dashed lines indicate $1/\xi_{\rm{shell}}$, while in the lower middle panel, they
    are proportional to $1/\Delta_w$ with an amplitude chosen to fit the values of $k_2 R_\ast$ at low $\vw$
    [see \Eq{k2R_fit} for fitted value of weak PTs].
    In the right column, the lower opacity regions indicate the
    peak as obtained using the double broken power-law fit of \Eq{eq:shape function}.
    We omit the values for weak PTs as they depend on the inclusion/exclusion of the exponential damping.}
    \label{fig:money plot}
\end{figure}

In obtaining the fit, we neglect the first
bin for simulations with $\tilde L/\vw = 20$ and the first two
bins for $\tilde L/\vw = 40$
to avoid the associated significant statistical scatter.
We cut the spectra in the UV where the fit including
the exponential damping deviates from the broken power law with no exponential damping.
While in \Fig{fig:fits} we show fits of \Eq{eq:shape function} to the spectra for different $v_{\rm{w}}$,
in \Fig{fig:money plot} we show the fitted spectral features 
$k_1$, $k_2$, and $k_\damp$ as functions of $v_{\rm{w}}$ for
weak ($\alpha = 0.0046$), intermediate ($\alpha = 0.05$), and strong ($\alpha = 0.5$) PTs.
We present these characteristic wave numbers in units of $1/\beta$
and $1/R_\ast$ to evaluate the resulting dependence on $\vw$ and
determine the scale characterizing the spectral knee and peak.
We note that the maximum value of the spectral shape used in
\Eq{eq:shape function} is located at $k_{\rm peak}$, which does not in general
exactly coincide with $k_2$ (see discussion in Ref.~\cite{Caprini:2024hue}) and their relation depends on the
fitting parameters.
Only when $k_2/k_1 \gg 1$, we can find the analytical relation $k_{\rm peak} = k_2 (-n_2/n_3)^{1/a_2}$.
However, this separation of scales
is only expected for thin fluid shells, i.e., $\vw \lesssim v_{\rm CJ}$, and only found in our simulations
for weak PTs.
We show the resulting spectral peaks obtained numerically from the fit in the
right columns of \Fig{fig:money plot}.

Extraction of the scale of exponential damping $k_\damp$ gives us a handle on the reliability of the measurement of other parameters and the peak; clearly, finding $k_2>k_\damp$ means we are in a regime where damping already dominates on scales larger than the peak in the spectrum. In this case, even though for weak transitions $k_2$ is found to track $1/\xi_{\rm{shell}}$ well above $k_2>k_\damp$, as can be appreciated from the middle column of \Fig{fig:money plot} (which means that we are potentially recovering a trend expected from physical considerations), caution should be taken in interpreting $k_2$ and $k_{\rm{peak}}$ as true physical parameters.
However, for intermediate and strong PTs, we do
not find any evidence for $k_2$ to be determined by $\xi_{\rm shell}$, as previously pointed out in Ref.~\cite{Jinno:2022mie}.
Using our numerical results with $\tilde L/\vw = 20$, which present
better resolution in the UV, averaged over $\vw$,
we find the following values for $k_2$,

\be
\frac{k_2 \,  R_\ast}{2 \pi} \simeq
\begin{cases}
(0.49 \pm 0.024) / \Delta_w \,,  &  \alpha = 0.0046\,, \\
0.93\pm0.13 \,,  & \alpha = 0.05\,, \\
0.45\pm0.042\,,  & \alpha = 0.5\,, 
\end{cases}
\label{k2R_fit}
\ee
where the indicated uncertainty corresponds to the range
in the measurements among the {\em reference} simulations, and
$\Delta_w = \xi_{\rm shell}/\max(\vw, \cs)$ is the normalized
sound-shell thickness.
Sample variance from the 10 nucleation histories, shown by
the error bars in \Fig{fig:money plot},
is generally of the order of the scatter with wall velocity.
On the other hand, the numerical values at the knee, expected to be related
to the fluid perturbations scale $R_\ast$, are found to be
\be
\frac{k_1 \,  R_\ast}{2 \pi} \simeq 0.39 \pm 0.1 \,.
\label{k1_fit}
\ee
We note that both scales $k_1 R_\ast$ and $k_2 R_\ast$ (for intermediate
and strong transitions)
present
very small variability with $\vw$, indicating a rather universal
behavior.
For weak PTs, $k_2 \, R_\ast \, \Delta_w$
is also almost independent of $\vw$, as expected from the
sound-shell model.
The values of $k_2 R_\ast$ and $k_1 R_\ast$ used in Ref.~\cite{Caprini:2024hue} are based on the numerical
results of Ref.~\cite{Jinno:2022mie}.
For weak PTs, we find $k_2 R_\ast \Delta_w$ consistent with
the values used in Ref.~\cite{Caprini:2024hue}, while
we find the value of $k_1 R_\ast$ to be
twice the one used in Ref.~\cite{Caprini:2024hue}.
We note that the extraction of the knee $k_1$ in the IR part of our
spectra is more sensitive to statistical variance, under-resolution,
and the duration of our simulations.

\subsubsection*{Time evolution of the spectral shape in the simulations}

\begin{figure*}[t]
    \centering
    \includegraphics[width=0.32\columnwidth]{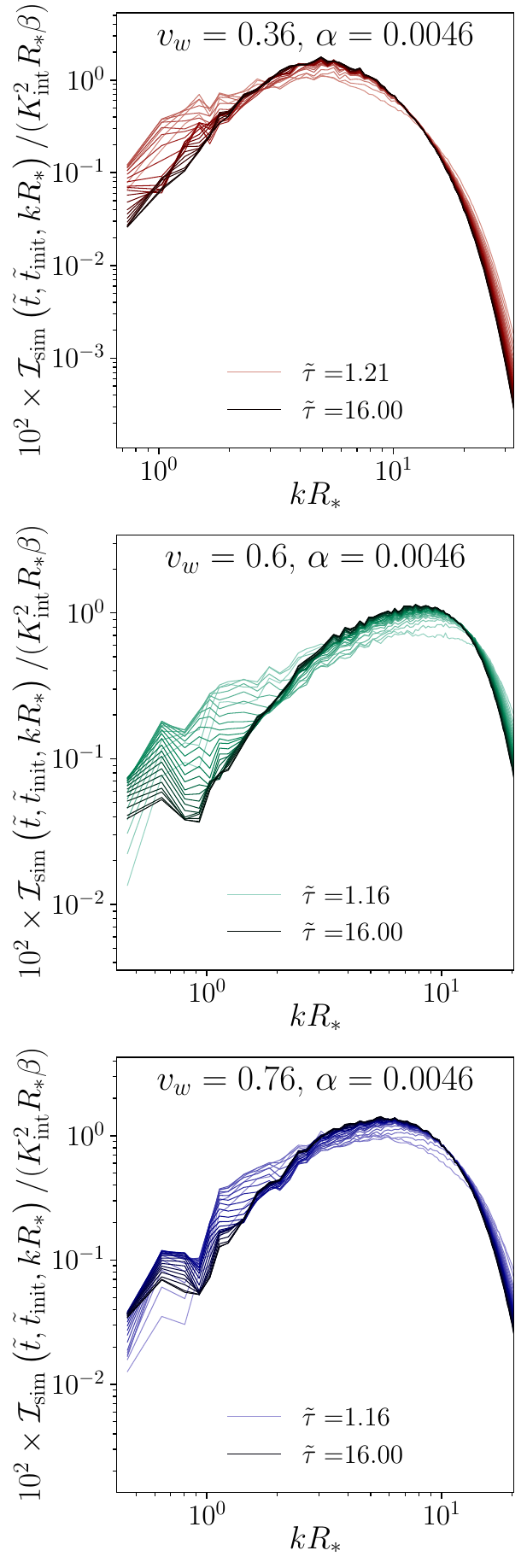}
    \includegraphics[width=0.32\columnwidth]{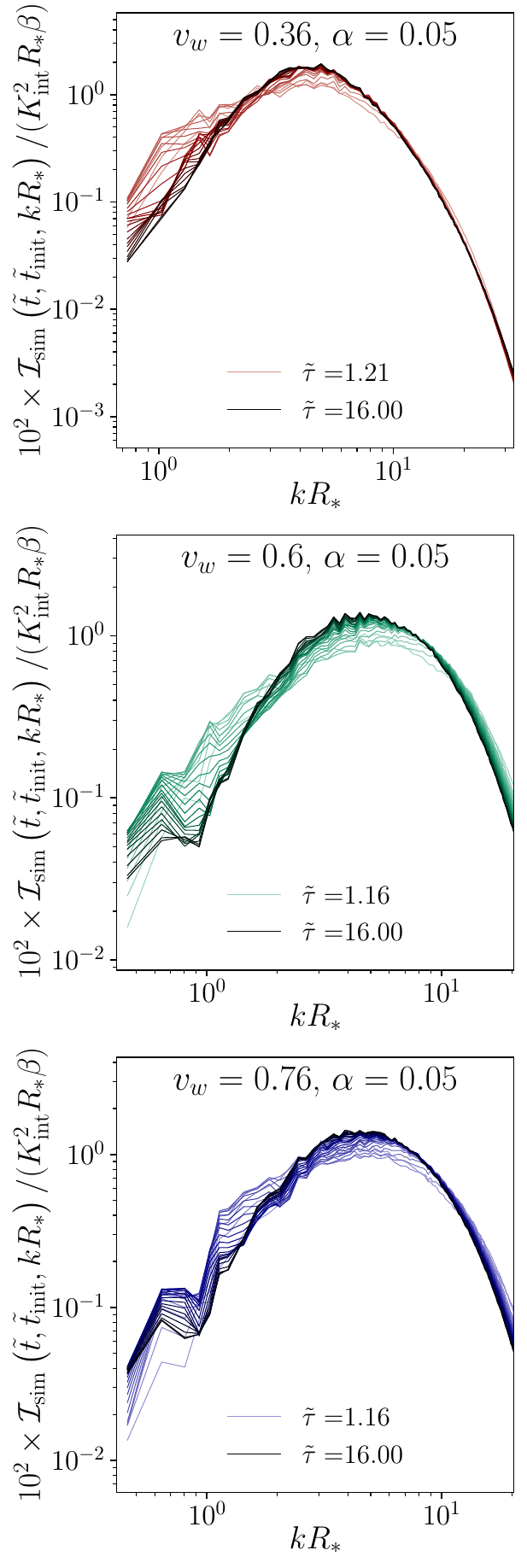}
    \includegraphics[width=0.32\columnwidth]{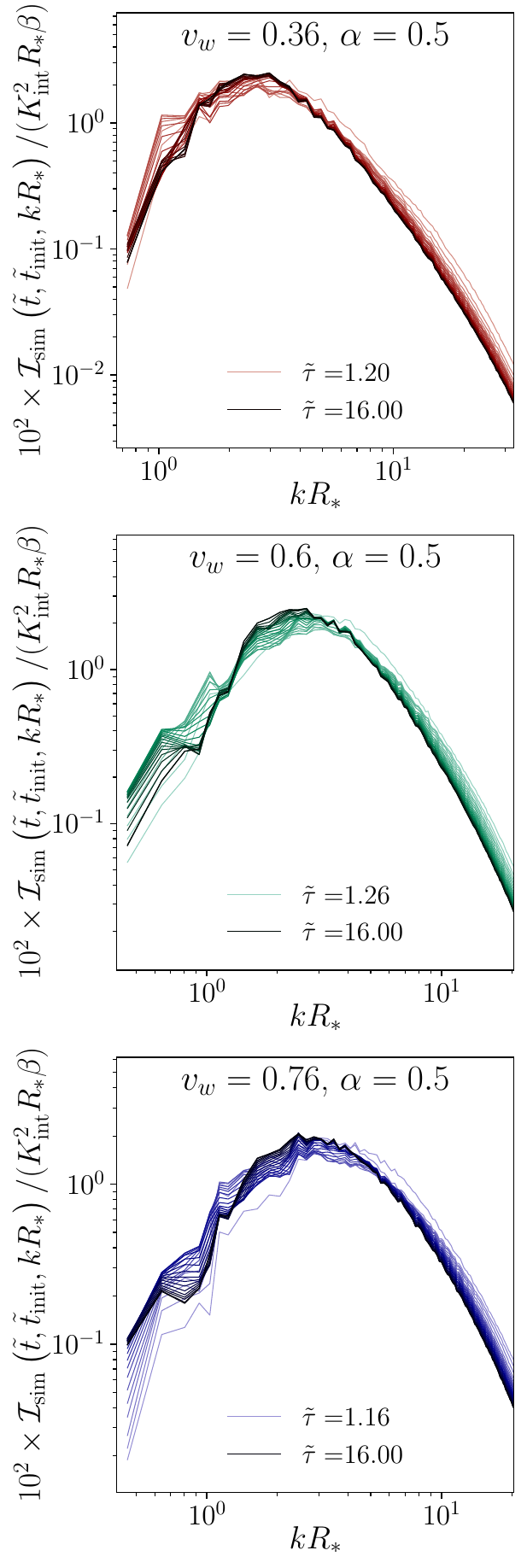}
    \caption{
    Time evolution of the GW spectral shape ${\cal I}(\tilde t_{\rm init}, \tilde t, \tilde k)$
    evaluated at times $\tilde t \in [17, 32]$ with $\tilde t_{\rm init} = 16$, for weak (left columns), intermediate (middle columns), and strong (right columns) PTs, for $\vw = 0.36$ (upper panel), 0.6 (middle panel), and 0.76 (lower panel).
    These simulation times correspond to source durations
    $\tilde \tau \equiv \tilde t - \tilde t_{\rm init} \in [1, 16]$.
    The numerical resolution is $N = 512$ and the
    size of the simulation box is $\tilde L/\vw = 40$.
    The GW spectra are normalized by the reference value $\tilde \Omega_{\rm GW} \simeq 10^{-2}$ and the expected scaling $K_{\rm int}^2\,R_\ast \beta$ as in \Fig{fig:GWgrowth} [see \Eq{OmGW_general}].
    }
    \label{fig:GWgrowth_spec}
\end{figure*}

We show in \Fig{fig:GWgrowth_spec} the GW spectrum
${\cal I}_{\rm sim} (\tilde t_{\rm init},
\tilde t, \tilde k)$ for different values of $\tilde t$ of the
simulation, indicating its dependence with the source duration $\tilde \tau = \tilde t - \tilde t_{\rm init}$.
The resulting spectral shape at the end of the simulation, $\tilde t_{\rm end} = 32$, is then shown in \Fig{fig:fits} and used to provide the
fits of the spectral shape presented above.

We find in general that the causal tail, proportional to $k^{3}$
at small $k$,
is present from early times, and a more complex structure
seems to develop below the peak as time advances. 
Although not very clear from our results, we remark that the bump feature developing in particular at high $\vw$
is potentially consistent
with analytical work \cite{RoperPol:2023dzg} and numerical
simulations \cite{Sharma:2023mao}.
Within the updated sound-shell model of Refs.~\cite{RoperPol:2023dzg,Sharma:2023mao}, this feature would correspond to the value of $k R_*$ after which the assumptions of the old sound-shell model \cite{Hindmarsh:2016lnk,Hindmarsh:2019phv} are valid.
This occurs around $0.1\lesssim k_{\rm ft} R_*\lesssim 1$ (the reasons behind the  position of this feature, and a consequent more precise estimate, require further study).
Depending on the duration of the GW sourcing $\tau_{\rm fin}$, Ref.~\cite{RoperPol:2023dzg} finds that this feature would be preceded either by the causal $k^3$ tail if
$1 \lesssim \tau_{\rm fin}/R_*\lesssim 10$, or by a linear tail if
$\tau_{\rm fin}/R_* \gtrsim 10$ (see also discussion in \Sec{GW_sw}). 
It is important to point out that this linear tail has nothing to do with the linear increase between $k_1$ and $k_2$.
For the simulations, indeed, $\tau_{\rm end}=(\tilde t_{\rm end}-\tilde t_{\rm init})/\beta=16/\beta$
so that $\tau_{\rm end}/R_*=16/(\beta R_*)=16/[(8\pi)^{1/3}{\max(\vw, \cs)]\lesssim 10}$.
Therefore, the simulations results seem to be consistent with the predictions of the revised sound-shell model of Ref.~\cite{RoperPol:2023dzg} (at least for weak PTs), predicting a  $k^3$ increase at low $k$, a feature at $0.1\lesssim k_{\rm ft} R_*\lesssim 1$, and a double peak structure at $k R_*\gtrsim 1$ (see Fig.~13 of Ref.~\cite{RoperPol:2023dzg}).
If the simulation would continue for longer times and the GWs would continue to be sourced with no
decay of the kinetic energy (e.g., for weak PTs), then we would
expect the linear regime in $k$ to develop in the IR tail \cite{RoperPol:2023dzg}.

Furthermore, we find that the growth of the GW amplitude with the source duration is faster than
linear at small wave numbers and early simulation times.
This is also consistent with the predictions of Ref.~\cite{RoperPol:2023dzg}, which, when no significant
decay of $K$ occurs, finds a
quadratic growth
with the source duration (i.e., with $\tilde t$ in the present case), 
that transitions to
a linear growth 
at later times the smaller is $k$.
As discussed in \Sec{sw_extended}, the
modeling presented and validated
for the integrated GW spectrum, based on the assumptions of a
locally stationary UETC and its small compact support, is only expected to hold at wave
numbers $k R_\ast \gg \beta R_\ast/(\tilde t - \tilde t_{\rm init})$.
Then,
using the GW spectral shape as measured at $\tilde t_{\rm end}$ as the one that ultimately enters the proposed model in \Eq{OmGW_general}, effectively implies the assumption that all wave numbers evolve
with the source duration in the same way as the overall amplitude, until $\tilde t_{\rm fin} > \tilde t_{\rm end}$ is
reached.
However,
different time evolutions than those validated for the
integrated amplitude, even if occurring at wave numbers that
do not significantly contribute to the integrated
amplitude
and/or at times after the end of the simulation,
could still potentially affect the
resulting spectral shape of the GW spectrum.
This can occur within the sound-shell model in the IR regime, as shown in Ref.~\cite{RoperPol:2023dzg}, where a transition from the linear
towards a quadratic growth with the source duration
is expected at small $k$ in the stationary case.
Moreover, modifications can also occur as the assumption of a
stationary (or its extension to locally stationary
in \Sec{sw_extended} to include decaying sources) UETC
is no longer valid due to,
for example,
the potential development of
non-linear fluid perturbations and vortical motion.
In the latter case, the
resulting GW spectrum is expected to have a different time evolution
than the one for compressional motion
\cite{Gogoberidze:2007an,Kosowsky:2001xp,Caprini:2009yp,RoperPol:2019wvy,RoperPol:2021xnd,RoperPol:2022iel} and we 
expect that the GW modes would reach their saturation amplitudes
in this regime.
Finally, we note that this spectral shape is computed in flat Minkowski space-time;
accounting for the expansion of the Universe could modify it,
for source durations comparable or longer than the Hubble time.
However, based on the results of Ref.~\cite{RoperPol:2023dzg}, we expect the modifications
to occur in the IR tail.

%%%%%%%%%%%%%%%%%%%%%%%%%%%%%%%%%%%%%%%%%%%%%%%%%%%%%%%%%%%%%%%%%%%%%%
\subsubsection*{GW spectral slopes}

In general, we find a clear $n_1 = 3$ slope at the smallest
frequencies of our simulations, consistent with the expected causal tail, $S(k) \sim k^3$ \cite{Caprini:2009fx,RoperPol:2023dzg}.
At intermediate wave numbers, we fix $n_2 = 1$, although this
range of $k$ is not large enough to have a clear prediction of
the exact intermediate slope. 
As already mentioned, this intermediate slope bears no relation with the linear increase found in the IR tail in Ref.~\cite{RoperPol:2023dzg} that could appear at later times
if the system remains in the linear regime (as argued above).
However, it is clear that a smoothing with respect to the $k^3$
occurs in this range that eventually leads to the decrease
$k^{n_3}$ with $n_3 < 0$ at large wave numbers $k > k_2$.
In this regime, the sound-shell model predicts a
slope $n_3 = -3$ \cite{Hindmarsh:2013xza,Hindmarsh:2015qta,Hindmarsh:2016lnk,Hindmarsh:2017gnf,Jinno:2020eqg,Jinno:2022mie,RoperPol:2023dzg,Sharma:2023mao}, and our simulations show a clear $n_3 \approx -3$
whenever $k_2 \ll k_\damp$.
However, we allow $n_3$ to be a parameter in our fits, since
deviations might occur, potentially due to the development of non-linearities.
From \Fig{fig:fits} it is apparent that, for strong PTs, the simulations offer sufficient dynamical range to sample the UV slope of the GW spectrum. 
This is particularly interesting, since for strong PTs, we expect a departure from $n_3 = -3$
if
non-linearities lead to a cascade of energy into the UV, thus modifying the slope towards a Kolmogorov turbulence
spectrum with $n_3 = -8/3$ \cite{Kosowsky:2001xp,Caprini:2006jb,Gogoberidze:2007an,Caprini:2009yp,
Niksa:2018ofa,RoperPol:2019wvy,RoperPol:2022iel}, or a shallower acoustic turbulence spectrum \cite{1973SPhD...18..115K,Dahl:2024eup}.
Hence, we allow $n_3\geq-3$ when deriving the fit for strong and intermediate PTs, while we fix $n_3 = -3$ for weak PTs, since the dynamical depth is typically insufficient to recover the UV behavior in these cases. 
Indeed, for weak PTs,
$k_2$ is determined by the sound-shell thickness [see \Eq{k2R_fit}], leading to $k_2 \gtrsim k_\damp$ in general.

\begin{figure}[t]
    \centering
    \includegraphics[width=0.6\columnwidth]{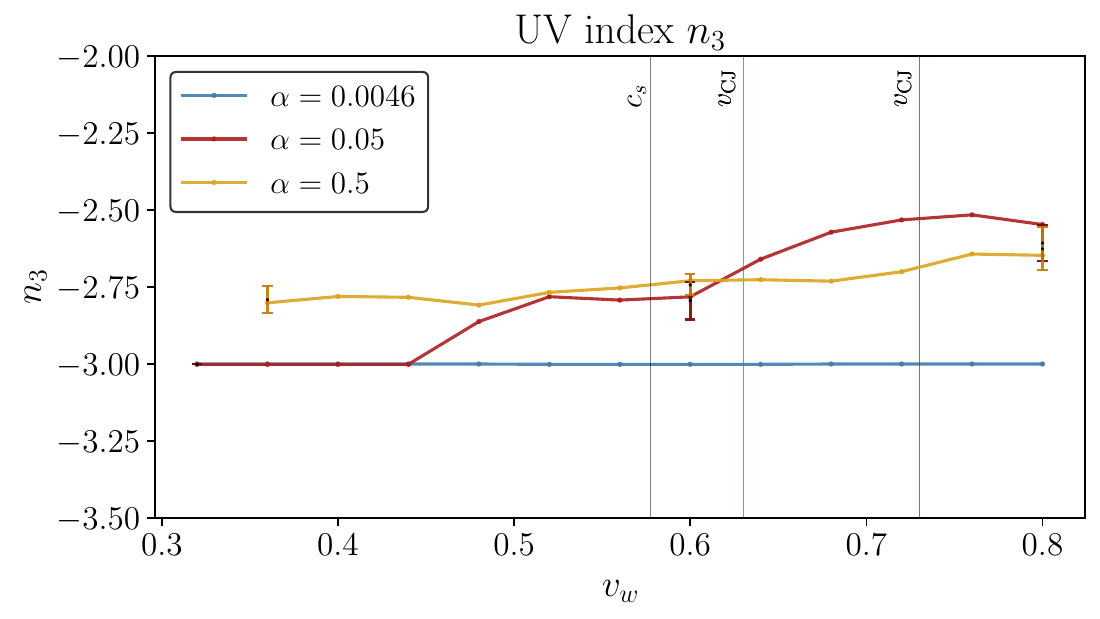}
    \caption{
    Fitted UV index $n_3\geq-3$ for the spectral shape
    of weak (blue), intermediate (red), and strong (orange) PTs, found for simulations with $N = 512$ and $\tilde L/\vw = 20$.
    We note that $n_3 = -3$ is fixed for weak PTs.
    The error bars show the standard deviation from 10 different
    bubble nucleation histories.
    The black vertical line indicates the sound speed $\cs$,
    while colored lines the Chapman-Jouguet $v_{\rm CJ}$ for
    weak and intermediate PTs ($v_{\rm CJ} \simeq 0.89$ for
    strong PTs is out of the plot).
    \label{fig:n_3}
}
\end{figure}

In \Fig{fig:n_3}, we plot the fitted
values of $n_3$.
For intermediate transitions, we observe a marginal increase in $n_3$ towards $-2.5$ as the wall velocity is increased. Strong transitions exhibit a similar trend, while also preferring an optimal $n_3\lesssim-2.75$ for small $\vw$.

%%%%%%%%%%%%%%%%%%%%%%%%%%%%%%%%%%%%%%%%%%%%%%%%%%%%%%%%%%%%%%%%%%%%%%
\subsubsection*{Smoothing/sharpening of the knees}

Introducing two new free parameters $a_1$ and $a_2$ obviously improves the fits to the numerical data, compared to using the simpler \Eq{eq:old_shape_function} \cite{Jinno:2020eqg,Jinno:2022mie}.
Since the peak of the GW spectrum $k_{\rm peak}$ is of greatest phenomenological interest, we adjust the parameters $a_1$ and $a_2$ to constants that universally recover the peak position well for all wall velocities and strengths. Empirically, we find that a slight sharpening of the knee and a slight smoothing of the peak typically improve the peak position recovery and yield good results for the fit overall. Measurements of $a_1$ benefit from simulations with more data points in the IR, and we use exclusively
simulations with $\tilde L/\vw = 40$
for its estimation, whereby $a_1=3.6$ (i.e., an increase from $a_1=2$ as used in Refs.~\cite{Jinno:2020eqg,Jinno:2022mie}) is found suitable.
Measurements of $a_2$, on the other hand, benefit from resolving the UV, for which we use exclusively simulations with $\tilde L/\vw = 20$, and find that $a_2 = 2.4$ (i.e., a reduction from $a_2=4$ as used in Refs.~\cite{Jinno:2020eqg,Jinno:2022mie}) is an adequate choice. We use these values for $a_1$ and $a_2$ throughout this study, but point out that in principle, 
the spectral fit could be improved by varying these parameters at the cost of a larger scatter in the parameter extraction (due to degeneracies).
The parameters $k_1$, $k_2$, $k_e$, and $k_{\rm peak}$ shown in \Fig{fig:money plot} have been obtained using
the aforementioned values of $a_1$ and $a_2$.
In the different rows, these parameters are expressed in terms of the physical length scales discussed in the 
last section.

%%%%%%%%%%%%%%%%%%%%%%%%%%%%%%%%%%%%%%%%%%%%%%%%%%%%%%%%%%%%%%%%%%%%%%
\section{Summary and conclusions \label{sec:summary}}
%%%%%%%%%%%%%%%%%%%%%%%%%%%%%%%%%%%%%%%%%%%%%%%%%%%%%%%%%%%%%%%%%%%%%%

We have conducted numerical simulations of cosmological
first-order phase transitions (PTs) using the Higgsless approach
\cite{Jinno:2022mie} to compute the fluid perturbations in
the primordial plasma induced by a PT and the resulting GW
spectra, for PT strengths $\alpha = 0.0046$ (weak),
$0.05$ (intermediate), and $0.5$ (strong); and a broad range of
wall velocities $\vw \in (0.32, 0.8)$.
These results extend the previous numerical results of Ref.~\cite{Jinno:2022mie} to strong PTs, and include a larger number
of numerical simulations for weak and intermediate PTs.

We present for the first time
results of the GW amplitude and spectral shape sourced by
fluid perturbations from strong PTs with $\alpha = 0.5$.
We have slightly updated the numerical code, although with no
significant impact on the numerical results.
We have compared the results of our simulations to those
obtained assuming
a stationary unequal time
correlator (UETC) for the GW source, an assumption
usually made in analytical computations of compressional motion (e.g., sound waves in the limit of linear perturbations)
based on the hypothesis
that the GWs are produced by a stationary
superposition of sound waves.
This assumption is also commonly used
to extrapolate/interpret
the results of numerical simulations.
We confirm that the stationary UETC assumption well describes the results of our simulations of weak PTs.
However, this assumption only holds if the kinetic energy fraction $K$ is constant in time, while
we find strong numerical evidence for the decay of
$K$ with time for intermediate PTs with
highly confined profiles, and for strong PTs. 
We also find a clear deviation
with respect to the linear
dependence of the GW amplitude with the
source duration, found in previous 
analytical studies based on the stationary UETC, 
and commonly assumed in the GW templates used
in the literature.
We associate this deviation to the numerically
found decay of the kinetic energy
fraction $K$.
Consequently, we propose a novel model to account for this decay, namely, we
extend the stationary UETC model to a locally stationary UETC.

The numerical results presented in this work have allowed us
to test the validity of the locally stationary UETC model
when non-linearities develop in the system.
In particular, we were able to find numerically the relevant scales that enter
in the resulting
GW amplitude and spectral shape; see \Eq{OmGW_general}.
One important difference with the stationary UETC model is that the linear dependence with the source duration
$\tau_{\rm fin}$ 
[see \Eq{OmGW_stat2}] is substituted with an integral in time of the squared kinetic energy fraction
[see \Eq{OmGW_stat4}].
We have tested the model prediction for the
dependence of the GW energy density with the
source duration with simulations.
In the case of weak PTs, we have confirmed that the GW energy density depends linearly on the source duration. 
Within the sound-shell model, this is then assumed to coincide with the time of development of non-linarities, $\tau_{\rm sh}$.
On the other hand,
we have shown that, for strong and some intermediate PTs,
the GW production does not stop at the time when non-linearities develop,
but it keeps increasing within the simulations, following the integrated $K^2$.
We expect that the observed increase after non-linearities develop is universal and would also
affect the GW amplitude of weak and some intermediate PTs for which non-linearities occur at times after
the end of our simulations.
At later times, we expect the GW amplitude to
saturate.
As the ultimate saturation stage is not present in our simulations,
we present our results as a function of the GW source duration,
treating it as a free parameter.
However, for strong and some intermediate PTs, we also find
that the decay rate $b$ of the kinetic energy squared
becomes larger than $1/2$ (see \Fig{fig:decay_2}), leading to a converged value of
the GW amplitude in the limit of long duration.
Hence, in these cases, the dependence of the GW amplitude with
the source duration is
weaker
and the error due to
a possible wrong choice for the source duration is significantly
reduced (see \Fig{fig:GWmodel}).
In any case,
it remains in general of paramount
importance to determine the exact value of the
resulting saturated GW amplitude
to further improve the accuracy of the predicted GW spectra expected from
weak to strong first-order
phase transitions.
The modeling of the remaining stage, until saturation is reached,
will also require numerical
simulations as it is deep in the non-linear regime and
potentially dominated by vortical turbulence.

In the following, we summarize our numerical results by
providing a template
that can be used by the community to estimate the GW amplitude from
first-order phase transitions, validated for the duration
of our simulations and extrapolated to later times,
taking into account that some of the values presented
might be sensitive to numerical
uncertainty.
The resulting template will be publicly available via \href{https://github.com/CosmoGW/CosmoGW}{\sc CosmoGW} \cite{cosmogw}.
It extends (and includes,
when appropriate)
previous templates based on the stationary UETC assumption of the sound-shell model,
which correctly describes the
stage of GW sourcing dominated by linear fluid perturbations.
Based on the model presented in \Sec{sw_extended} and validated with our numerical simulations in \Sec{GW_spec_time}, we find the following parameterization
of the GW spectrum when the Universe expansion can be ignored
\be
    \OmGW (k)  = \, 3 \, {\cal T}_{\rm GW} \, \tilde \Omega_{\rm GW} \, (H_\ast/\beta) \, K_{\rm int}^2 \ H_\ast R_\ast \ S(k R_\ast) \, ,
    \label{eq:base_normalization}
\ee
where $S(k)$ denotes the shape function of the spectrum that is normalized to $\int d\ln k \, S(k) =1$, and $K_{\rm int}^2$
is the integrated kinetic energy fraction $K^2$ over $\tilde t \equiv t \beta$, such that
$(H_\ast/\beta)\, K_{\rm int}^2$ reduces to $K^2 \, H_\ast \tau_{\rm fin}$ when $K$ is
constant, being $\tau_{\rm fin}$ the GW source duration.
Therefore, \Eq{eq:base_normalization} is a generalization of the 
parameterization used in the stationary UETC assumption previously
tested with numerical simulations \cite{Hindmarsh:2013xza,Hindmarsh:2015qta,Hindmarsh:2017gnf} and
usually assumed for sound-wave sourcing of GWs \cite{Caprini:2015zlo,Hindmarsh:2016lnk,Caprini:2019egz,Hindmarsh:2019phv,RoperPol:2023dzg,RoperPol:2023bqa,Caprini:2024hue}
that predicts a linear dependence with the GW source duration when $K$ does not decay
with time.

Our most robust result to prove the validity of \Eq{eq:base_normalization}
is the almost independent value
of $\tilde \Omega_{\rm GW}$ with the PT parameters.
This is obtained when the typical bubble separation $R_\ast$,
which determines the length scale of fluid perturbations, is given by the front of the expanding bubbles \cite{Caprini:2019egz}
\be
R_* \beta = (8 \pi)^{1/3} \, {{\rm max} (\vw, \cs)} \, ,
\ee
where $\beta^{-1}$ parameterizes the duration of the PT, $\vw$ is the 
wall velocity, and $\cs$ the speed of sound. This way,
the residual dependence on the wall velocity in $\tilde \Omega_{\rm GW}$
is quite limited and we estimate from our numerical simulations values for the GW efficiency $\tilde \Omega_{\rm GW} \sim {\cal O} (10^{-2})$ for a range of PTs [see \Fig{fig:GW_efficiency} and \Eq{eq:Omegatilde}],
\be
10^2 \, \tilde \Omega_{\rm GW} = 
\begin{cases}
1.04^{+0.81}_{-0.67}\,,  & \quad \textrm{for} \quad \alpha = 0.0046\,; \\ 
1.64^{+0.29}_{-0.13}\,,  & \quad \textrm{for} \quad \alpha = 0.05\,; \\
3.11^{+0.25}_{-0.19}\,,  & \quad \textrm{for} \quad \alpha = 0.5\,, 
\end{cases}
\ee
consistent with
previous numerical simulations \cite{Hindmarsh:2013xza,Hindmarsh:2015qta,Hindmarsh:2017gnf} for weak and intermediate PTs, and with the sound-shell model \cite{Hindmarsh:2016lnk,Hindmarsh:2019phv}
for weak PTs.
For intermediate and strong PTs, we find larger efficiencies and much
less dependence with $\vw$ than for weak PTs,
clearly showing a departure with respect to
the predictions of the sound-shell model (see \Fig{fig:GW_efficiency}).

We also provide an estimate of the relevant kinetic energy
fraction ${\cal K}_0$ at the end of the PT using our numerical results
[see \Fig{fig:kappa_eff_kappa} and \Eq{K0_fit}], given in units of the single-bubble $K_\xi$
[see \Eq{Kxi} and values in \Tab{tab:kappas}], which, averaged over wall velocities, becomes
\begin{equation}
    {\cal K}_0 = 0.84^{+0.24}_{-0.29} \, K_\xi\,.
\end{equation}
As a function of $\vw$, we generally find
that $K_\xi$ might slightly underestimate ${\cal K}_0$ for the smallest
$\vw$, while it tends to overestimate it for larger $\vw$
(see \Fig{fig:kappa_eff_kappa}).
This might be a consequence of the energy transfer between thermal
and kinetic energies during the phase of collisions and
the expected development of the sound-wave regime \cite{Hindmarsh:2016lnk,Hindmarsh:2019phv,RoperPol:2023dzg}.

We have studied the decay of the kinetic energy fraction $K$ with
time $t$ in \Sec{decay_K2}, and provide a power-law fit $K(t>t_0) = K_0 (t/t_0)^{-b}$, with $b \geq 0$ indicating the decay rate, that accurately reproduces the numerical results
(see \Figs{fig:E_kin_evolution}{fig:decay_2}).
For small or vanishing values of $b$, one can directly use
\begin{equation}
    K_{\rm int}^2 (b = 0) \to {\cal K}_0^2 \, \beta \,
    \tau_{\rm fin} \to 
    {\cal K}_0^{3/2} \, \beta R_\ast \label{Kint_const}
\end{equation}
in \Eq{eq:base_normalization},
assuming that the duration of the GW
sourcing is given by the shock formation time
$\beta \tau_{\rm fin} \sim \beta \tau_{\rm sh} = \beta R_\ast/\sqrt{K}$, when non-linearities are expected to develop.
In general, we find $b \ll 1$ when the shock
time is larger than our final simulation time $\beta \tau_{\rm sh}
\gg \beta t_{\rm end} = 32$
(as it is the case for weak PTs and some intermediate ones, see values in \Tab{tab:kappas}).
In these cases, the transition towards non-linearities
is not explored in our simulations and, to improve the accuracy
of the saturated GW amplitude, simulations covering
the possible development of the non-linear regime are required in the future.

For non-negligible values of $b$, we find that the decay of $K$ occurs
within the duration of our simulations, potentially indicating that we
are already modeling the GW production in the non-linear regime.
We indeed find that this might be the case as the shock
formation time
is included in the duration of our simulations for some intermediate PTs and for
strong ones, where we find larger values of $b$.
For these PTs, we find that the integrated $K_{\rm int}^2$ becomes
\begin{equation}
    K_{\rm int}^2 (b, \tau_{\rm fin}) \to {\cal K}_0^2 \, \beta \, \Delta t_\ast \, 
    \frac{(1 + \tau_{\rm fin}/\Delta t_\ast)^{1 - 2b} - 1}
    {1 - 2b}\,, \label{Kint_decaying}
\end{equation}
when one uses the power-law fit for $K(t)$ and assumes that the
GW production roughly starts at the time $\beta
\Delta t_\ast \simeq
\beta  \Delta t_0 \simeq 10$ (note that the actual value of $\beta
\Delta t_0$ only
appears as a consequence of our particular fit; see discussion in \Sec{sw_extended}).
It is unclear what should be the final time of GW sourcing in these cases,
as the simulations seem to already be modeling the non-linear regime, so we leave $\tau_{\rm fin}$ as a free parameter.
We note that this is an indication that the GW spectrum keeps growing proportional to \Eq{Kint_decaying}
once non-linearities develop in the fluid,
such that the use of \Eq{Kint_const} would in general underestimate the GW production.
We expect this additional growth to also occur for weak PTs for which non-linearities are
expected to occur at times not covered in our simulations.
We compare in \Fig{fig:GWmodel} the numerical
dependence of the GW amplitude with the source duration $\tilde \tau_{\rm fin} \equiv 
\beta \tau_{\rm fin}$ found in the simulations to the one obtained
using \Eq{Kint_decaying}, extending
the analytical fit beyond the time when the simulations end.
As mentioned above, as the decay rate increases (and it
is larger than 1/2), the GW amplitude saturates, alleviating its dependence
with the source duration.
Therefore, we expect that for strong PTs, the GW amplitudes obtained at the end
of our simulations are close to the saturated ones.

{As a final remark on the integrated GW amplitude, we note that so far the
Universe expansion has been ignored, which is not justified for
long source durations.
Taking into account that the fluid equations are conformally invariant
if the fluid is radiation-dominated \cite{Brandenburg:1996fc,RoperPol:2025lgc},
our simulations in Minkowski space-time would correctly model the fluid evolution
after the end of the PT at $\tilde t > \tilde t_0 \simeq 10$
(see \Sec{sec:numerical_setup} for more details on our numerical setup).
The simulations would also
correctly model the PT at $\tilde t < \tilde t_0$
as long as its duration, measured from the time
at which the first bubble nucleates at $\tilde t_n = 0.5$, is short ($\beta/H_\ast \gg 10$).
However, the GW production is not conformally invariant:
therefore, modifications in the GW spectrum are expected, with respect to the one obtained by numerical simulations (see discussion in \Sec{sec:expansion}).}
As a proxy to estimate the effect of the Universe expansion {on the GW spectrum amplitude}, we
can use the following value for $K_{\rm int}^2$ [see \Eqs{Kexp_fit}{OmGW_locally_stat_exp}]
\begin{equation}
    K_{\rm int, exp}^2 \to {\cal K}_0^2 \, \Upsilon_b
    (\HH_\ast \delta \eta_{\rm fin})\,,
\end{equation}
which generalizes the suppression factor
$\Upsilon = \HH_\ast \delta \eta_{\rm fin}/(1 + \HH_\ast \delta \eta_{\rm fin})$
when the source does not decay \cite{Guo:2020grp,RoperPol:2023bqa} to any decay rate $b$ using 
\Eqss{Kexp_fit}{hyper}
for the power-law fit of $K(\eta)$,
for which we have fixed $\Delta \tilde \eta_\ast = \Delta \tilde \eta_0 \simeq 10\,H_\ast/\beta$ to compute ${\cal K}_0$ and $b$
in the simulations.
Then, the GW spectrum becomes [see \Eq{OmGW_locally_stat_exp}]
\begin{equation}
    \Omega_{\rm GW, exp} (k)  = \, 3 \, {\cal T}_{\rm GW} \, \tilde \Omega_{\rm GW} \, K_{\rm int, exp}^2 \ \HH_\ast \RR_\ast \ {S_{\rm exp} (k_c \RR_\ast)} \,.
    \label{Kint_exp_decay}
\end{equation}
We also compare in \Fig{fig:GWmodel} the expected evolution of the integrated GW amplitude
with the source duration according to \Eq{Kint_exp_decay} for $\beta/H_\ast = 100$ and $1000$.
We note that when one associates $\delta \eta_{\rm fin}$ to the
shock time $\delta \eta_{\rm sh} = \RR_\ast/\sqrt{K}$,
they should correspond to conformal time intervals, instead of cosmic time, due to the conformal invariance of the fluid equations.

Regarding the spectral shape $S(kR_\ast)$ in \Eq{eq:base_normalization}, 
we find that the following template fits accurately the numerical results (see \Sec{sec:shape}),
\be
S(k,\,k_1,\,k_2)=S_0\times \left(\frac{k}{k_1}\right)^{n_1}\left[1+\left(\frac{k}{k_1}\right)^{a_1}\right]^{\frac{-n_1+n_2}{a_1}}\left[1+\left(\frac{k}{k_2}\right)^{a_2}\right]^{\frac{-n_2+n_3}{a_2}} \, ,
\label{eq:spec_shape_conc}
\ee
with $n_1\simeq3$, $n_2\simeq1$, $a_1 \simeq 3.6$, and $a_2 \simeq 2.4$.
To compare with the numerical results we have included 
an exponential damping $e^{-(k/k_\damp)^2}$ in \Sec{sec:shape} [see \Eq{eq:shape function}], effective at $k > k_\damp$, but we omit
it here as we expect it to correspond to numerical viscosity and not have
physical relevance.
The slope of the UV tail
is $n_3\simeq -3$ for weak PTs, and intermediate ones with small wall velocities $\vw \lesssim \cs$.
The slope
becomes slightly shallower (up to $-2.5$) for intermediate PTs with supersonic $\vw$ and strong 
PTs (see \Fig{fig:n_3}).
This effect should not play a major role in phenomenological studies, but a
more detailed description is given in \Sec{sec:shape}.
Furthermore, this shallower GW spectral slope, together with the
decay of the kinetic energy,
{would be consistent with the presence of a forward energy cascade, and therefore}
indicate the development of non-linearities.
To confirm this statement would require a detailed study of the kinetic spectrum properties.
For now,
we present a preliminary study of the vorticity production in
our simulations in \App{sec:vort}.
Note that the spectral shape in \Eq{eq:spec_shape_conc}, fitted from the simulations, might be different from the one of the GW spectrum today. 
For example,
when the fluid perturbations are still linear at the
end of our simulations (e.g., for weak PTs),
further sourcing of GWs would be expected, and this would change their spectral shape,
mainly in the IR tail (see, e.g., Ref.~\cite{RoperPol:2023dzg} and
the discussion in \Sec{sec:shape}). 
Furthermore, the expansion of the Universe would also lead to changes in the spectral shape (see Ref.~\cite{RoperPol:2023dzg} and discussion in \Sec{sec:expansion}).
In this work, we only provide a fit for
the spectral shape derived from the simulations in flat space-time $S(k)$, and defer an evaluation of $S_{\rm exp} (k_c \RR_\ast)$ of \Eq{Kint_exp_decay} to a future study.

The most relevant feature of the spectrum is the position of the peak,
determined by $k_2$. 
In this regard, we find a distinction between weak and intermediate/strong PTs (as already previously
seen in Ref.~\cite{Jinno:2022mie}).
For weak PTs, the peak follows the 
thickness of the fluid shells, $\xi_{\rm shell}$, as given in \Eq{eq:xi_shell},
while for intermediate and strong PTs the dependence on the wall velocity is 
much weaker.
As shown in \Eq{k2R_fit} and \Fig{fig:money plot}, we find the following results for $k_2$,
averaged over all wave numbers and 10 different nucleation histories:
$k_2 R_\ast \simeq \pi/\Delta_w$ for weak PTs,
$k_2 R_\ast \simeq 2 \pi$ for intermediate PTs, and $k_2 R_\ast \simeq \pi$ for strong ones,
where $\Delta_w = \xi_{\rm shell}/\max(\vw, \cs)$ is the normalized
sound-shell thickness, which is only found to enter in $k_2$ for
weak PTs.
Finally, the position of the knee that relates to the typical 
size of the bubbles does not even depend on the strength of the PT
and, quite generally, we find
$k_1 R_\ast \simeq 0.4 \times 2 \pi$ (see \Eq{k1_fit} and \Fig{fig:money plot}).

\acknowledgments

This research was supported in part through the Maxwell computational resources operated at Deutsches Elektronen-Synchrotron DESY, Hamburg, Germany. TK and IS acknowledge support by the Deutsche Forschungsgemeinschaft (DFG, German Research Foundation) under Germany's Excellence Strategy – EXC 2121 “Quantum Universe” - 390833306. HR is supported by the Deutsche Forschungsgemeinschaft under Germany's Excellence Strategy EXC 2094 `ORIGINS'. (No.\,390783311). IS acknowledges support by the Generalitat Valenciana through the Programa Prometeo for Excellence Groups, grant CIPROM/2022/69 ``Sabor y origen de la materia.''
ARP acknowledges support by the Swiss National Science Foundation
(SNSF Ambizione grant \href{https://data.snf.ch/grants/grant/208807}{182044}).
ARP and CC acknowledge the working space provided during
the program on the ``Generation, evolution, and observations of cosmological magnetic fields'' at the Bernoulli Center in Lausanne.
ARP and IS acknowledge the hospitality of the Centro de Ciencias de Benasque
Pedro Pascual during ``The Dawn of Gravitational Wave Cosmology'' conference.
CC was supported by the Swiss National Science Foundation (SNSF Project Funding grant \href{https://data.snf.ch/grants/grant/212125}{212125}) during the development of this project. 
RJ is supported by JSPS KAKENHI Grant Numbers 23K17687, 23K19048, and 24K07013.

%%%%%%%%%%%%%%%%
\appendix

%%%%%%%%%%%%%%%%%%%%%%%%%%%%%%%%%%%%%%%
\section{Computation of the gravitational wave production in simulations}
\label{sec:GW_prod}
%%%%%%%%%%%%%%%%%%%%%%%%%%%%%%%%%%%%%%%

The tensor-mode perturbation $h_{ij}$ enters in the metric line element as
\begin{equation}
    \dd s^2 = a^2 \bigl[- \dd \eta^2 + (\delta_{ij} + h_{ij}) \dd x^i \dd x^j\bigr]\,.
\end{equation}
To conform with simulations, we neglect the expansion of the Universe during the time of the GW generation,
which is in general justified when the sourcing duration, $\tau_{\rm fin} = t_{\rm fin} - t_\ast$,
is shorter than the Hubble time $H_\ast^{-1}$.
In the following, time is therefore denoted by $t$.
The solution of the GW equation 
while the source is active, 
$t < \tfin$, is
\cite{Caprini:2018mtu}
\begin{equation}
    h_{ij} (t < \tfin, \kk) = \frac{6 H_*^2}{k} \int_{t_*}^{t}
    \Pi_{ij} (t', \kk) \sin k (t - t') \dd t',
\end{equation}
where $t_*$ is the initial time at which the tensor anisotropic
stresses, $\Pi_{ij} = \Lambda_{ijlm} T_{lm}/\bar \rho$, start to source GWs, being $\bar \rho = 3 H^2 \Mpl^2$ the critical energy density
and $\Lambda_{ijlm} = P_{il} P_{jm} - \half P_{ij} P_{lm}$ the traceless and transverse projector,
with $P_{ij} = \delta_{ij} - \hat k_i \hat k_j$.
$\Mpl$ is the reduced Planck mass,
$\Mpl = (8\pi G)^{-1/2} \simeq 2.4 \times 10^{18}$ GeV.

The final time of GW sourcing $\tfin$
is the time at which the source stops operating. 
After the sourcing has ended,
the solution is
\begin{equation}
    h_{ij} (t \geq \tfin, \kk) = \frac{6 H_*^2}{k} \int_{t_*}^{\tfin}
    \Pi_{ij} (t', \kk) \sin k (t - t') \dd t'\,,
    \label{sol_der_strains}
\end{equation}
such that 
all GW modes propagate as free plane waves, with an amplitude that is determined by the sourcing process.

From \Eq{sol_der_strains}, the time derivatives of the strains $h_{ij}$ are
\begin{equation}
\partial_t h_{ij} (t \geq \tfin, \kk) = 6 H_*^2 \int_{t_*}^{\tfin}
    \Pi_{ij} (t', \kk) \cos k (t - t') \dd t'\,,
    \label{der_strains}
\end{equation}
which can be used to find the
fractional energy density at present time $t_0$ \cite{Caprini:2018mtu},
\begin{equation}
 \OmGW (t_0) = \frac{\rho_{\rm GW}}{\rho_{\rm tot}^0}  = \frac{(a_*/a_0)^4}{12 H_0^2} \bra{\partial_t h_{ij} (t_0, \xx) \, \partial_t h_{ij} (t_0, \xx)} \,.
\end{equation}
We consider the GW spectrum $\OmGW (t_0, k) \equiv \dd \Omega_{\rm GW} (t_0)/\dd \ln k$, which describes
the two-point correlation
function of the statistically homogeneous
and isotropic strain derivatives,
following the notation of Ref.~\cite{RoperPol:2018sap},
\begin{equation}
\frac{(a_*/a_0)^4}{12 H_0^2} \bra{\partial_t h_{ij} (t_0, \kk) \, \partial_t h_{ij}^* (t_0, \kk')} = (2 \pi)^6 \, \delta^3 (\kk - \kk') \, \frac{\OmGW (t_0, k)}{4 \pi k^3}\,,
\label{two_point_GW}
\end{equation}
such that $\OmGW (t_0) = \int \OmGW(t_0, k) \dd \ln k$.
Substituting \Eq{der_strains} into \Eq{two_point_GW}, we find
\begin{equation}
    \OmGW (t_0, k) = 3 k\, {\cal T}_{\rm GW} \, H_\ast^2 \int_{t_*}^{\tfin} \int_{t_*}^{\tfin}
    E_\Pi (t_1, t_2, k) \cos k(t_0 - t_1) \, \cos k (t_0 - t_2)
    \dd t_1 \dd t_2\,,
    \label{OmGW_today_k}
\end{equation}
where
\begin{equation}
    {\cal T}_{\rm GW} \equiv \biggl(\frac{a_*}{a_0}\biggr)^4 \biggl(\frac{H_*}{H_0}
    \biggr)^2 \simeq 1.6 \times 10^{-5} \, \biggl(\frac{g_\ast}{100}\biggr)^{-1/3}\hspace{-5mm},
    \label{transf_func}
\end{equation}
is the transfer function,
with
\begin{equation}
    \frac{a_\ast}{a_0} \simeq 8 \times 10^{-16} \, 
    \frac{100 {\rm \, GeV \,}}{T_\ast} \biggl(\frac{g_{\ast \rm s}}{100}\biggr)^{-1/3}
    \hspace{-5mm},
\end{equation}
being the ratio of scale factors, and $g_\ast$ and $g_{\ast \rm s}$ denote respectively the relativistic and entropic
degrees of freedom at the time of GW production.
$E_\Pi (t_1, t_2, k)$ is the unequal-time correlator (UETC)
of the anisotropic stresses,
\begin{equation}
\bra{\Pi_{ij} (t_1, \kk) \, \Pi_{ij}^* (t_2, \kk')} = (2 \pi)^6 \, \delta^3 (\kk - \kk') \, \frac{E_\Pi(t_1, t_2, k)}
{4 \pi k^2}\,. \label{twop_Pi}
\end{equation}
At present time, for modes $k t_0 \gg 1$, we can average the product of Green's functions in \Eq{OmGW_today_k} over oscillations to find
\begin{align}
\OmGW (k) = &\, \frac{3 k}{2} {\cal T}_{\rm GW} \, H_\ast^2 \int_{t_*}^{\tfin} \int_{t_*}^{\tfin}
    E_\Pi (t_1, t_2, k) \cos k(t_1 - t_2) 
    \dd t_1 \dd t_2\,. \label{OmGW_aver}
\end{align}
Therefore, once we know the UETC
of the source of GWs, $E_\Pi (t_1, t_2, k)$, we can directly
compute the GW spectrum.
In the present scenario, the GW source corresponds to the fluid perturbations $\Pi_{ij} =  w \gamma^2 \Lambda_{ijlm} v_l v_m/\bar \rho$, modeled through the simulations.
Within a simulation,
the UETC can be computed by
approximating the ensemble average of the
anisotropic stresses in \Eq{twop_Pi} with the average
over spherical shells of radius $k$ in Fourier space,
\begin{equation}
E_\Pi (t_1, t_2, k) = \frac{k^2}
{2 \pi^2 V} \int_{\Omega_k} \frac{\dd \Omega_k}{4\pi} \, \Pi_{ij} (t_1, \kk) \, \Pi_{ij}^* (t_2, \kk) \,.
\label{EPi_Pi}
\end{equation}
Then, substituting \Eq{EPi_Pi} into \Eq{OmGW_aver}
and writing $\cos k (t_1 - t_2) = \cos kt_1 \cos kt_2 + \sin kt_1 \sin kt_2$,
the double integral over $t_1$ and $t_2$ can be expressed as the following product:
\begin{align}
    \OmGW (k) = \frac{3 k^3}{4 \pi^2 V} \, {\cal T}_{\rm GW} \, H_\ast^2 \int_{\Omega_k}
    \frac{\dd \Omega_k}{4\pi} \,  &\, \int_{t_\ast}^{\tfin}
    \Pi_{ij} (t_1, \kk) \, e^{ikt_1} \dd t_1  \nonumber \\ \times &\, \int_{t_\ast}^{\tfin}
    \Pi_{ij}^\ast (t_2, \kk) \, e^{-ikt_2} \dd t_2\,,
\end{align}
where we have used the fact that the resulting
$\Omega_{\rm GW} (k)$ is real.
Finally, defining the following integral over the stress-energy
tensor $T_{ij} = w \gamma^2 v_i v_j$,
\begin{equation}
T_{ij} (q, \kk,t_*,\tfin) = \int_{t_*}^{\tfin} T_{ij} (t, \kk) \, e^{iqt} \dd t \,, \label{Tij_qk}
\end{equation}
and using the property $\Pi_{ij} \Pi_{ij}^\ast = \Lambda_{ijlm} T_{ij} T_{lm}^\ast/\bar \rho^2$, we find the following expression
\begin{align}
\OmGW (k) = &\, \frac{3 k^3}{4 \pi^2 V \bar \rho^2} \, {\cal T}_{\rm GW} \, H_\ast^2 \int_{\Omega_k} \frac{\dd \Omega_k}{4\pi} \, \Lambda_{ijlm} (\hat \kk) \bigl[T_{ij} (q, \kk) \, T_{lm}^* (q, \kk)\bigr]_{q=k} \,.
\label{OmGW}
\end{align}
This is the expression that is implemented in the Higgsless simulations to evaluate the GW spectrum: it has been
previously used in Refs.~\cite{Jinno:2020eqg,Jinno:2022mie}, where it is referred
to as Weinberg's formula, due to its similarity with the expression obtained
for the power emitted by isolated deterministic binaries \cite{WeinbergGC}.
Each simulation produces one realization of the stochastic variables $T_{ij}$, and the assumption of statistically homogeneity and isotropy of the early Universe source is then implemented through the average
over shells for each wave number $k$.
The applicability of \Eq{OmGW} is limited to this
(very reasonable)
assumption.

We have demonstrated that
the expression in \Eq{OmGW} allows to compute the GW spectrum at present time, after performing an average over oscillations in time, for $t>\tfin$. 
However, in the simulations, one computes numerically the right hand side of \Eq{OmGW} only until
the end of the simulation at $t_{\rm end}$, which is,
in general, smaller than $t_{\rm fin}$.
Indeed, as explained in the main text, the GW sourcing by fluid perturbations can last several Hubble times, as the physical decay of the fluid bulk motion is determined by the kinematic viscosity, which is very small in the early Universe.
Therefore, the numerical result will correctly represent the physical GW
amplitude at a given mode $k$ only if the latter
has already reached its free-propagation regime by
the time $t_{\rm end}$, i.e., if its amplitude has already saturated.
Otherwise, the numerical result only provides the GW amplitude 
that would be obtained misleadingly assuming that the source is abruptly switched off at $t_{\rm end}$.

%%%%%%%%%%%%%%%%%%%%%%%%%%%%%%%%%%%%%%%
\section{Corrections to the kinetic energy for multiple bubbles
\label{sec:kinetic_ed}}
%%%%%%%%%%%%%%%%%%%%%%%%%%%%%%%%%%%%%%%

\begin{figure}
    \centering
    \includegraphics[width=0.44\linewidth]{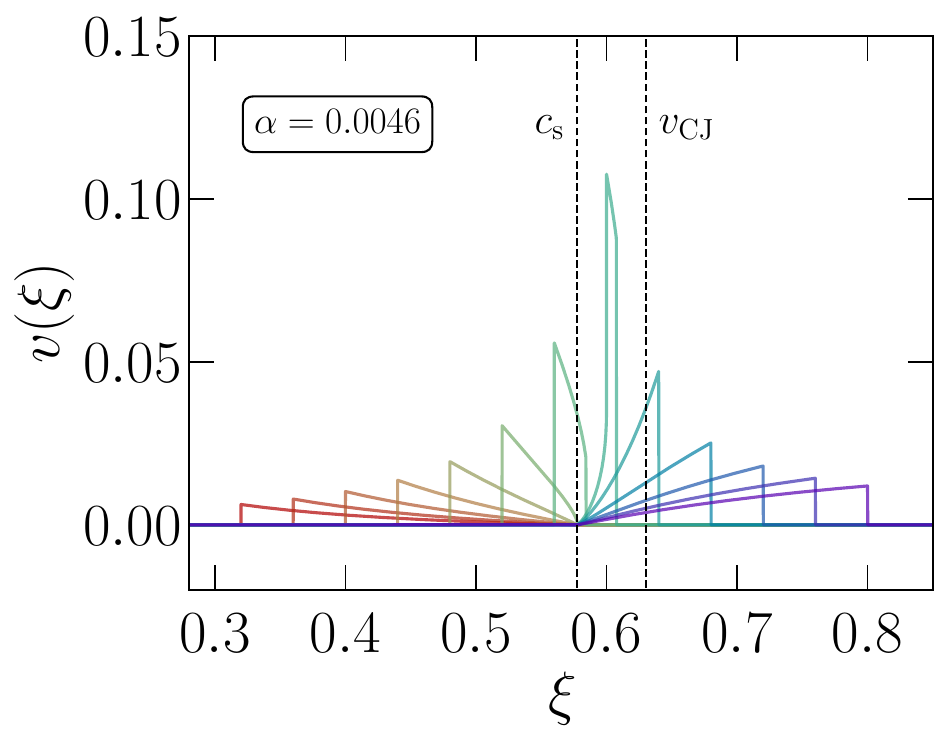}
    \includegraphics[width=0.5\linewidth]{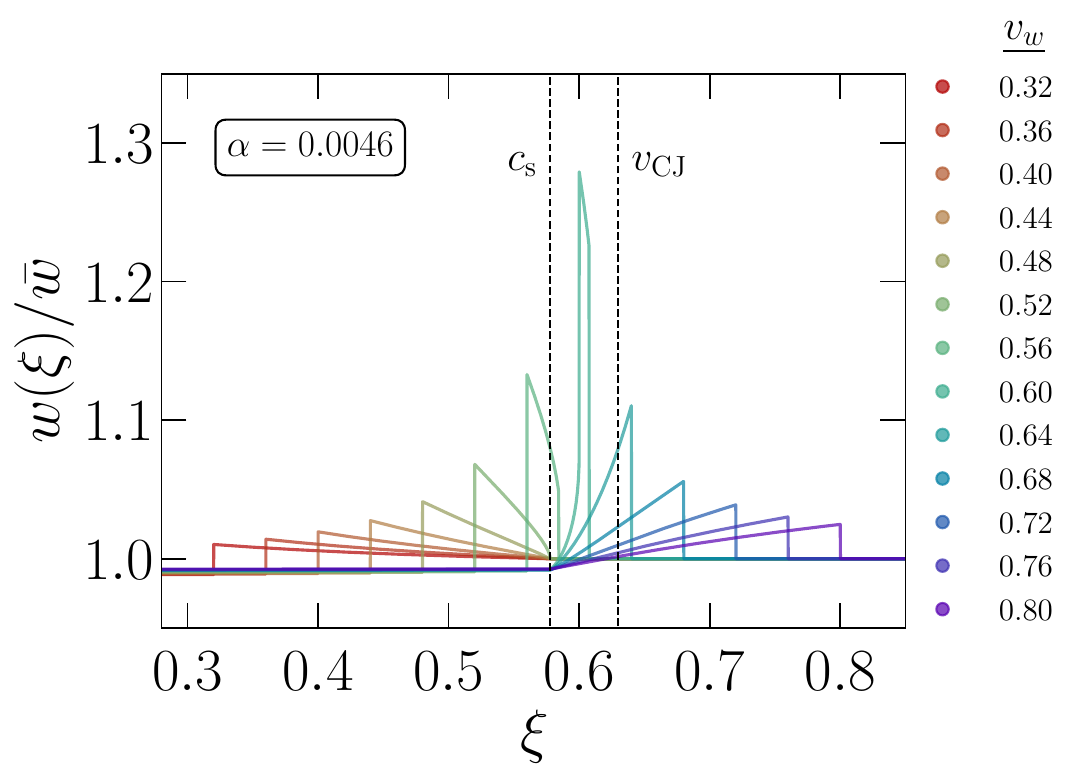}\\
    \includegraphics[width=0.42\linewidth]{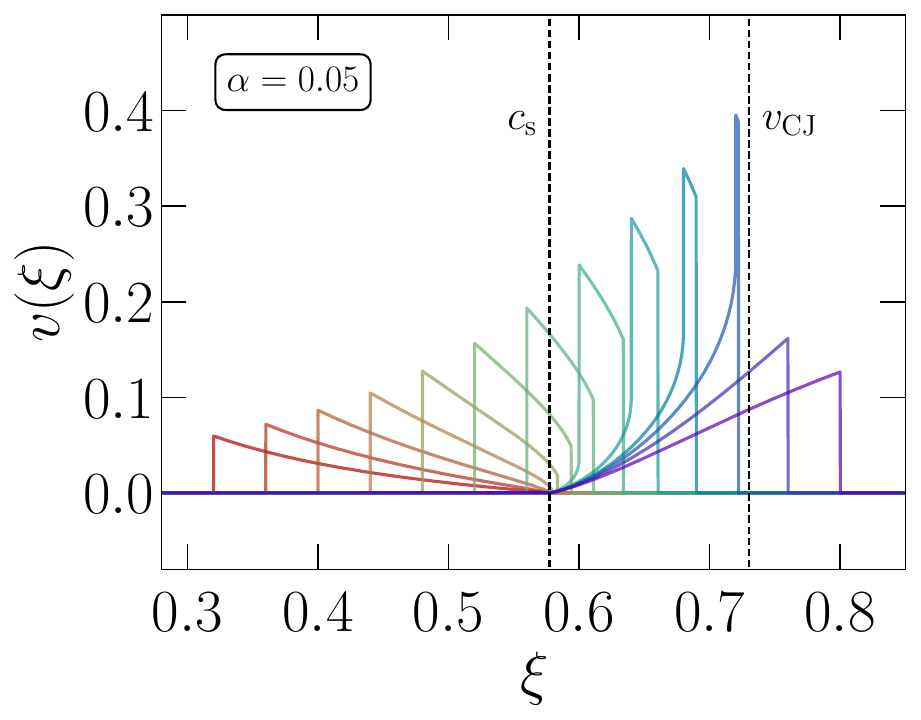}
    \includegraphics[width=0.48\linewidth]{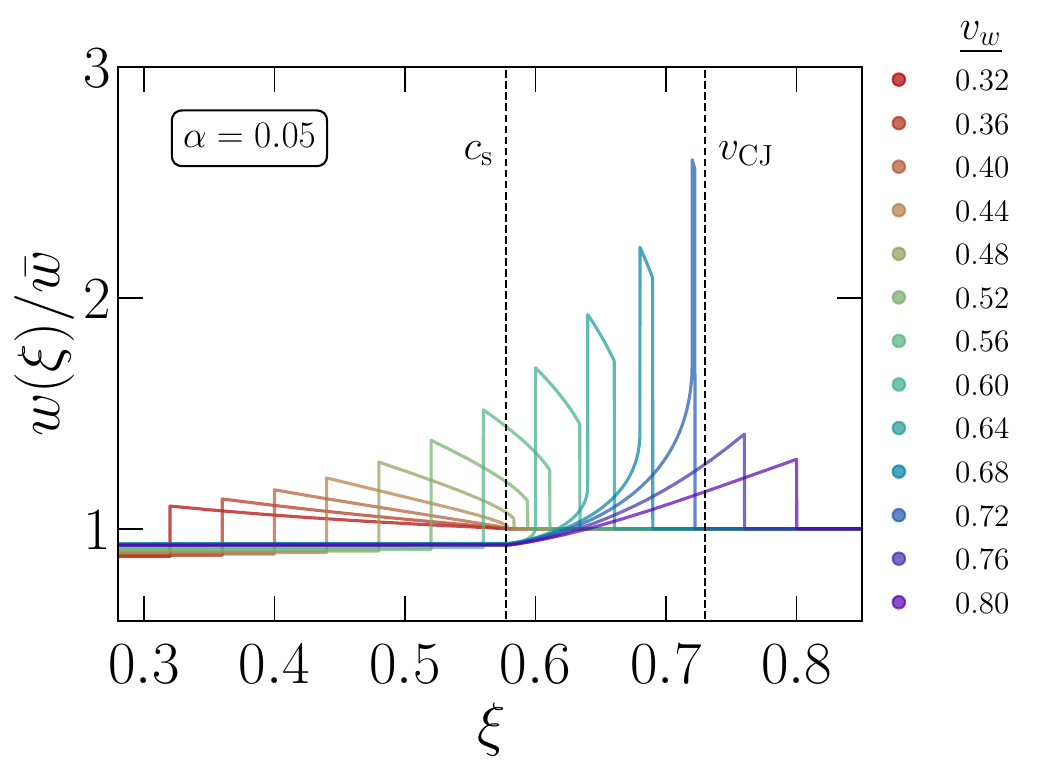}\\
    \includegraphics[width=0.43\linewidth]{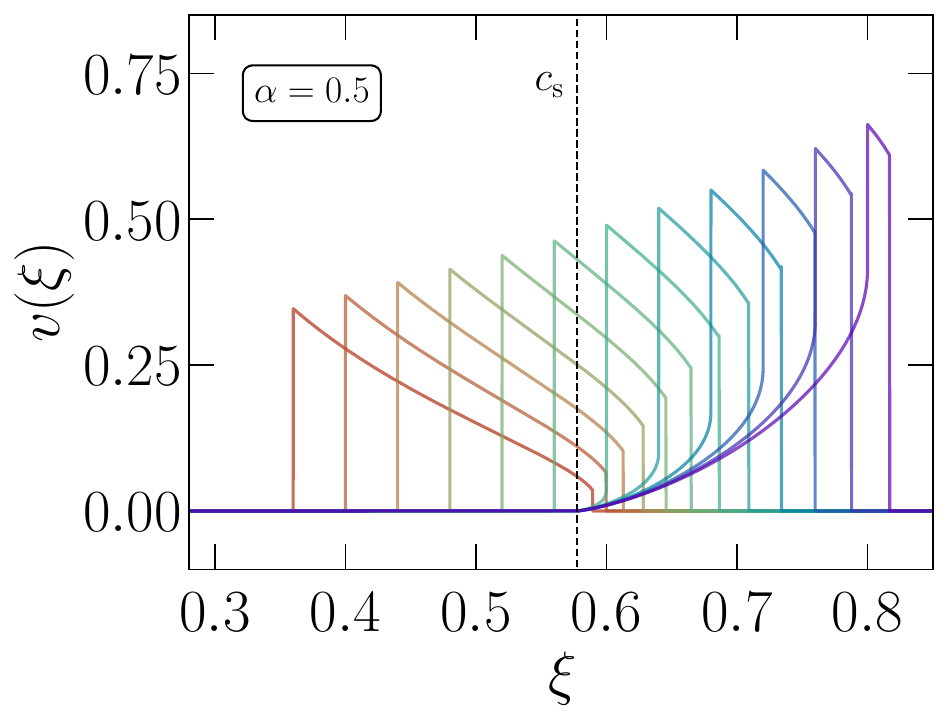}
    \includegraphics[width=0.47\linewidth]{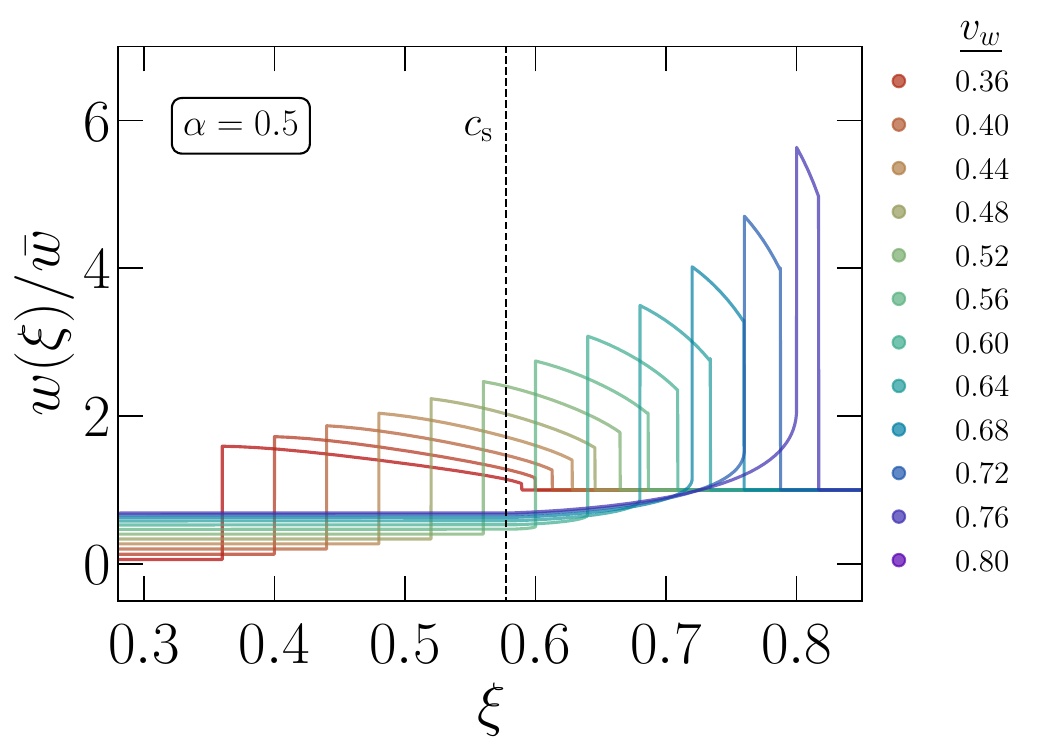}
    \caption{One-dimensional self-similar profiles of the fluid velocity (left columns)
    and enthalpy (right columns) perturbations for a single 
    bubble as a function of $\xi \equiv r/(t - t_n)$
    for weak (upper panel), intermediate (middle panel),
    and strong (lower panel) PTs, and for a range
    of wall velocities.
    The profiles are computed using {\sc CosmoGW} \cite{cosmogw}.
    Vertical dashed lines indicate the speed of sound $\cs$
    and the Chapman-Jouget velocity $v_{\rm CJ}$ \cite{Espinosa:2010hh}.
    These profiles correspond to the set of simulated PTs
    presented in this work (see \Tab{tab:simulation_summary}).}
    \label{fig:1d_profiles}
\end{figure}

In this section, we study in more detail the convergence of
the numerical {\em reference} simulations of multiple bubbles used for the analyses
of \Sec{sec:results} by
comparing it to that of single-bubble simulations. 
This is useful because, for the latter, we know
exact solutions to compare them to: indeed,
we expect the self-similar profiles described in Ref.~\cite{Espinosa:2010hh} to develop in the simulations.
For reference, we show in \Fig{fig:1d_profiles} these
profiles, for the PTs considered in the present work (see \Tab{tab:simulation_summary}).
In order to evaluate the effect of the numerical resolution, we compare the kinetic energy of a single bubble $K_\xi$ obtained by direct integration of the self-similar profiles, with the kinetic energy obtained numerically before collisions, both from single-bubble and multiple-bubble runs.
In \Tab{tab:kappas}, we also present for reference
the values
of the efficiency $\kappa_\xi$ and the kinetic energy
fraction $K_\xi \equiv \kappa_\xi\, \alpha/(1 + \alpha)$ for a subset of the parameters used
in the simulations.
We also provide the values of the shock formation time,
$\tilde \tau_{\rm sh} \equiv \beta R_\ast/\sqrt{K_\xi}$,
which allows us to estimate the time at which we expect non-linearities to develop in the system.

\begin{table}[b]
    \centering
    {\footnotesize
    \begin{tabular}{|c|c|c|c|c|} \hline
    $\alpha$ & $\vw$ & $\kappa_\xi$ & $K_\xi$ & $\tilde \tau_{\rm sh}$ \\ \hline
    0.0046  & 0.36 & $9.73\times 10^{-3}$ & $4.46\times 10^{-5}$ & 253.34 \\
            & 0.44 & $1.68\times 10^{-2}$ & $7.69\times 10^{-5}$ & 192.89 \\
            & 0.52 & $3.79\times 10^{-2}$ & $1.73\times 10^{-4}$ & 128.44 \\
            & 0.60 & $1.40\times 10^{-1}$ & $6.40\times 10^{-4}$ & 69.48 \\
            & 0.68 & $2.70\times 10^{-2}$ & $1.24\times 10^{-4}$ & 179.17 \\
            & 0.76 & $1.50\times 10^{-2}$ & $6.87\times 10^{-5}$ & 268.51 \\  \hline
    0.05    & 0.36 & $9.10\times 10^{-2}$ & $4.33\times 10^{-3}$ & 25.69 \\
            & 0.44 & $1.36\times 10^{-1}$ & $6.47\times 10^{-3}$ & 21.02 \\
            & 0.52 & $2.11\times 10^{-1}$ & $1.00\times 10^{-2}$ & 16.87 \\
            & 0.60 & $3.30\times 10^{-1}$ & $1.57\times 10^{-2}$ & 14.02 \\
            & 0.68 & $3.43\times 10^{-1}$ & $1.63\times 10^{-2}$ & 15.58 \\
            & 0.76 & $1.55\times 10^{-1}$ & $7.40\times 10^{-2}$ & 25.88 \\  \hline
    0.5     & 0.36 & $4.66\times 10^{-1}$ & $1.55\times 10^{-1}$ & 4.29 \\
            & 0.44 & $5.33\times 10^{-1}$ & $1.78\times 10^{-1}$ & 4.01 \\
            & 0.52 & $5.91\times 10^{-1}$ & $1.97\times 10^{-1}$ & 3.81 \\
            & 0.60 & $6.46\times 10^{-1}$ & $2.15\times 10^{-1}$ & 3.79 \\
            & 0.68 & $6.89\times 10^{-1}$ & $2.30\times 10^{-1}$ & 4.16 \\
            & 0.76 & $7.00\times 10^{-1}$ & $2.33\times 10^{-1}$ & 4.61 \\  \hline
    \end{tabular}
    }
    \caption{Values of the efficiency $\kappa_\xi$, the kinetic energy fraction
    $K_\xi \equiv \kappa_\xi\, \alpha/(1 + \alpha)$,
    and the estimated shock formation time $\tilde \tau_{\rm sh} \equiv \beta R_\ast/\sqrt{K_\xi}$
    of single bubbles for a subset of the parameters used in the simulations
    (see \Tab{tab:simulation_summary}).
    These values have been computed numerically using {\sc CosmoGW} \cite{cosmogw}.
    For comparison to $\tilde \tau_{\rm sh}$,
    the simulations run for durations $\tilde t_{\rm end}
    = 32$
    in units of $\beta^{-1}$ and the PT
    ends at the time $\tilde t_0 \simeq 10$.
    }
    \label{tab:kappas}
\end{table}

\begin{figure*}[t]
    \centering
    \includegraphics[width=0.32\columnwidth]{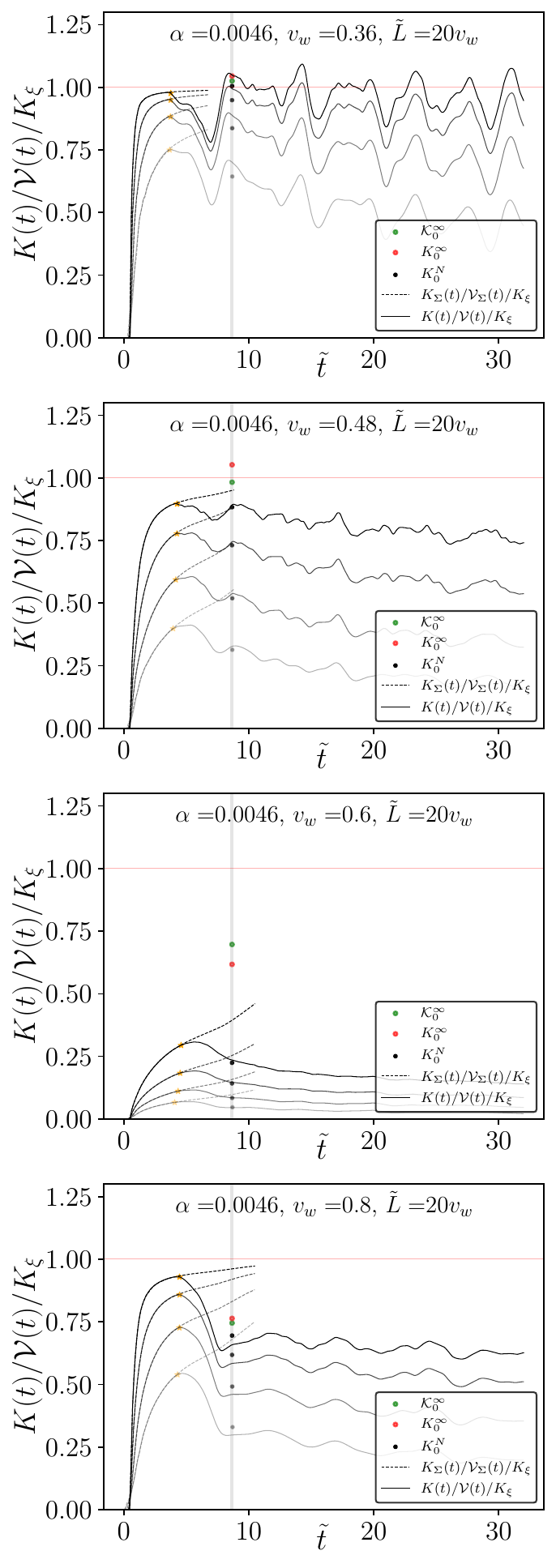}
    \includegraphics[width=0.32\columnwidth]{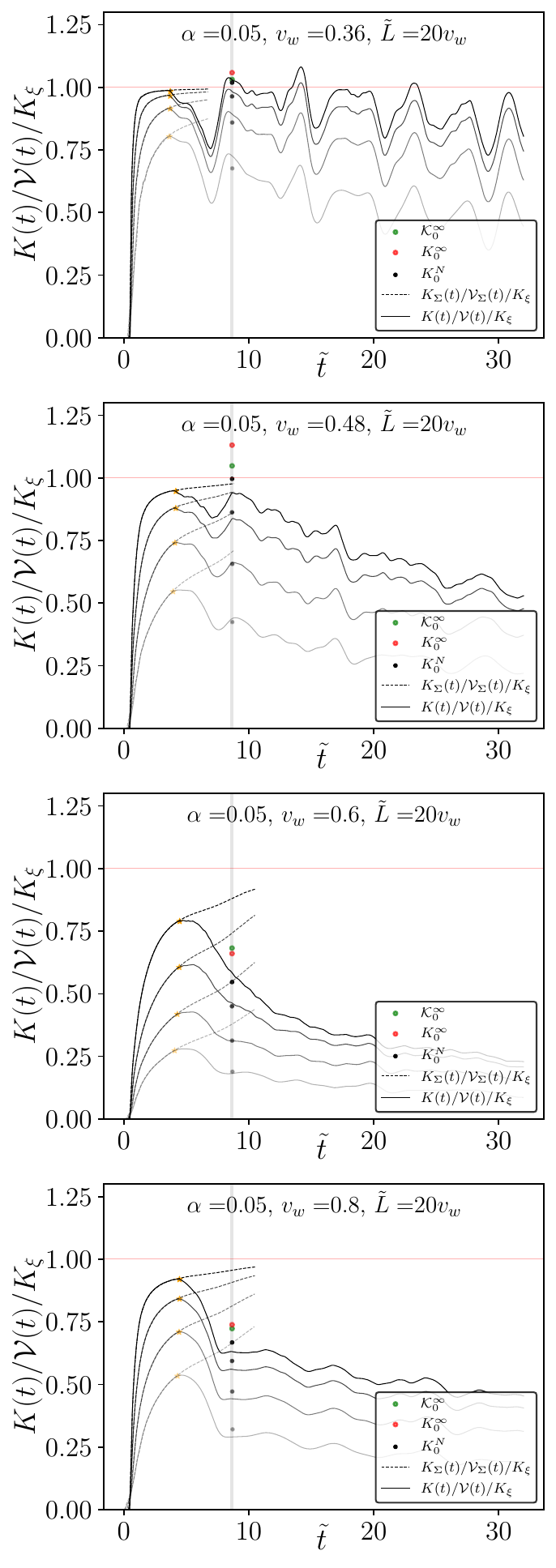}
    \includegraphics[width=0.32\columnwidth]{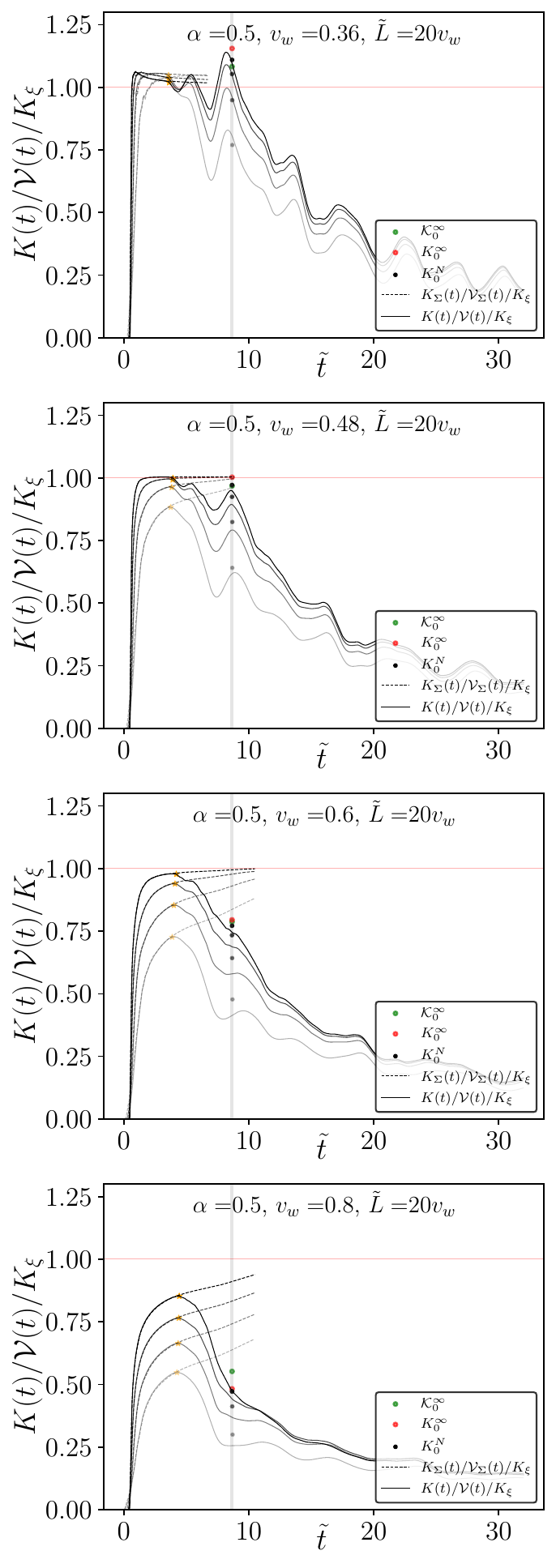}
    \caption{
    Time evolution of the kinetic energy fraction in the broken-phase volume $K(\tilde t)/{\cal V} (\tilde t)$ for multiple-bubble {\em reference} simulations (solid lines), normalized
    by the single-bubble $K_\xi$ (see values in \Tab{tab:kappas}),
    for different resolutions $N=\{64, 128, 256, 512\}$
    in increased opacity, and fixed box size $\tilde L/\vw = 20$.
    Results are shown for weak (left columns), intermediate (middle columns), and strong (right columns) PTs, and for a range of 
    wall velocities $\vw = \{0.36, 0.48, 0.6, 0.8\}$.
    Dashed lines correspond to the ratio $K_\Sigma(\tilde t)/{\cal V}_\Sigma (\tilde t)$ computed from the single-bubble simulations, such that the departure between the
    solid and dashed lines indicates the time $\tilde t_{\rm coll}$ when the first fluid-shell
    collision takes place in the multiple-bubble simulations.
    This time is marked with orange stars and the corresponding numerical self-similar profiles
    obtained are shown in \Fig{fig:1d_profiles_with_sims}.
    Black dots are the values of $K_0$ obtained
    from the fit $K(\tilde t) = K_0 (\tilde t/\tilde t_0)^{-b}$ studied in \Sec{decay_K2} for
    different $N$.
    Red and green dots correspond to the estimated values
    ${\cal K}_0$ [see \Eq{eq:calK_0 definition}] and $K_0^\infty$ (obtained
    from the convergence analysis of \Sec{sec:convergence}).
    Orange stars also correspond to the inverse of the factor ${\cal S}$ at the collision time $\tilde t_{\rm coll}$, used
    to compute ${\cal K}_0$ from $K_0$ [see \Eq{eq:calK_0 definition}].}
    \label{fig:KoK}
\end{figure*}

\begin{figure}
    \centering
    \includegraphics[width=1\linewidth]{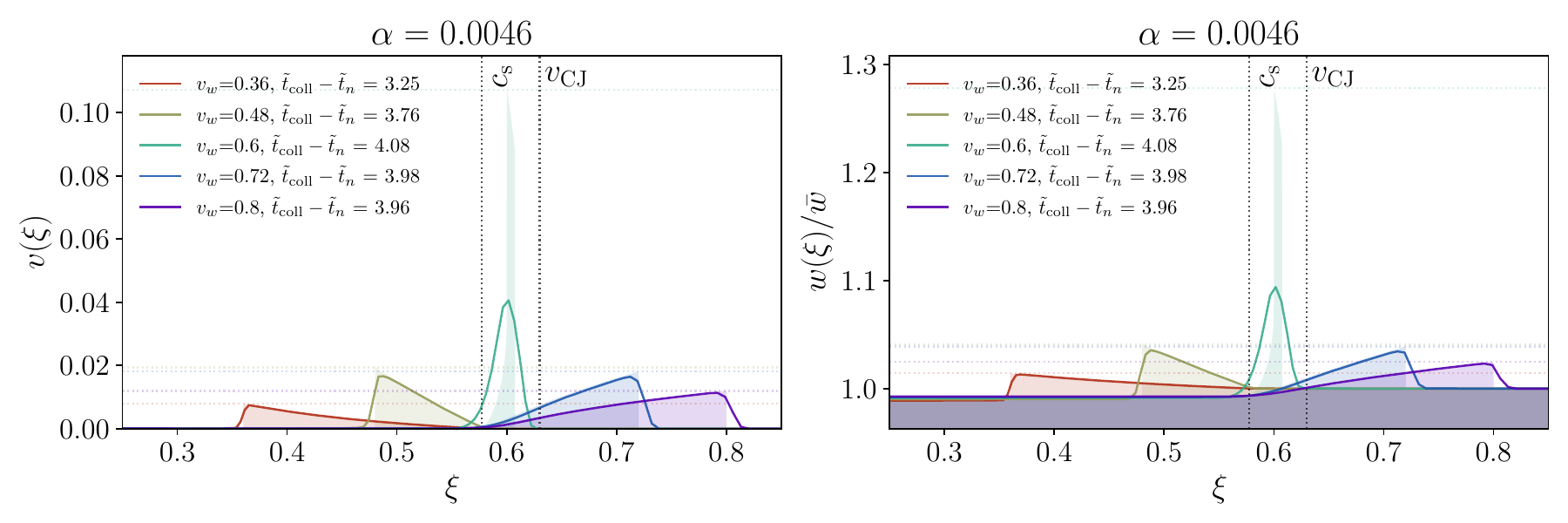}
    \includegraphics[width=1\linewidth]{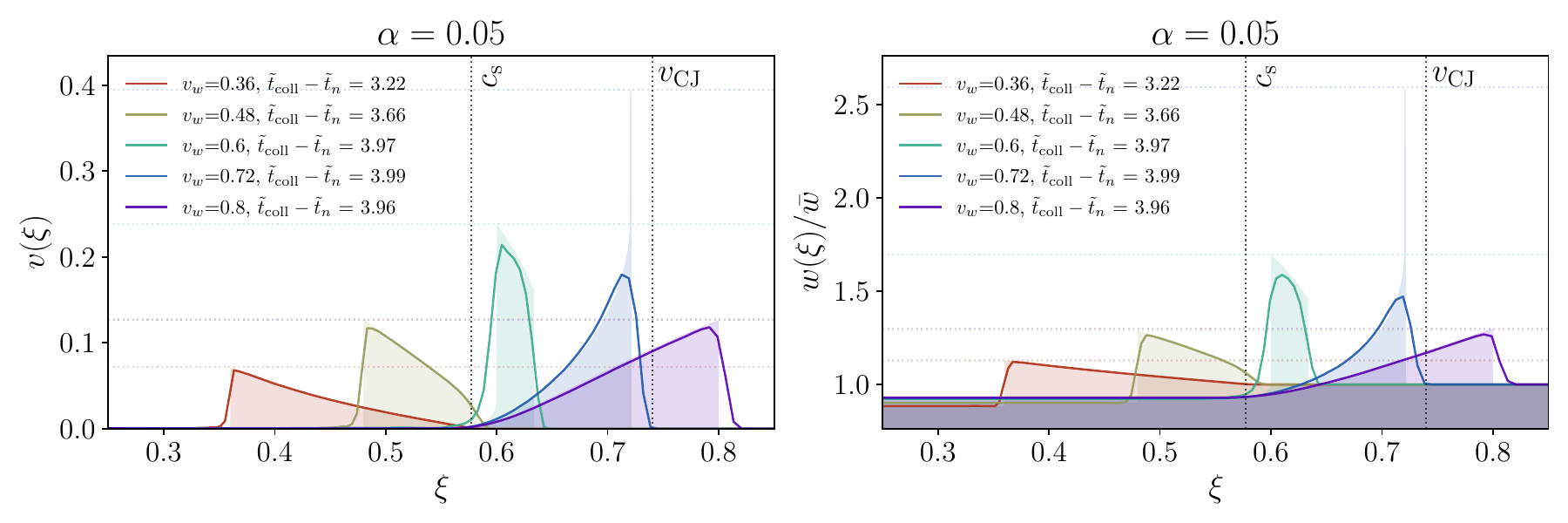}\\
    \includegraphics[width=1\linewidth]{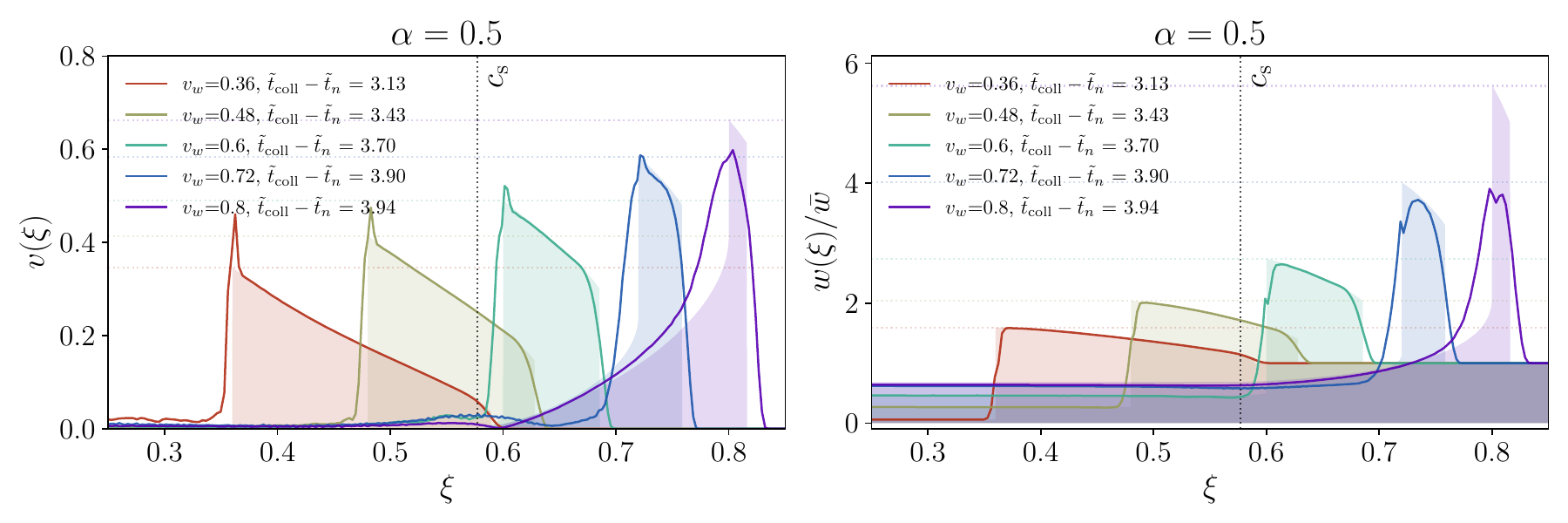}
    \caption{
    Self-similar profiles (filled regions) for fluid velocity (left columns)
    and enthalpy (right columns) perturbations  for a single bubble as a function of $\xi \equiv r/(t - t_n)$ for weak (upper panel), intermediate (middle panel),
    and strong (lower panel) PTs, for a selection of representative wall velocities, over-plotted with the fluid profiles achieved by the {\em single-bubble} and {\em reference}
    simulations at the collision time, $\tilde t_{\rm coll}$,
    when the kinetic energy fraction starts to be affected by collisions.
    The box size is $\tilde{L}/\vw=20$ and $N=512$.
    The self-similar profiles are a subset of those shown in \Fig{fig:1d_profiles}.}
    \label{fig:1d_profiles_with_sims}
\end{figure}

We have previously defined the kinetic energy fraction $K$ such
that $\bar \rho \, K(\tilde t) \equiv \bra{\rho_{\rm kin}(\xx, \tilde t)}$, where $\bra{\rho_{\rm kin}}$ corresponds to the kinetic energy
density averaged over the simulation volume $V$.
However, we note that $K_\xi$ for single bubbles, defined in \Eq{Kxi},
is taken as the average of the kinetic energy density fraction
induced by
a single bubble over the broken-phase volume.
Then, defining the ratio of the volume in the broken phase (bp) to the total volume,
\begin{equation}
    \mathcal{V}(\tilde{t})=\frac{V_\mathrm{bp}}{V}\,,
\end{equation}
we can define the analog of $K_\xi$ for multiple bubbles
as the ratio $K(\tilde t)/{\cal V}(\tilde t)$.
Before fluid shells collide,
and in the limit of infinite resolution, this ratio should be identical to $K_\xi$ after a very short transient period, over which the fluid profiles develop.
Deviations from $K_\xi$ before collisions are thus artifacts due to numerical inaccuracy.
We plot the ratio $K(\tilde t)/[{\cal V}(\tilde t) K_\xi]$
as solid lines in \Fig{fig:KoK} for all four resolutions $N\in\{64,\,128,\,256,\,512\}$.
Orange stars mark the time of first collision $\tilde t_{\rm coll}$,
where we assume that the maximum possible degree of convergence of the self-similar profiles is reached, as at later times collisions affect
the development of the fluid-shell profiles.
From \Fig{fig:KoK}, one can appreciate that $K(\tilde t)/{\cal V}(\tilde t)$ largely differs from $K_\xi$ even before the collision time, for the parameter sets for which the self similar profiles approach thin hybrids, which is a clear effect of under-resolution.
We show in \Fig{fig:1d_profiles_with_sims} the single-bubble profiles found in the simulations at
$\tilde t_{\rm coll} - \tilde t_n$, where $\tilde t_n$ is the time of nucleation,
compared to the expected ones in the limit of infinite resolution,
already shown in \Fig{fig:1d_profiles}.
We note that in the {\em single-bubble} runs, the accuracy of the self-similar profiles keeps improving
at times after $\tilde t_{\rm coll}$, as can be seen
by the fact that the dashed lines in \Fig{fig:KoK} become closer to one 
(see also discussion in Ref.~\cite{Jinno:2022mie}).

In the multiple-bubble runs, we can similarly define $\bar \rho \, K_i (\tilde t) = \bra{\rho_{{\rm kin}, i} (\xx, \tilde t)}$ as the kinetic energy fraction
of each of the single bubbles $i$ before their corresponding first collision,
and define the ratio $K_i(\tilde t)/{\cal V}_i (\tilde t)$,
where ${\cal V}_i$ corresponds to the fractional
broken-phase volume occupied by each
bubble $i$.
To monitor the time-dependence of $K_i$, we simulate single bubbles nucleated at the center of the simulation box (see {\em single-bubble} runs in \Tab{tab:simulation_summary}).
As the convergence of the single-bubble profiles depends on the resolution
in $\xi \equiv r/(t - t_i)$, where $t_i$ is the nucleation time of the bubble $i$ and $r$ the radial distance to the nucleation center,
we empirically find that doubling the resolution from $N$ to $2N$ is
equivalent
to evaluating the profile at time $2(\tilde t - \tilde t_i)$ (see also discussion in Ref.~\cite{Jinno:2022mie}).
Hence,
the kinetic energy of single-bubble simulations (with $t_i = 0$) obeys $K_i^{2N}(\tilde t)/\mathcal{V}_i(\tilde{t}) = K_i^N(2\tilde t)/\mathcal{V}_i(\tilde{t})$ to an excellent degree
and it suffices to run single-bubble simulations for the largest resolution
$N=512$.
These simulations are run approximately until the front of the fluid profile collides with its own mirror image at the edge of the simulation box,
which occurs around $\tilde{t}_{\mathrm{end}}^{\rm sb} = \tilde{L}v_w/[2\,\max(c_{\mathrm{s}}, v_{\mathrm{w}})]$
and determines the end of the dashed lines in \Fig{fig:KoK}.

Then, in the full simulations and before fluid shells collide, the state of the simulation is exactly the superposition of single bubbles nucleated at times $\tilde{t}_i < \tilde t$ in the bubble nucleation history. We thus construct the sum
\begin{equation}
\label{eq:K_Sigma}
    K_{\Sigma}(\tilde{t}) \equiv \sum_{i : \{\tilde{t}_i < \tilde{t}\}}{ K_i(\tilde{t}-\tilde{t}_i)}\,,
\end{equation}
which corresponds to the expected kinetic energy fraction for multiple-bubble simulations in the hypothetical case that no
bubble would collide, following the
bubble nucleation history up to time $\tilde{t}$.
Then, before the first fluid-shell collision occurs,
we have that $K_{\Sigma}(\tilde{t}) = K(\tilde t)$,
while $K(\tilde t)$ starts to deviate from $K_\Sigma (\tilde t)$ after the
first collision at $\tilde t_{\rm coll}$.
Similarly, we can construct the fractional broken-phase volume
occupied by the superposition of single bubbles as 
\begin{equation} 
    \mathcal{V}_{\Sigma}(\tilde{t}) \equiv \sum_{i : \{\tilde{t}_i < \tilde{t}\}}{ \mathcal{V}_i(\tilde t-\tilde t_i)} \,,
\end{equation}
which can become larger than one, as it ignores interactions between bubbles.
However, the ratio $K_\Sigma/{\cal V}_\Sigma$ is bounded by $K_\xi$.\footnote{This is the case for all the considered PTs, with the
exception of strong PTs with $v_w=0.36$ and $v_w=0.4$, where
values $K_\Sigma/\mathcal{V}_\Sigma \gtrsim K_\xi$ are found before collisions due to numerical inaccuracy
(see \Fig{fig:KoK}).}

We plot the time evolution of the ratio $K_\Sigma/{\cal V}_\Sigma$
as dotted lines in \Fig{fig:KoK}, using the nucleation history of the
{\em reference} multiple-bubble simulations with $\tilde L/\vw = 20$.
As for $K(\tilde t)/{\cal V}(\tilde t)$, the under-resolution of the self-similar profiles for the parameter sets leading to hybrids causes the ratio $K_\Sigma/{\cal V}_\Sigma$ to be very different from $K_\xi$ in these cases.
However, this quantity
indicates the global degree of 
convergence of the full multiple-bubble simulations in the hypothetical case
that all bubbles keep evolving without interacting with each other.
Therefore, contrary to $K(\tilde t)/{\cal V}(\tilde t)$, this ratio does approach $K_\xi$ as the simulation proceeds after collisions and the resolution in $\xi$ improves.
Furthermore, as expected, the
ratio $K/{\cal V}$ computed in the multiple-bubble simulations 
is initially identical to $K_\Sigma/{\cal V}_\Sigma$
at times $\tilde t < \tilde t_{\rm coll}$.
However, as collisions take place,
we clearly see in \Fig{fig:KoK} that
these quantities
deviate from each other,
as a consequence of mainly four phenomena: {\em (1)} the self-similar profiles
stop converging towards the expected ones when collisions take place;
{\em (2)} oscillatory conversion between thermal and kinetic energy due to collisions and the development of sound waves;
{\em (3)} upon and after collisions, the fluid self- and inter-shell interactions may be non-linear and result in the kinetic energy decay studied in \Sec{decay_K2}, which is again affected
by the previous two effects; and {\em (4)} numerical viscosity also
results in damping of the kinetic energy.
The first and last phenomena are purely numerical, while the remaining two
are physical effects, which might also be affected by the numerical accuracy.

Because of {\em (1)}, and since the kinetic energy of the uncollided bubbles is in general
underestimated with respect to $K_\xi$ because of under-resolution,
the value of $K(\tilde t)/\mathcal{V}(\tilde t)$ found
after a short transient period dominated by collisions is in general underestimated.
This quantity is very relevant for us, since it determines the GW production in the period of time in which we calculate it, namely in the last part of the simulation from $\tilde t_{\rm init} = 16$ to $\tilde t_{\rm end} = 32$. 
The convergence analysis performed in this appendix allows us to find a way to
attempt to compensate for the underestimation of the kinetic energy fraction due to insufficient resolution at $\tilde t > \tilde t_{\rm coll}$: namely,
multiplying $K(\tilde t)$
by the factor ${\cal S} = {\cal V} (\tilde t_{\rm coll}) \, K_\xi/K (\tilde t_{\rm coll})$,
effectively correcting it to the expected value $K_\xi$ at the time when
collisions take place.
In particular, the kinetic energy fraction at the time when the PT ends,
$\tilde t_0$, can be corrected to the following value
\be\label{eq:calK_0 definition}
\mathcal{K}_0 = \mathcal{S} \, K_0 = \frac{{\cal V}(\tilde t_{\rm coll})\, K_\xi}{K(\tilde t_{\rm coll})} \, K_0 \,.
\ee

In \Fig{fig:kappa_eff_kappa}, we plot $\mathcal{K}_0$ obtained
for numerical resolutions $N\in\{64,\,128,\,256,\,512\}$.
The relative differences between $\mathcal{K}_0$ in the $N = 512$ simulations and $\mathcal{K}_0$ in the $N = 256$ simulations are smaller than the respective ones
for $K_0$. 
Therefore, ${\cal K}_0$ seems to be a more robust estimate of the actual value of $K_0$
in the continuum limit.
We note that the computation of ${\cal K}_0$ takes into account the expected convergence
of the numerical results of the self-similar profiles.
Furthermore, the resulting values of ${\cal K}_0$ are closer to $K_\xi$ than $K_0^\infty$
and therefore are more conservative estimates, reducing potential deviations with respect to $K_\xi$ that
could originate from our extrapolation scheme.
However, one needs to keep in mind that the under-resolution at the time of
collisions might strongly affect the following evolution of the kinetic
energy, when non-linearities dominate the dynamics.

%%%%%%%%%%%%%%%%%%%%%%%%%%%%%%%%%%%%%%%
\section{Preliminary results for vorticity} \label{sec:vort}
%%%%%%%%%%%%%%%%%%%%%%%%%%%%%%%%%%%%%%%

In this section, we present some preliminary measurements of
the fluid vorticity in our numerical simulations.

\subsection{Vorticity on the lattice}
\label{sec:Vorticity on the lattice}
The vorticity is computed as 
\begin{equation}
\nabla \times \mathbf{v}=\left(\frac{\partial  v_z}{\partial y}-\frac{\partial v_y}{\partial z}\right) \hat{\boldsymbol{x}}+\left(\frac{\partial v_x}{\partial z}-\frac{\partial v_z}{\partial x}\right) \hat{\boldsymbol{y}}+\left(\frac{\partial v_y}{\partial x}-\frac{\partial v_x}{\partial y}\right) \hat{\boldsymbol{z}}\,.
\end{equation}
On the lattice, we approximate the derivatives
using first-order central differences,
\begin{equation}
\label{eq:central derivative}
\frac{\partial v_i}{\partial \tilde x_j} (\tilde \xx) \simeq
\frac{v_i\left[\tilde \xx + \delta \tilde x \, \hat{x}_j\right]
- v_i \left[\tilde \xx - \delta \tilde x \, \hat{x}_j \right]
}{2 \, \delta \tilde x}\,,
\end{equation}
where $\delta \tilde x$ is the uniform grid spacing.

With this choice of derivative operator, the magnitude of the curl $|\tilde \nabla \times \mathbf{v}|$ is computed at every grid point. 2D simulation slices at different times are shown in  the lower panels of \Fig{fig:bubbles}.
Note that the definition of the numerical derivative operator of
\Eq{eq:central derivative} inevitably introduces potentially large vorticity at points where the velocity field varies considerably from lattice site to lattice site. This occurs, e.g., around the bubble shock fronts where discontinuities are present.
Ideally, the velocity gradients are aligned with the radial direction, in which case no vortical component is present. 
However, on the lattice, artifacts may arise from the discretization, causing rather strong vorticity to appear at and just around the bubble walls.
This is clearly seen in the lower left frame of \Fig{fig:bubbles}.
The numerical nature of this vorticity is nevertheless clear from the observation that the vortical structure, as we traverse around bubble wall, is seen to inherit the symmetry of the lattice (see, in particular, the largest bubble in the center).
Furthermore, mostly small but spurious oscillations of the fluid velocity occur at the bubble wall interface.
These oscillations additionally give rise to extremely local but very steep velocity gradients, potentially showing up as spurious vorticity with large amplitude, confined to very small scales. 

In the lower panel of \Fig{fig:bubbles}
it is observed that production of vorticity occurs, upon collisions,
at the interface of a sound shell from one bubble crossing over the bubble wall of another, as seen, e.g., immediately to the right of the top section of the central bubble in the left column. The velocity field in the upper panel is included to make vorticity production possible to correlate with the velocity field by eye. In this sense, the resulting vorticity pattern initially appears to track the \emph{sweeping} of this sound-shell-bubble-wall-crossing interface over time.

We have described the expected and observed presence of spurious vorticity components associated with the choice of the derivative operator and the lattice structure, and spurious osculations around the bubble wall interface. These are very localized effects and do not contribute meaningfully to large-scale vorticity correlated over macroscopic scales. Therefore, vorticity components that emerge from numerically induced oscillations and limited grid resolution
are expected to contribute mostly to the UV part of the velocity spectra.
The development of a transverse vortical component should
be visible in the spectra of the velocity fields
decomposed into longitudinal and transverse contributions, as we present in the following (see \Fig{fig:vert}).
In particular, physical macroscopic contributions should distinguish themselves from numerical contributions through a separation of scales.

\subsection{Velocity power spectra}

\begin{figure*}
    \centering
    \includegraphics[width=0.4\columnwidth]{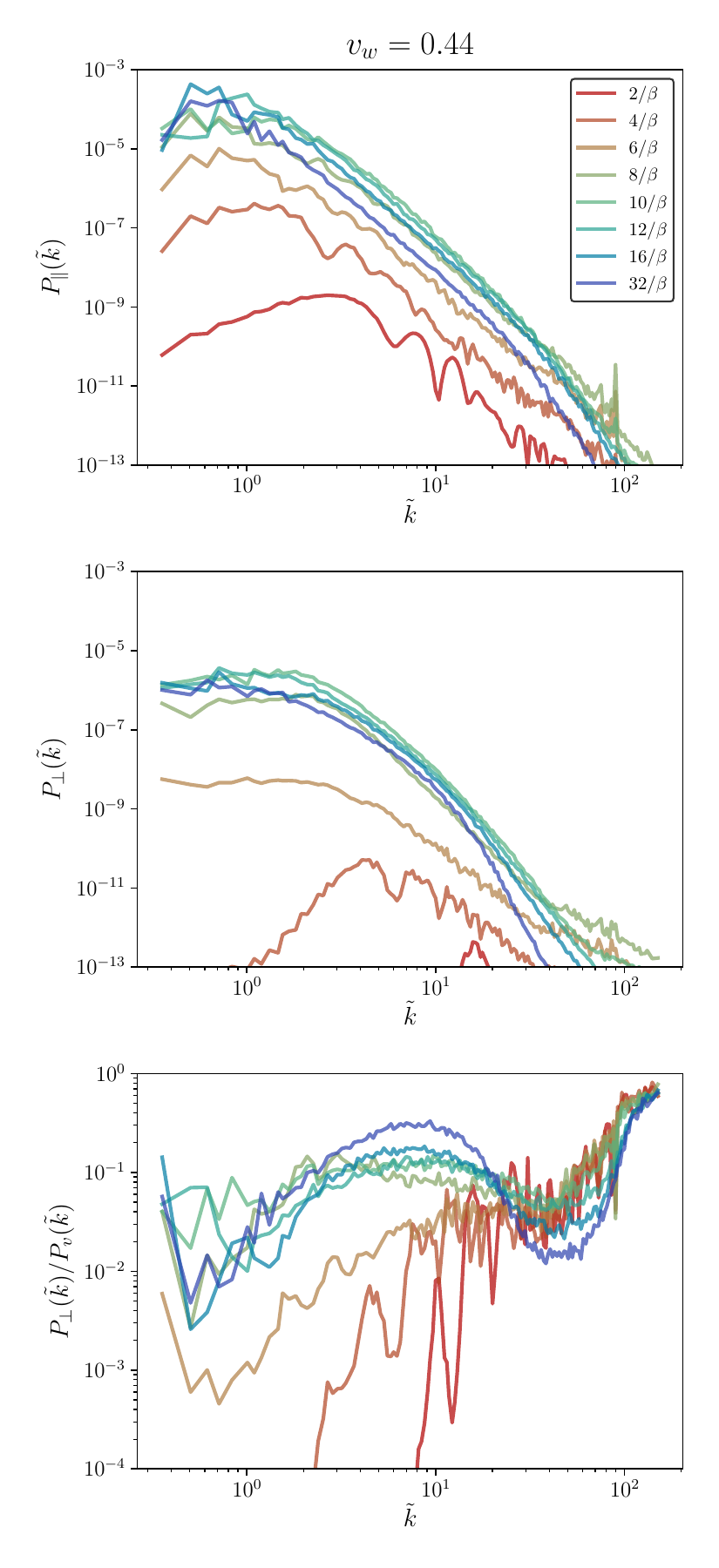}
    \includegraphics[width=0.4\columnwidth]{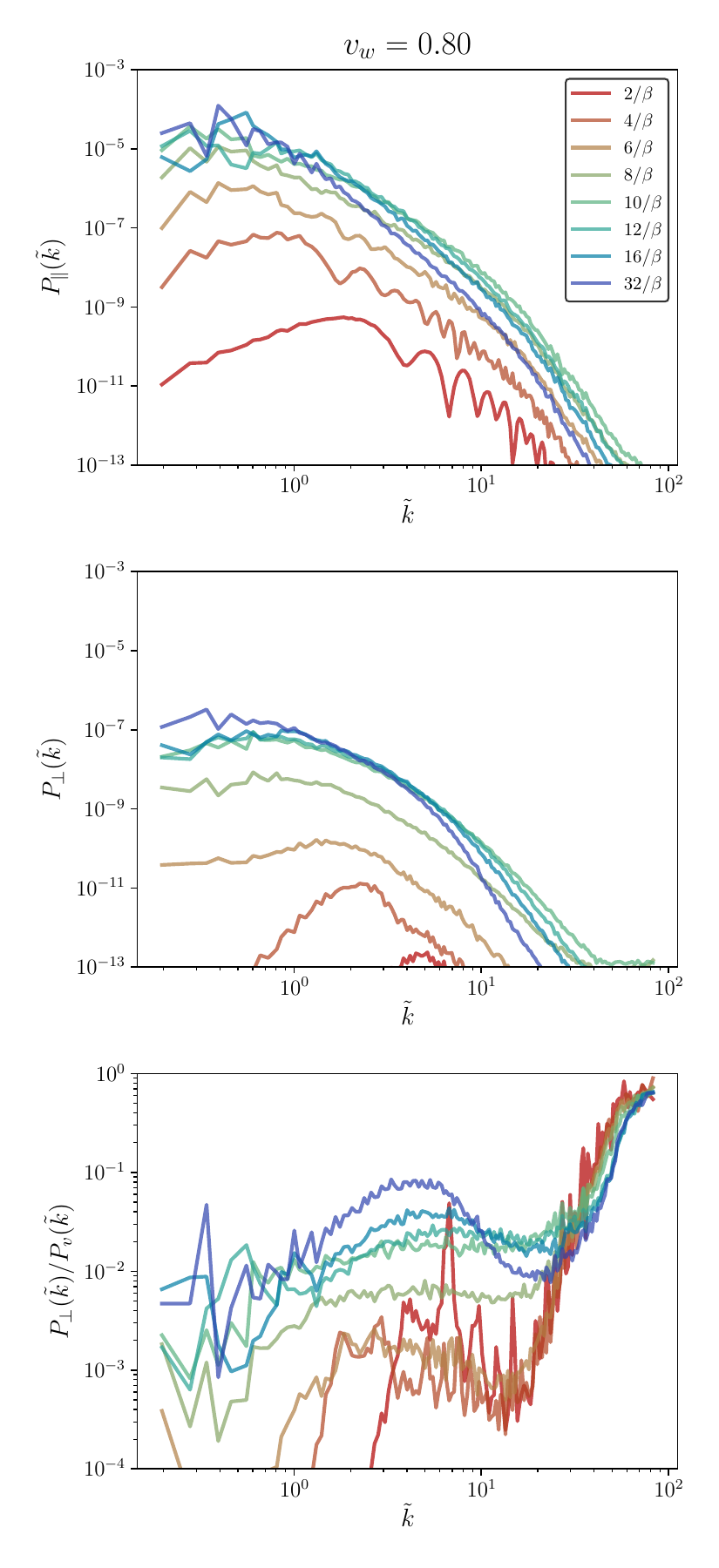}
    \caption{
      Transverse $P_\perp (k)$ and longitudinal $P_\parallel (k)$ velocity power spectra (in arbitrary units) of the fluid for a strong PT ($\alpha=0.5$) with wall velocities $v_w = 0.44$ (left columns) and $\vw = 0.8$ (right columns).
      The panels show the power in longitudinal (top) and vortical (middle) modes as 
well as the fraction of power in vortical modes (bottom), $P_\perp(k)/P_v(k)$. The different lines correspond to different times in 
the simulation $\tilde t \in (2, 32)$.
For reference, we remind that the first bubbles nucleate around $\tilde t \simeq 0$, the first collisions happen around $\tilde t \simeq 5$, and 
the simulation is filled with the broken phase at $\tilde t_0 \simeq 10$.
    \label{fig:vert}
    }
\end{figure*}

We generate the Fourier transform $v(\kk)$ of the fluid velocity field and then 
extract the power spectral density from the two-point
correlation in Fourier space \cite{MY75}
\be
\bra{v_i (\kk) \, v_i^\ast (\kk')} = (2 \pi)^3 \, \delta^3(\kk - \kk') \, P_v (k)\,,
\ee
where the ensemble average is performed over
momenta with the same absolute value $|\kk| = |\kk'|$ due
to statistical homogeneity and isotropy.
We also construct the power spectral density
of the longitudinal modes 
\be
\bra{\hat k_i \, v_i (\kk) \, \hat k_j' \, v_j^\ast (\kk')}
= (2 \pi)^3 \, \delta^3(\kk - \kk') \, P_\parallel (k)\,,
\ee
where $\hat{\kk} = {\rm saw}(\kk) / |{\rm saw}(\kk)|$ is a unit vector according to the {\em saw} description, as discussed in \Sec{subsec:Updates to the simulation} [see \Eq{eq:momenta mapping}].
Likewise, we extract the
vortical component of the spectra
\begin{equation}
    \Bigl\langle\bigl[\hat \kk \times \vv (\kk) \bigr]_i\bigl[\hat \kk' \times \vv^\ast (\kk')\bigr]_i\Bigr\rangle
    = (2 \pi)^3 \delta^3(\kk - \kk') \, P_\perp (k)\,,
\end{equation}
such that $P_v (k) = P_\perp(k) + P_\parallel(k)$.

\FFig{fig:vert} shows the power in longitudinal and vortical modes as 
well as the fraction of power in vortical modes for strong PTs.
The power in longitudinal modes builds up when the 
first bubbles nucleate while vorticity requires collisions, as expected.
There are some artifacts 
in the UV even before collisions, which result from the discretization of space on a grid. 
Overall, the power in vorticity can reach values of around $P_\perp/P_v \simeq 0.3$ for the deflagration ($v_w = 0.44$),
while this fraction is somewhat smaller, $P_\perp/P_v \simeq 0.1$, for the hybrid ($v_w = 0.80$).
Still, a sizable fraction of vorticity is observable in both cases. 
Moreover, the power spectra in vortical modes appear somewhat less steep than the ones in longitudinal modes.
All this is consistent with the hypothesis that the energy loss observed 
over time for strong PTs is due to the decay of fluid kinetic energy into vortical motion and 
eventually turbulence. This point deserves further attention in the future.
We also have obtained velocity power spectra for weak and intermediate PTs, finding that $P_\perp/P_v < 10^{-3}$ for those scenarios. This indicates that turbulence becomes progressively more important the stronger the PT, and that for strong PTs ($\alpha=0.5$) it already has a dominant role in determining the hydrodynamical evolution after the PT ends. This observation underlines the importance of fully non-linear 3D simulations.

\bibliographystyle{JHEP}
\bibliography{biblio.bib}

\end{document}